\documentclass[letterpaper,8pt]{extarticle}  
\usepackage[authoryear,round]{natbib}

\usepackage{rotating}
\usepackage{pdflscape}
\usepackage{float}
\usepackage{import}
\usepackage{local}
\usepackage{amsfonts}
\usepackage{amsmath}
\usepackage{amssymb}
\usepackage{aas_macros}
\usepackage{physics}
\usepackage{longtable}
\usepackage{booktabs}
\usepackage[left=.65in,top=.65in,bottom=.65in,right=.65in]{geometry}
\usepackage{lipsum} 

\title{Giant Radio Galaxies from the Third Data Release of the LOFAR Two-metre Sky Survey (LoTSS-DR3): Physical Characteristics and Environments}

\author{%
\textbf{Souvik Manik \orcidlink{0000-0002-6794-7405},\textcolor{Accent}{\textsuperscript{1,3}} %
Netai Bhukta\orcidlink{0000-0002-1829-847X}\textcolor{Accent}{\textsuperscript{2}} 
Sabyasachi Pal \orcidlink{0000-0003-2325-8509},\textcolor{Accent}{\textsuperscript{1}} %
Sushanta K. Mondal\orcidlink{0000-0003-4353-8487}\textcolor{Accent}{\textsuperscript{2}} }\\
\begin{small}\textcolor{Accent}{\textsuperscript{1}}Midnapore City College, Kuturia, Bhadutala, Paschim Medinipur, West Bengal, 721129, India
\\ 
\textcolor{Accent}{\textsuperscript{2}}
Department of Physics, Sidho Kanho Birsha University, Ranchi Road, Purulia, 723104, India
\\ 
\textcolor{Accent}{\textsuperscript{3}}
Space, Planetary \& Astronomical Sciences \& Engineering (SPASE), IIT Kanpur, 208016, India
\\ 
\end{small}
}
\date{}
\begin{document}
\justifying
\maketitle


\section{Abstract}
Giant radio galaxies (GRGs) are radio-loud active galactic nuclei (AGN) with projected linear sizes exceeding 0.7 Mpc, offering insights into the evolution of large-scale radio jets and their environments. We present 2333 GRGs from the LOFAR Two-metre Sky Survey Data Release 3 (LoTSS-DR3), with projected sizes of 0.7--2.9 Mpc, redshifts of $0.126 \leq z \leq 3.517$, and 150-MHz radio luminosities of $1.72 \times 10^{24}$--$3.04 \times 10^{28}$ W Hz$^{-1}$. Of these, 726 ($\sim 31\%$) lie at $z \geq 1$. We investigate their radio properties, lobe energetics, equipartition pressures, black hole masses, mid-infrared colours, and environments. Equipartition pressure decreases with increasing projected size, consistent with declining lobe energy density as radio structures expand, while its positive correlation with redshift persists after controlling for radio power and projected size. WISE colours highlight the diversity of the host-galaxy population, with approximately 75\% of sources with available photometry falling within the empirical AGN/HERG region. For BCG-hosted GRGs, physical projected size shows no significant dependence on cluster mass. Although the normalized projected half-size, $D/(2r_{500})$, decreases with increasing cluster mass, this trend is consistent with the expected self-similar scaling of cluster radius rather than an independent suppression of radio-source growth in massive clusters. The substantial high-redshift population provides an opportunity to investigate GRGs at early cosmic epochs, although surface-brightness dimming and inverse-Compton losses against the cosmic microwave background can limit their detectability.
\begin{doublespacing}

\textit{Keywords}: {Active galactic nuclei; Giant Radio Galaxies; Radio sources; Active galaxies; Radio astronomy; Radio continuum emission}
\begin{multicols*}{2} 
\section{Introduction}
\label{sec:intro}
Radio galaxies (RGs) are powered by active galactic nuclei (AGN) that launch bipolar, relativistic, magnetised jets into the ambient medium \citep{Blandford2019}. Their extended radio emission provides a powerful probe of the connection between AGN activity, host galaxy properties, and the large-scale environment \citep{Magliocchetti2022}. RGs are commonly classified into two morphological classes following the Fanaroff--Riley scheme, with FRII radio galaxies characterised by well-collimated jets terminating in compact hotspots that produce edge-brightened radio lobes, whereas FRI sources display edge-darkened radio emission generated by jets that gradually decelerate with distance from the nucleus \citep{FR74}. At the extreme end of the radio galaxy size distribution are giant radio galaxies (GRGs), conventionally defined as sources with projected linear sizes exceeding 700~kpc \citep{Willis1974, Kuzmicz2018, reviewgiant}. Their enormous physical extent makes them valuable laboratories for investigating the long-term evolution of relativistic jets, AGN life cycles, and the role of the large-scale environment in shaping radio-source evolution.

Wide-area radio surveys, including FIRST \citep{FIRST1995}, WENSS \citep{WENSS1997}, NVSS \citep{NVSS1998}, SUMSS \citep{SUMSS2003}, TGSS \citep{intematgss}, and RACS \citep{RACS2020}, have enabled the discovery of several thousand giant radio galaxies over the past two decades \citep{Is99, Ku12, Dab17, Kuzmicz2018, Da20a, Ku21, An22, bhukta24, Andernach25, manik25}. The advent of the LOw-Frequency ARray \citep[LOFAR;][]{LOFAR2013}, with its exceptional sensitivity to diffuse, low-surface-brightness emission and high angular resolution at low frequency, has transformed studies of giant radio galaxies by substantially expanding the known population. In particular, the LOFAR Two-metre Sky Survey \citep[LoTSS;][]{Shimwell2019, Shimwell2022} has increased the number of known GRGs to nearly $10^4$ \citep{Da20b, Si22, Oei2023a, mos24}, including several exceptional systems extending over multiple megaparsecs, such as \textit{Alcyoneus} ($\sim$5 Mpc) \citep{Alcyoneus} and \textit{Porphyrion} ($\sim$7 Mpc) \citep{Porphyrion}. While this dramatic increase in sample size has enabled the first robust statistical studies of the GRG population, the physical conditions that allow only a small fraction of radio galaxies to reach such enormous sizes remain poorly understood.

The rapid growth of homogeneous GRG samples has enabled the first statistical investigations of their population properties. Using a sample of 2060 GRGs at $z < 0.4$, \citet{Oei2023a} showed that the linear size distribution follows a steep cumulative power law (tail index $\xi = 3.5 \pm 0.5$), indicating that GRGs are intrinsically rare in the local Universe. Extending these studies to higher redshift, \citet{Si22} and \citet{mos24} found that giant radio galaxies become progressively more abundant relative to local expectations. Furthermore, \citet{Si22} demonstrated that the size distributions of both GRGs and smaller radio galaxies are well described by similar exponential functions, suggesting that GRGs are more likely to represent the extreme end of the general radio galaxy population rather than a physically distinct class.

The growth of giant radio galaxies is thought to be governed by a combination of jet power, source lifetime, host galaxy properties, and the density and structure of the surrounding environment \citep{Turner2015, Hardcastle2018, Turner2023, Giri2025a, Giri2025b}. GRGs are also observed to possess intrinsically more powerful radio jets than typical radio galaxies, which may supply the additional thrust needed to drive lobe expansion to megaparsec scales \citep{Wiita1989}. Evolutionary models indicate that giant sizes are generally favoured when powerful and/or long-lived jets propagate through relatively low-density environments, allowing radio lobes to expand to megaparsec scales over extended timescales; for example, \citet{Ma08} inferred jet advance speeds of $\sim$0.1c--0.2c combined with an exceptionally low ambient density in the record-holding giant J1420$-$0545 ($\sim 4.7$ Mpc). Sustained or recurrent AGN activity has also been proposed as an important factor in driving the exceptional sizes of these systems \citep{Da20a, Dabhade2025ddrg}. 
 
Recent three-dimensional relativistic MHD simulations by \citet{Giri2025a, Giri2025b} investigated the evolution of radio jets over a wide range of jet powers and environmental conditions, reproducing several key features of the observed radio continuum and polarization properties of their lobes. Their results show that FR~II jets propagating through steep density gradients near the outskirts of the ambient medium exhibit enhanced longitudinal expansion, allowing them to reach megaparsec scales within $\sim10^{8}$~yr. The simulations also show that GRG lobes remain overpressured relative to the unshocked ambient medium, while a shocked gas shell regulates their expansion. A change in the expansion behaviour appears at a one-sided distance of $\sim350$~kpc, marking a possible dynamical transition between normal radio galaxies and GRGs. These results indicate that the emergence of megaparsec-scale radio structures does not depend on jet power alone, but on the combined effect of jet energetics, the density structure of the environment, and the evolutionary stage of the source. Synthetic radio continuum and polarization maps derived from the same simulations show that the observed properties of GRGs retain signatures of their dynamical history. The morphology of the lobes and cocoons depends strongly on the surrounding environment: asymmetric environments produce widened lobes, wings, and asymmetric cocoon structures, while higher-power jets remain more collimated. Jet decollimation also produces short-lived filamentary structures within the cocoon. Enhanced radio-emitting regions can appear away from the jet head, which does not necessarily indicate restarted jet activity. The simulations further show that radio luminosity cannot be related to jet power through a single universal relation, since the observed emission depends on the individual growth history of each GRG. In addition, inverse-Compton losses, entrainment of ambient matter, and increasing turbulence affect how long diffuse emission persists and lead to a gradual decline in fractional polarization as the source evolves.

Observations indicate that giant radio galaxies inhabit a wide range of environments, from relatively low-density regions to galaxy groups and clusters \citep[e.g.][]{Da20a, Da20b, An22, bhukta24, manik25}. \citet{Lan2021} compared the environmental properties of GRGs and smaller radio sources and found broadly similar galaxy populations and surface number densities around the two classes, suggesting that an unusually underdense environment is not a prerequisite for reaching giant sizes. In contrast, \citet{oei24} reported that nearly 31\% of GRGs at $z < 0.16$ are associated with galaxy clusters, demonstrating that giant radio sources can also develop in relatively dense environments. This is consistent with simulations showing that jets launched from central, cluster-like positions within the ambient medium can still evolve to giant scales \citep{Giri2025a}. The asymmetric morphologies observed in many GRG radio lobes have likewise been linked to inhomogeneities in the surrounding medium \citep{Machalski2011, Malarecki2013, Malarecki2015}. On larger scales, the cosmic web may further influence the growth and orientation of giant radio sources. For example, \citet{manik25} found that at least 13\% of their giant radio quasar sample resides in galaxy groups or clusters, with a comparable lower-bound fraction located near large-scale filamentary structures. They further found enhanced arm-length asymmetry in GRQs at $z\geq1$ compared with non-giant extended radio quasars, potentially reflecting the influence of asymmetric environments. More recently, \citet{Mahato2026} showed that the growth of GRGs is linked not simply to their proximity to cosmic-web filaments, but to the orientation of their radio jets relative to the filamentary structure. GRGs preferentially exhibit large jet--filament orientation angles, consistent with jets propagating into lower-density, void-facing regions where environmental resistance is reduced. Together, these results highlight the importance of anisotropic large-scale environments in regulating the growth and asymmetry of extended radio sources. Despite these advances, the large-scale environments of extended radio sources at higher redshifts remain comparatively less explored.

Despite growing observational and theoretical efforts, the physical conditions that enable radio galaxies to attain giant scales remain poorly understood. In particular, the connections between the physical properties of radio sources, their central black holes, and their host galaxies remain relatively unexplored in large, homogeneous GRG samples. Here, we present a new sample of GRGs identified in the third data release of the LOFAR Two-meter Sky Survey  \citep[LoTSS-DR3;][]{Shimwell2026}. We characterise their radio physical properties, estimate their central black hole masses, and investigate their host galaxies, including their association with brightest cluster galaxies (BCGs). A detailed analysis of the large-scale environments of these GRGs will be presented in a forthcoming paper.

This paper is organized as follows. Section \ref{sec:data} describes the LoTSS-DR3 dataset and the methodology adopted to identify radio sources and their host galaxies. Section \ref{sec:analysis_results} presents the physical properties of the GRGs identified in LoTSS-DR3, along with their environmental characteristics. Section \ref{sec:discussion} discusses the main findings, focusing on their physical properties and their dependence on the surrounding environments. Finally, Section \ref{sec:conclusion} summarizes the key results of this study.

Throughout this work, we adopt a flat $\Lambda$CDM cosmology with $H_0 = 67.8 \, \rm km \, s^{-1} \, Mpc^{-1}$, $\Omega_{\mathrm{M},0} = 0.308$, and $\Omega_{\Lambda,0} = 0.692$ \citep{Ag18}.

\section{Identification of giant radio Galaxies}
\label{sec:data}
\subsection{The LOFAR Two-metre Sky Survey third data release (LoTSS-DR3)}

In this work, we used data from the LOFAR Two-meter Sky Survey \citep[LoTSS;][]{Shimwell2017, Shimwell2019, Shimwell2022, Shimwell2026}, a deep 120--168 MHz continuum survey conducted with the LOw-Frequency ARray \citep[LOFAR;][]{LOFAR2013}. Earlier LoTSS releases, including the preliminary release and the first two full-quality data releases (DR1 and DR2), covered approximately 27\% of the northern sky \citep{Shimwell2017, Shimwell2019, Shimwell2022}. The recent LoTSS-DR3 \citep{Shimwell2026} extends the coverage to nearly 88\% of the northern sky using approximately 12,950 hours of observations obtained over 10.5 years. The DR3 continuum mosaics provide angular resolutions of $\sim6^{\prime\prime}$ at declinations above $10^\circ$, increasing to $\sim9^{\prime\prime}$ between $0^\circ$ and $10^\circ$. The median rms sensitivity of LoTSS-DR3 is $\sim92~\mu$Jy beam$^{-1}$, reaching $\sim68~\mu$Jy beam$^{-1}$ at high elevations. 
 LoTSS-DR3 consists of 2551 individual pointings observed prior to the LOFAR 2.0 upgrade. Fields within approximately $10^\circ$ of extremely bright radio sources, including Cassiopeia A, Cygnus A, Virgo A, Taurus A, and Hercules A are excluded because of strong contamination effects in the surrounding regions \citep{Shimwell2026}.\\

\begin{figure*}[ht]
    \centering
    \includegraphics[width=17cm,angle=0,origin=c]{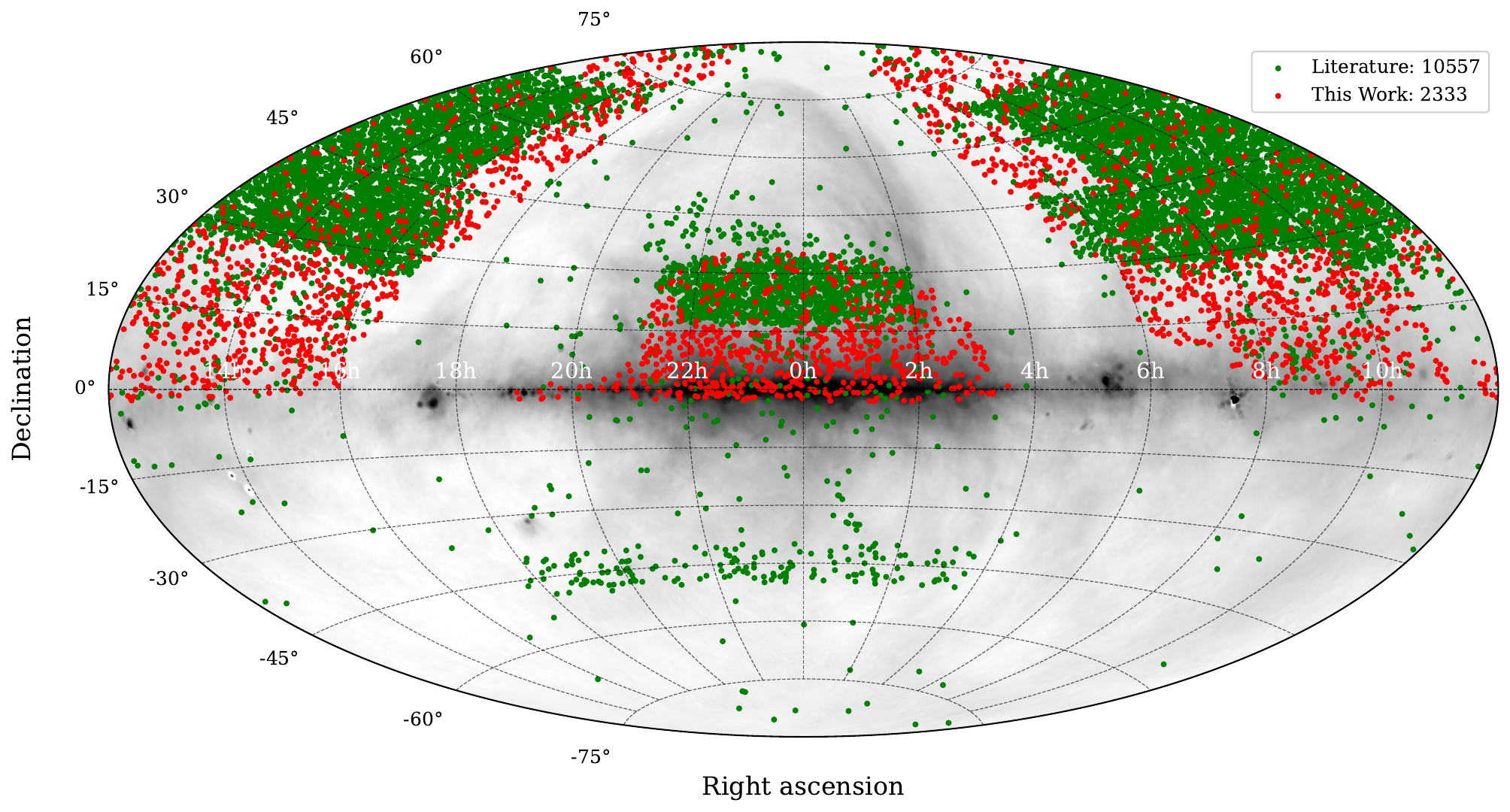}
   \caption{Sky distribution of the giant radio galaxies (GRGs) identified in this work and previously reported in the literature, shown in an Aitoff projection. The previously reported GRGs are taken from the compilation of \citet{mos24}, which includes sources identified in earlier surveys and studies, including those from TGSS ADR1 \citep{bhukta24,manik25}. The background represents the Milky Way specific intensity at an observing frequency of $150\,$MHz, based on \citet{Zheng17skymodel}. Red and green markers indicate the GRGs identified in this study and those reported previously, respectively.}
	\label{fig:sky-dist}
\end{figure*}

\subsection{Sample selection from LoTSS-DR3}
\label{subsection:search}

We visually inspected the radio maps from the third data release of the LOFAR Two-metre Sky Survey \citep[LoTSS-DR3;][]{Shimwell2026} to identify candidate giant radio galaxies (GRGs). The methodology adopted to identify and characterise the final GRG sample is summarised below.

\begin{enumerate}
	
	\item We performed a manual visual search (MVS) of the LoTSS-DR3 mosaic images to identify candidate extended radio sources, with particular attention to double-lobed morphologies and sources exhibiting large angular lengths  $\geq 20\arcsec$.
	
	\item For each candidate, we constructed multiwavelength overlays of the LoTSS-DR3 high- and low-resolution radio images, VLASS component positions, and DESI imaging to identify the radio core and its corresponding optical host. For fields without DESI imaging, we used the \(r\)-band images from the Panoramic Survey Telescope and Rapid Response System 1 (Pan-STARRS1) survey. We visually inspected these overlays to identify the most plausible host galaxy by assessing the spatial association between compact radio emission, where present, and an optical counterpart. The radio morphology was examined simultaneously to distinguish genuine lobes from unrelated or misassociated radio components and to exclude components that were not physically connected to the candidate source. For each confirmed source, we measured the integrated flux density at 150 MHz from the LoTSS-DR3 low-resolution images, which are more sensitive to extended, low-surface-brightness emission.
	
	We determined the largest angular size (LAS) from the full observed extent of the radio emission associated with each source. For sources with clearly identifiable terminal hotspots, particularly classical FR II sources, the LAS was measured as the angular separation between the brightness peaks of the two outermost hotspots. When significant radio emission extended beyond the hotspot peaks, the outermost detected emission was instead used to define the source extent. For FR I, diffuse, bent, and morphologically irregular sources without well-defined terminal hotspots, the LAS was measured as the straight-line angular separation between the two outermost regions of emission associated with the source, using the (3$\sigma$) contour of the corresponding LoTSS image as the primary reference for the detectable extent. This was adopted for bent sources, avoiding assumptions about their unknown three-dimensional geometry. Where robust, spatially coherent diffuse emission was identified beyond the (3$\sigma$) contour, we included this emission in the LAS measurement after visually verifying that it was physically connected to the source rather than arising from noise or an unrelated radio structure.
	
	\item After identifying the most plausible radio core and associated optical host, we obtained the redshift of the host galaxy from the available SDSS spectroscopic \citep{sdss16} and reliable good quality (with reliability flags $f_{clean}=1$ and $f_{clean} = 1$) DESI photometric redshifts from \citet{duncan22}. Using the identified host position, the angular extent of the radio source, and the corresponding redshift, we calculated the radio source's projected linear size. We classified sources with projected linear sizes $D_{\rm proj} > 0.7$~Mpc as giant radio galaxies (GRGs).
	
	\item In the final selection step, we cross-matched the identified host galaxies with previously published GRG catalogues and the literature to identify sources that had already been reported as GRGs. We adopted an angular search radius of $5''$ around the host position and visually inspected the matched sources to confirm their association. Sources previously reported as GRGs were excluded from the final sample. After applying all the selection criteria described above, we obtained our final sample of newly identified GRGs from LoTSS-DR3.
	
\end{enumerate}

\subsection{GRG Population from LoTSS-DR3}

\begin{figure*}
    \centering
    \includegraphics[width=\columnwidth]{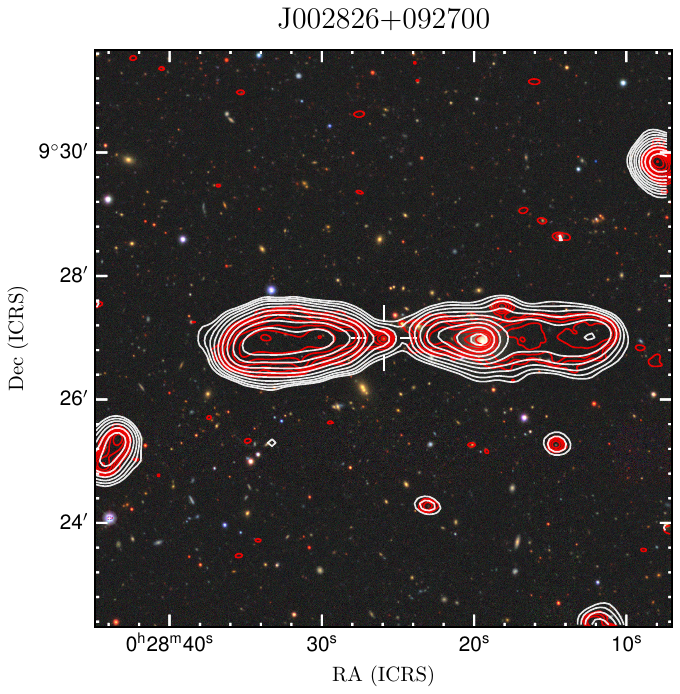}
    \caption{A representative example of a giant radio galaxy (GRG) identified from LoTSS-DR3. The optical background is an RGB composite of the DESI $g$, $r$, and $z$-band images, overlaid with white and red contours representing the low- and high-resolution LoTSS-DR3 radio images, respectively. For each radio image, ten logarithmically spaced contour levels are plotted, starting at $3\sigma$, where $\sigma$ denotes the background rms noise. The position of the radio core is indicated by a plus sign (+). The complete set of corresponding images for all newly identified GRGs will be made available as online supplementary material.}
    \label{fig:grg_montage}
\end{figure*}

Using the aforementioned selection criteria, we identified 2333 GRGs with high confidence, which constitute our final sample. The GRGs in the sample span a redshift range of $0.126 \leq z \leq 3.517$. The sky distribution of the newly identified GRGs is shown in Figure~\ref{fig:sky-dist}. For comparison, we also show the sky distribution of previously reported GRGs compiled by \citet{mos24}, which includes GRGs identified in earlier studies, including those discovered from the TIFR GMRT Sky Survey Alternative Data Release 1 \citep{bhukta24,manik25}. The comparison provides an overview of the current sky coverage of known GRGs and the contribution of the present sample to the overall population. The angular jet length of the GRGs ranges from 1.34$'$ to 12$'$. We find that nearly 31\% (726/2333) of the GRGs in our sample are located at redshift $z \geq 1$. In the following sections, we investigate the physical and environmental properties of these newly identified GRGs. The new sample of grgs along with their physical properties is presented in Table \ref{tab:grg_properties}. An example of the radio morphology of the newly identified GRGs is shown in Fig.~\ref{fig:grg_montage}.

\section{Analysis \& Results}
\label{sec:analysis_results}

\subsection{Projected Jet Length}
\label{subsubsec:projected_size}

The projected jet length of each GRG was estimated from it's measured LAS and the redshift of the associated host galaxy. For a source at redshift $z$, the projected jet length was estimated as

\begin{equation}
D (Mpc)= \frac{\theta_{\rm LAS} \times D_{\rm CO}}{1+z}\left(\frac{\pi}{10800}\right),
\end{equation}
where $\theta_{\rm LAS}$ is the largest angular size of the radio source in arcminutes and $D_{\rm CO}$ is the comoving distance of the source in Mpc. The conventional projected jet length for GRGs is 0.7 Mpc \citep[e.g.][]{Da20b, Oei2023a,bhukta24,manik25}. The estimated projected jet length of GRGs in our sample ranges from 0.7 to $\sim$ 2.9 Mpc, with a mean and median value of $\sim$0.95 and 0.86 Mpc.

\subsection{Radio Power}
\label{subsec:radio_power}

We estimated the 150-MHz radio power of the GRGs from their integrated LoTSS-DR3 flux densities. The radio power was calculated as
\begin{equation}
P_{150} = 4\pi D_{\rm L}^{2} S_{150}(1+z)^{\alpha-1},
\end{equation}
where $D_{\rm L}$ is the luminosity distance, $S_{150}$ is the integrated flux density at 150 MHz, and $\alpha$ is the radio spectral index defined by $S_{\nu}\propto\nu^{-\alpha}$. We adopted a typical spectral index of $\alpha=0.75$ for all GRGs to estimate their radio powers at 150 MHz, consistent with the median spectral index reported for GRGs in previous studies \citep[e.g.][]{Da20b,bhukta24,manik25} and also used in previous GRG studies\citep[e.g.][]{Si22,si24}. Since individual GRGs can exhibit a range of spectral indices owing to differences in their jet, lobe, and radiative histories, the derived radio powers should be regarded as estimates based on a common spectral-index assumption. We note, however, that moderate variations in the adopted value of $\alpha$ have only a limited effect on the resulting radio luminosity.

The GRGs in our sample span a broad range of 150-MHz radio powers, $P_{150}=1.72\times10^{24}$--$3.04\times10^{28}$~W~Hz$^{-1}$, with a median value of $2.56\times10^{26}$~W~Hz$^{-1}$. This broad distribution encompasses both relatively low-power giant radio sources and highly luminous systems, demonstrating that the GRG population is not restricted to a narrow range of radio power. The median radio power is broadly consistent with values reported for GRGs in previous studies \citep[e.g.][]{Da20a, Da20b,bhukta24,manik25}, while our sample extends to substantially higher radio powers.

For comparison with the classical Fanaroff--Riley (FR) dichotomy, we adopt $P_{150}=10^{26}$~W~Hz$^{-1}$ as a fiducial FR division \citep{FR74}. On this basis, 688 sources (29.50\%) lie below the fiducial division, while 1644 sources (70.50\%) lie at or above it. Thus, the majority of the GRGs in our sample occupy the high-radio-power regime traditionally associated with FR~II radio galaxies. This division, however, should not be interpreted as a strict morphological boundary. Deep low-frequency surveys have demonstrated substantial overlap between the radio-power distributions of FR~I and FR~II sources. In particular, the LoTSS study of \citet{mingo19} identified a significant population of low-luminosity FR~II sources extending up to three orders of magnitude below the conventional $P_{150}\sim10^{26}$~W~Hz$^{-1}$ FR division. This indicates that radio luminosity alone is not a reliable predictor of FR morphology and that the FR dichotomy is also influenced by the jet power, host-galaxy properties, and surrounding environment \citep{mingo19}. Therefore, the fractions quoted above represent the distribution of our GRGs relative to a fiducial luminosity division rather than the actual FR~I/FR~II morphological fractions.

\subsection{Jet Kinetic Power}
\label{subsec:jetpower}

Recent three-dimensional magnetohydrodynamic simulations show that the propagation of powerful AGN jets over megaparsec scales depends not only on jet power, but also on jet collimation, thrust, and magnetic stabilization \citep{Giri2026}. The large sizes of giant radio galaxies (GRGs) therefore provide useful laboratories for investigating the transport of AGN energy into the surrounding environment. For example, the recently reported Porphyrion extends over a projected distance of $\sim$7~Mpc \citep{Porphyrion}.

We estimate the characteristic time-averaged jet kinetic power of the GRGs in our sample using the size-dependent radio-luminosity--mechanical-power relation of \citet{ShabalaGodfrey2013}:
\begin{equation}
\frac{Q_{\rm jet}}{10^{36}\ {\rm W}} =
1.5
\left(\frac{L_{151}}{10^{27}\ {\rm W,Hz^{-1}}}\right)^{0.8}
(1+z)
\left(\frac{D}{\rm kpc}\right)^{0.58},
\label{eq:qjet}
\end{equation}
where $Q_{\rm jet}$ is the characteristic jet power, $L_{151}$ is the rest-frame 151-MHz radio luminosity, $z$ is the source redshift, and $D$ is the projected linear size of the radio source. Unlike luminosity-only relations, this formulation incorporates the source size as an empirical proxy for the evolutionary state of the radio source and accounts, in an approximate manner, for radiative and evolutionary losses affecting the observed radio luminosity \citep{ShabalaGodfrey2013}. This is particularly relevant for GRGs, which represent evolved radio sources with large-scale lobes. Since the relation was calibrated using FR-II radio galaxies, we treat the resulting $Q_{\rm jet}$ values as characteristic, time-averaged estimates and apply them with caution to FR-I and hybrid sources.

The estimated jet kinetic powers span $Q_{\rm jet}=5.47\times10^{35}-3.94\times10^{39}$~W, with a median value of $4.94\times10^{37}$~W, corresponding to $\sim4.94\times10^{44}$~erg~s$^{-1}$. This range is broadly consistent with previous estimates for GRGs. For example, \citet{Da20b} discussed jet powers of $\sim10^{41}-10^{45}$~erg~s$^{-1}$ for GRGs, while \citet{Ursini2018} reported $Q_{\rm jet}\sim10^{42}-10^{44}$~erg~s$^{-1}$ for a sample of 14 GRGs. Our median value lies within these previously reported ranges, while the upper end of our distribution extends to $\sim4\times10^{46}$~erg~s$^{-1}$.

The relatively broad range of jet powers in our sample reflects the diversity in the kinetic output of GRG jets and the evolutionary stages of their large-scale radio structures. Previous studies have suggested that GRGs may represent evolved radio galaxies whose jets were more powerful during earlier stages of their growth and subsequently appear less powerful as the lobes undergo radiative and adiabatic losses \citep{Da20b, Ursini2018}. In this context, the size-dependent formulation of \citet{ShabalaGodfrey2013} is particularly useful because it partially accounts for the evolutionary effects associated with the growth of the radio source. Nevertheless, the derived values should be regarded as characteristic estimates rather than direct measurements of the instantaneous jet kinetic power.
\begin{figure*}[ht!]
    \centering
    \includegraphics[width=0.9\textwidth,height=13cm]{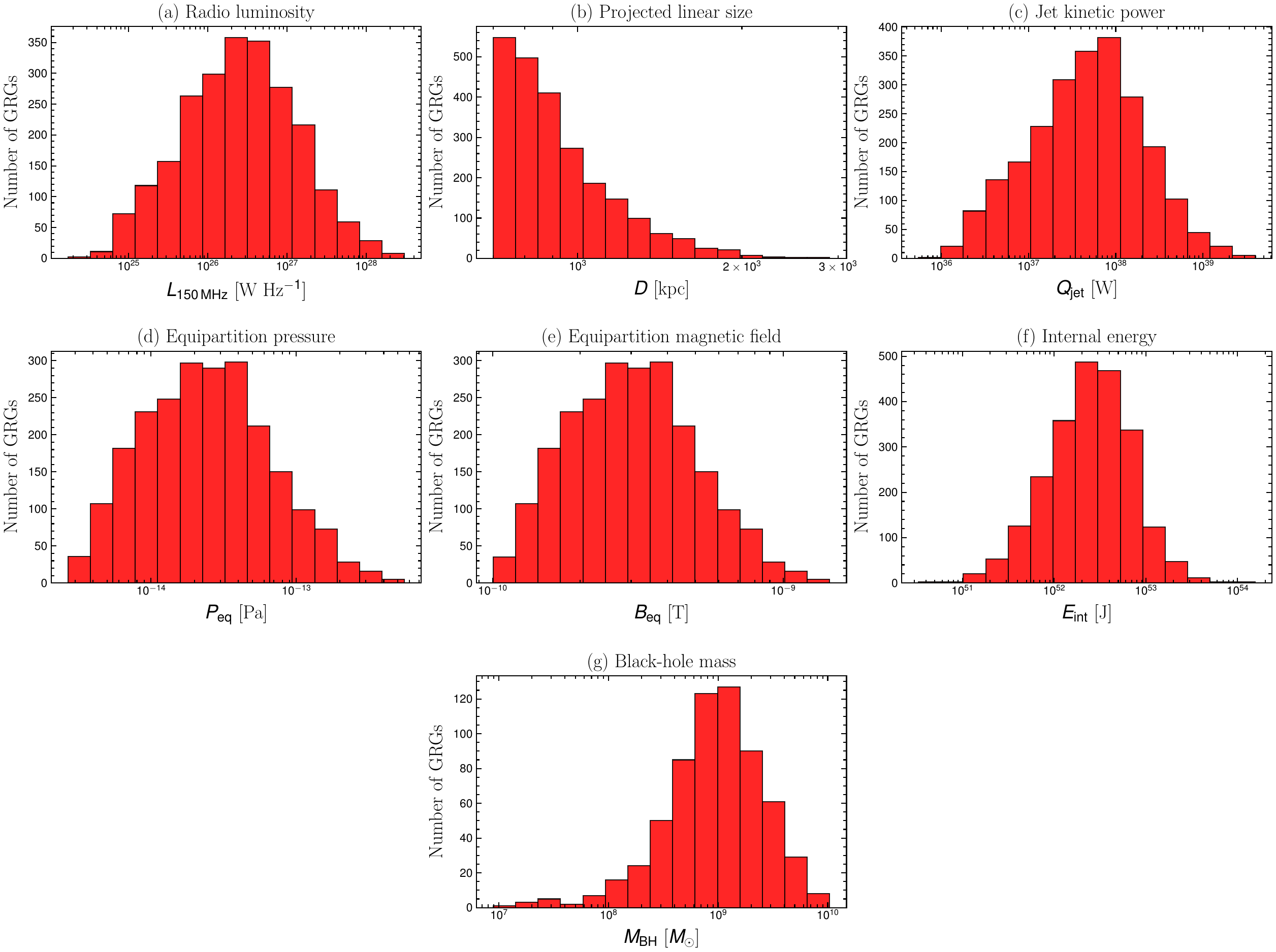}
    \caption{Distributions of the derived physical properties of the radio galaxies in our sample. The panels show the distributions of (a) 150-MHz monochromatic radio luminosity ($L_{150}$), (b) projected linear size ($D$), (c) jet kinetic power ($Q_{\rm jet}$), (d) equipartition pressure ($P_{\rm eq}$), (e) equipartition magnetic field strength ($B_{\rm eq}$), (f) total internal energy stored in the radio lobes ($E_{\rm int}$), and (g) black hole mass ($M_{\rm BH}$). The histograms illustrate the ranges and overall distributions of these physical properties used in the subsequent analysis.}
    \label{fig:grg_distributions}
\end{figure*}

\subsection{Lobe Energetics and Magnetic Fields}
\label{subsec:lobeenergetics}

From the measured radio flux densities and estimated lobe volumes of the GRGs, we derive characteristic magnetic-field strengths, internal pressures, and total internal energies of the radio-emitting plasma. We approximate the radio lobes as cylindrically symmetric structures and estimate their combined volume as $V_{\rm tot}=\pi r_{\rm lobe}^{2}D$,
where $r_{\rm lobe}$ is the characteristic radius of an individual lobe and $D$ is the projected jet length. This expression assumes two approximately symmetric lobes, each approximated as a cylinder with radius $r_{\rm lobe}$ and length $D/2$.

We assume that the relativistic electron population follows a power-law distribution in Lorentz factor, $N(\gamma)\propto\gamma^{-p}$, between $\gamma_{\min}$ and $\gamma_{\max}$. We adopt $\gamma_{\min}=10$, $\gamma_{\max}=10^{4}$, and $p=2$, and assume a negligible contribution from relativistic protons, $\kappa=0$, and a unity filling factor, $\phi=1$. Using the standard synchrotron equipartition formalism \citep[e.g.][]{Hardcastle1998, BeckKrause2005}, we estimate the equipartition magnetic-field strength $B_{\rm eq}$ and the corresponding internal energy density of the radio-emitting plasma.

Under the assumption of a relativistic plasma with adiabatic index $\Gamma=4/3$, the internal pressure is related to the total internal energy density by $P_{\rm eq}=u_{\rm eq}/3$. The total internal energy stored in the radio lobes is therefore
\begin{equation}
E_{\rm int}=3P_{\rm eq}V_{\rm tot}.
\end{equation}
These quantities provide characteristic estimates of the magnetic field, pressure, and energy content of the radio-emitting lobes and allow us to characterize the energetics of the GRGs in our sample.

The radio lobes of the GRGs exhibit a broad range of inferred equipartition properties. The equipartition pressure, $P_{\rm eq}$, ranges from $2.68\times10^{-15}$ to $5.55\times10^{-13}$~Pa, with a median value of $2.36\times10^{-14}$~Pa. The corresponding equipartition magnetic-field strength, $B_{\rm eq}$, spans $0.10-1.45$~nT, with a median of $0.30$~nT. The magnetic-field distribution is broadly consistent with previous studies of giant radio sources. \citet{Is99} reported $B_{\rm eq}=0.06-1.25$~nT, with a median of $0.20$~nT, for a compilation of 53 giant radio sources, while \citet{Konar2004} found $B_{\rm eq}=0.08-0.70$~nT, with a median of $0.25$~nT, for a sample of 17 largely giant radio sources. Our median value of $0.30$~nT is therefore close to those reported previously, although our sample extends to somewhat higher magnetic-field strengths.

The equipartition pressures are also comparable to those derived by \citet{Malarecki2013}, who obtained $P_{\rm eq}=1.1\times10^{-15}$--$2.0\times10^{-14}$~Pa for 12 nearby GRGs, although our distribution extends to substantially higher pressures. At the low-pressure end, \citet{oei23} derived $P_{\rm eq}\simeq6\times10^{-16}$~Pa for the lobes of NGC~6185, illustrating that particularly extended, low-surface-brightness GRGs can reach pressures below those sampled here.

The total internal energy of the radio lobes in our sample spans $3.32\times10^{50}$--$1.55\times10^{54}$~J, with a median value of $2.57\times10^{52}$~J. This median is comparable to the total equipartition lobe energy of approximately $2\times10^{52}$~J inferred for NGC~6185 by \citet{oei23}, although the broader range in our sample reflects the diversity in radio luminosity, source size, and lobe volume. Overall, the equipartition magnetic-field strengths, pressures, and internal energies of our GRGs are broadly consistent with previous measurements of giant radio sources, while the substantially larger sample provides a wider view of the physical conditions and energetic content of GRG lobes.

\subsection{Black Hole Mass}
\label{subsubsec:black_hole_mass}

We estimated the black hole masses ($M_{\rm BH}$) of the GRG host galaxies using two approaches, depending on the availability of the required measurements: (i) the stellar velocity dispersion and (ii) virial estimates from the SDSS DR16 quasar spectral-properties catalogue \citep{wu22}.

For sources with reliable measurements of the stellar velocity dispersion ($\sigma_\star$), we estimated the black hole mass using the $M_{\rm BH}$--$\sigma_\star$ relation of \citet{KH13},
\begin{equation}
\log\left(\frac{M_{\rm BH}}{M_\odot}\right)
=
\alpha_{\sigma}
+
\beta_{\sigma}
\log\left(\frac{\sigma_\star}{200~{\rm km,s^{-1}}}\right),
\end{equation}
where $\alpha_{\sigma}=8.49\pm0.05$ and $\beta_{\sigma}=4.38\pm0.29$ \citep{KH13}. The uncertainty in the estimated $M_{\rm BH}$ was calculated by propagating the measurement uncertainty in $\sigma_\star$ together with the quoted uncertainties in $\alpha_{\sigma}$ and $\beta_{\sigma}$. 

For GRGs classified as quasars in the SDSS DR16 quasar catalogue, we used the corresponding black hole mass estimates from the DR16Q spectral-properties catalogue. These masses are based on single-epoch virial estimates using the widths and luminosities of broad emission lines. We adopted the catalogue values and their reported uncertainties.

The estimated black hole masses span $8.9\times10^{6}-1.05\times10^{10}~M_\odot$, with a median value of $9.77\times10^{8}~M_\odot$ for the 631 sources with available estimates. The median black hole mass is in good agreement with that reported by \citet{Da20b}, who found a median $M_{\rm BH}=8.4\times10^{8}~M_\odot$ for 164 GRGs with reliable SDSS stellar velocity-dispersion measurements. Our median value is therefore consistent with the characteristic black hole masses previously inferred for GRG hosts, while the broader range in our sample extends from $\sim10^{7}$ to $\sim10^{10}~M_\odot$. Earlier studies of individual GRG samples have likewise found that their hosts generally contain massive black holes with $M_{\rm BH}\sim10^{8}$--$10^{9}~M_\odot$, with a subset reaching masses above $10^{9}~M_\odot$ \citep{Dab17}. The substantially larger number of estimates in our sample provides a broader view of the black hole mass distribution of GRG hosts.

\subsection{WISE colour--colour properties}
\label{subsec:wise_colour}

The mid-infrared (MIR) emission of radio-galaxy hosts provides a useful probe of their dust content and nuclear activity. Emission in the WISE bands can contain contributions from stellar emission, dust heated by star formation, and warm dust associated with AGN activity. We therefore use the WISE colour--colour plane to examine the MIR properties of the GRG hosts and to compare them with a larger reference sample.

\begin{figure*}
    \centering
    \includegraphics[width=0.6\linewidth]{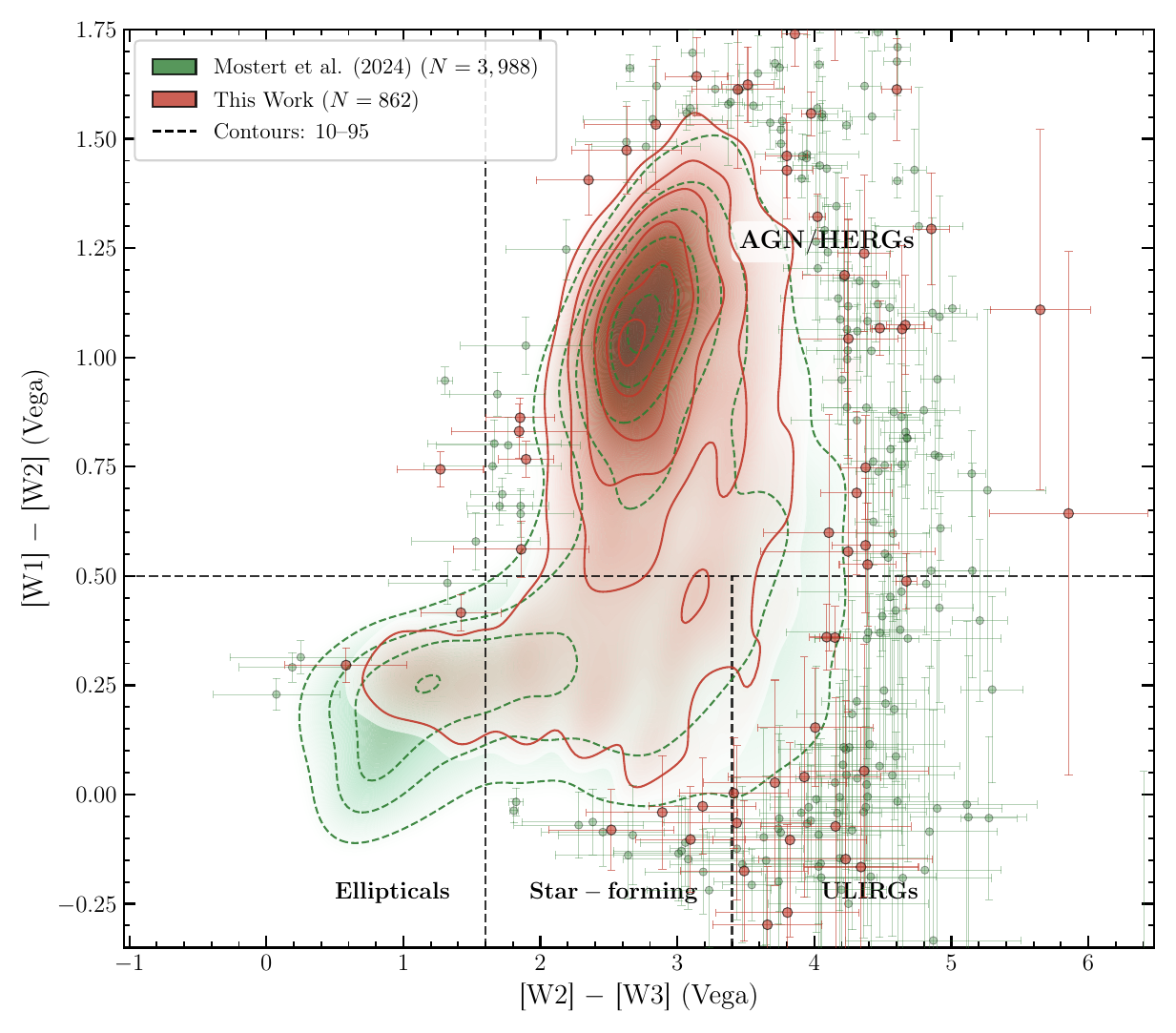}
    \caption{WISE mid-infrared colour--colour diagram showing $[W2]-[W3]$ as a function of $[W1]-[W2]$ for the GRGs in our sample (red) and the reference sample of \citet{mos24} (green). The solid red and dashed green contours show KDE density levels of 10, 25, 40, 60, 80, and 95 per cent of the respective peak densities. Individual sources in the low-density regions are shown with their photometric uncertainties. The shaded regions indicate the empirical WISE colour classifications following \citet{gurkan14} and subsequently adopted by \citet{Da20b,bhukta24,Manik26}: AGN/HERG, LERG-like elliptical galaxies, star-forming galaxies (SFGs), and ultraluminous infrared galaxies (ULIRGs). All colours are calculated using Vega magnitudes.}
    \label{fig:wisecolorcolor}
\end{figure*}

We crossmatched our GRG sample and the literature sample of \citet{mos24} with the AllWISE data using a crossmatch radius of 2$\arcsec$, obtaining usable $W1$, $W2$, and $W3$ photometry for 862 GRGs in our sample and 3988 sources from \citet{mos24}. Figure~\ref{fig:wisecolorcolor} shows the $[W2]-[W3]$ versus $[W1]-[W2]$ colour--colour distribution of the two samples. All colours are calculated using Vega magnitudes. Sources with missing photometry or non-positive photometric uncertainties in any of the three WISE bands were excluded from the analysis.

To characterize the distributions of the two populations, we estimated their two-dimensional density distributions using Gaussian kernel density estimation (KDE). The KDEs were evaluated on a common $[W2-W3]$--$[W1-W2]$ grid and normalized independently to their respective peak densities. The dashed green and solid red contours represent KDE density levels of 10, 25, 40, 60, 80, and 95 per cent of the corresponding peak density for the \citet{mos24} and GRG samples, respectively. These contours indicate relative density levels and do not represent the corresponding fractions of sources enclosed by each contour. Individual sources lying in the low-density regions of the respective distributions are additionally shown with their photometric uncertainties.

The two populations show substantial overlap in WISE colour space, with the highest-density regions occurring around $\rm [W1]-[W2]\simeq0.8$--$1.3$ and $[W2]-[W3]\simeq2.5$--$3.2$. The GRG distribution shows a more extended and structured morphology, with significant density over a broad range of MIR colours. In particular, the GRGs populate both the relatively blue $[W1]-[W2]$ regime associated with galaxy-dominated MIR emission and the redder $[W1]-[W2]$ regime where the contribution from nuclear activity can be more prominent. This distribution indicates that the GRG hosts do not occupy a single, narrowly defined MIR colour population.

To further characterize the GRG population, we indicate the empirical WISE colour regions used to distinguish different classes of radio and infrared galaxies. Following the classification proposed by \citet{gurkan14} and subsequently adopted by \citet{Da20b,bhukta24, Manik26}, sources with $[W1]-[W2]\geq0.5$ and $[W2]-[W3]<5.1$ are associated with the AGN/high-excitation radio-galaxy (HERG) region. Sources with $[W1]-[W2]<0.5$ are further separated according to their $[W2]-[W3]$ colours: LERG-like elliptical galaxies occupy $0<[W2]-[W3]<1.6$, star-forming galaxies (SFGs) occupy $1.6\leq[W2]-[W3]<3.4$, and ultraluminous infrared galaxies (ULIRGs) occupy $[W2]-[W3]\geq3.4$. These regions provide an empirical framework for comparing the MIR properties of the GRG hosts with those of other radio-galaxy populations.

Applying these colour criteria, 643 of the 862 GRGs (74.59\%) fall within the AGN/HERG region, while 46 (5.34\%), 139 (16.13\%), and 32 (3.71\%) occupy the LERG-like elliptical, SFG, and ULIRG regions, respectively. Only two GRGs (0.23\%) fall outside the four defined regions. In comparison, the \citet{mos24} sample contains 2099 AGN/HERG-region sources (52.63\%), 673 LERG-like ellipticals (16.88\%), 911 SFGs (22.84\%), and 302 ULIRGs (7.57\%), with only three sources (0.08\%) outside the defined regions. Thus, the GRG sample has a substantially larger fraction of sources in the AGN/HERG colour region, accompanied by smaller fractions in the LERG-like elliptical, SFG, and ULIRG regions. For the combined sample of 4850 sources, the corresponding fractions are 56.54\%, 14.82\%, 21.65\%, and 6.89\%, respectively, for the AGN/HERG, LERG-like elliptical, SFG, and ULIRG regions.

The large concentration of GRGs in the AGN/HERG colour region, particularly around $[W1]-[W2]\sim0.8-1.3$ and $[W2]-[W3]\sim2.5-3.2$, is consistent with the qualitative distribution seen in Figure~\ref{fig:wisecolorcolor}. At the same time, the presence of GRGs in the LERG-like elliptical, SFG, and ULIRG regions demonstrates that their MIR properties are not homogeneous. The substantially higher AGN/HERG-region fraction in our GRG sample than in the \citet{mos24} reference sample (74.59\% versus 52.63\%) indicates that the GRGs in our sample preferentially occupy the redder $[W1]-[W2]$ regime. Conversely, the lower fractions in the LERG-like elliptical (5.34\% versus 16.88\%), SFG (16.13\% versus 22.84\%), and ULIRG (3.71\% versus 7.57\%) regions further distinguish the MIR colour distribution of the GRGs from that of the reference population.

Larger $[W2]-[W3]$ colours can arise from enhanced emission by dust associated with star formation and/or AGN-heated dust. Since the WISE measurements represent integrated emission from the host and nucleus, the colours alone do not uniquely determine the dominant physical origin of the MIR emission. The differences between the two samples may therefore reflect variations in host-galaxy properties, star-formation activity, dust content, and the relative contribution of the active nucleus to the MIR emission.

The WISE colour classification should nevertheless be regarded as an empirical indication rather than a definitive spectroscopic classification of the AGN excitation state. In particular, the designation of the $[W1]-[W2]\geq0.5$ region as AGN/HERG-like does not imply that all sources in this region are spectroscopically confirmed HERGs. Robust separation of HERGs and LERGs requires complementary information, such as optical emission-line diagnostics or other multiwavelength indicators of the accretion mode.

\subsection{Environmental Analysis of BCG-hosted GRGs}
\label{subsec:environmental_analysis}

\begin{figure*}[htpb]
    \centering
    \includegraphics[scale=0.3]{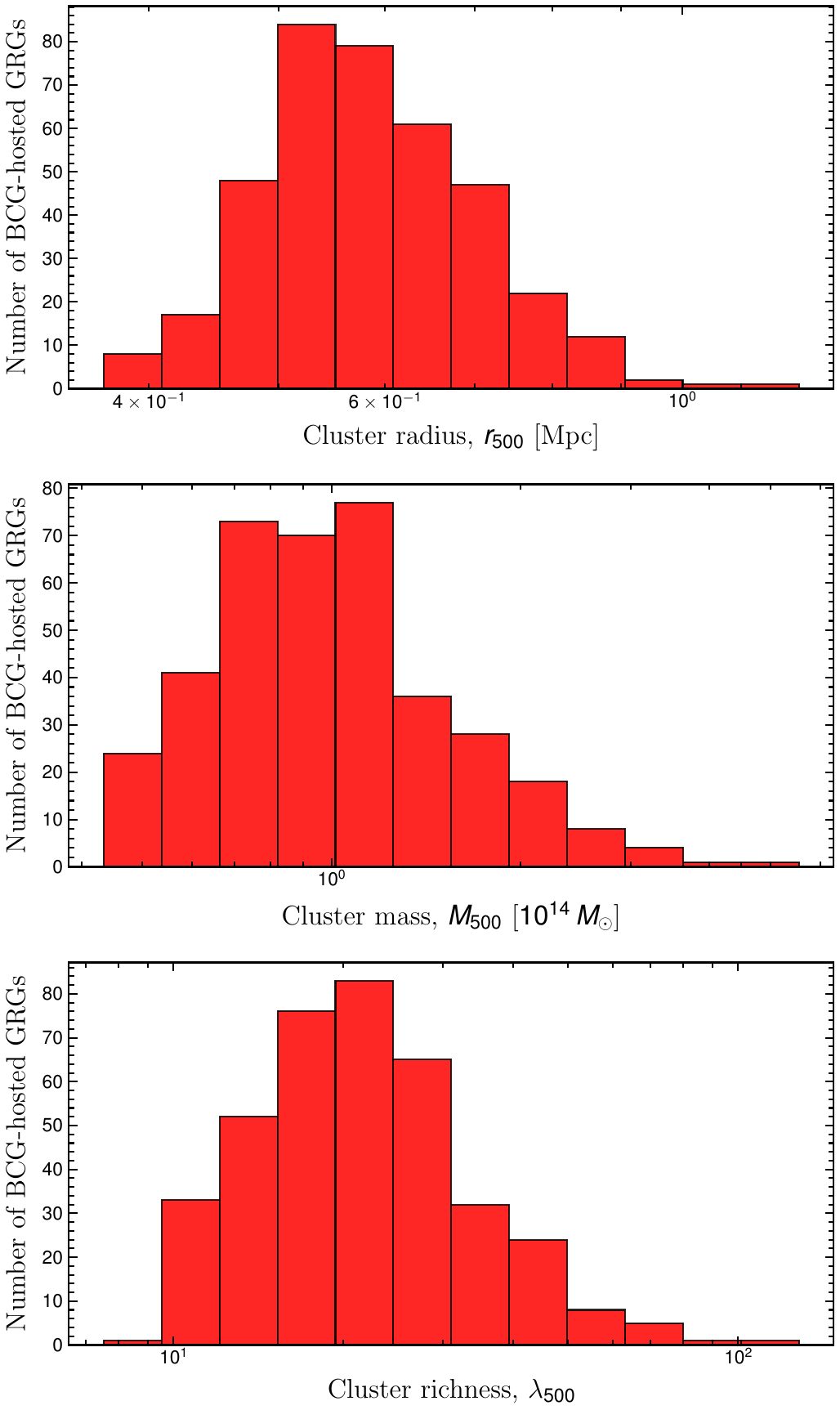}
    \caption{Distributions of cluster properties for the BCG-hosted giant radio galaxies (GRGs): (a) cluster radius, $r_{500}$, (b) cluster mass, $M_{500}$, and (c) cluster richness, $\lambda_{500}$. The histograms show the number of BCG-hosted GRGs as a function of each cluster property. Cluster masses are expressed in units of $10^{14}\,M_\odot$.}
    \label{fig:bcg_grg_cluster_distributions}
\end{figure*}

To investigate the environments of giant radio galaxies (GRGs), we cross-matched our LoTSS DR3 GRG catalogue with the available galaxy-cluster catalogues, including \citet{WH15,WH22,WH24}. The cross-matched catalogues provide cluster properties, including the mass ($M_{500}$), characteristic radius ($r_{500}$), and richness ($\lambda_{500}$), which we used to examine the relationship between GRG extent and cluster environment. We identified 382 BCG-hosted GRGs (around 16\%) with the required environmental measurements and used this sample for the subsequent analysis. The distributions of cluster properties for the BCG-hosted GRG subsample are presented in Fig.~\ref{fig:bcg_grg_cluster_distributions}. These include the cluster radius ($r_{500}$), mass ($M_{500}$), and richness ($\lambda_{500}$), which collectively characterise the environments hosting these radio galaxies. The corresponding environmental measurements and BCG-related properties are summarised in Table~\ref{tab:cluster_properties}.

The projected linear sizes of these sources range from 0.70 to 2.90 Mpc, with 94 (24.6\%) having $D>1$ Mpc. The redshift distribution is dominated by sources at $z<1$: 213 (55.76\%) lie at $z<0.5$, 164 (42.93\%) lie at $0.5\leq z<1.0$, and five (1.31\%) lie at $1.0\leq z<1.5$. 

We examine and discuss the relationships between projected linear size, normalized size, and cluster properties in the following subsections \ref{subsec:environment1} and \ref{subsec:bcg_nonbcg_comparison}.

\section{Discussion}
\label{sec:discussion}

\subsection{Radio Power--Linear Size Relation}
\label{sec:pd_diagram}

\begin{figure*}[htpb]
\vbox{
\centerline{
    \includegraphics[scale=0.5]{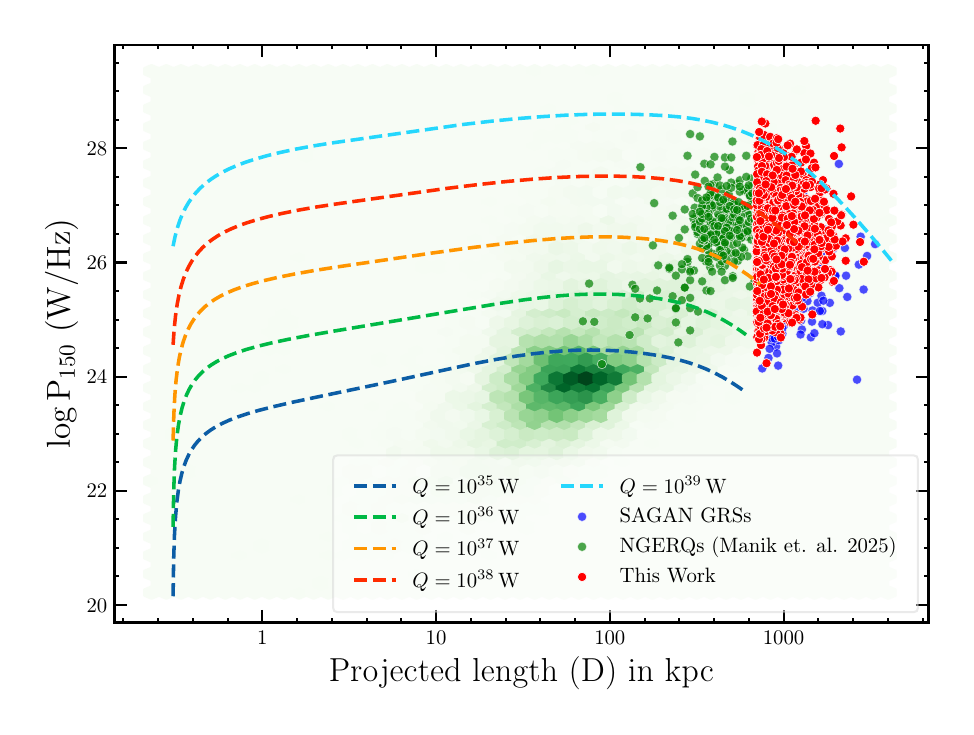}}}
\caption{Radio power--linear size ($P$--$D$) diagram for the 2333 high-confidence GRGs in our sample. The green hexagonal density distribution represents the radio-AGN population from \citet{Hardcastle2019}, while previously reported giant and extended radio sources are shown for comparison \citep[e.g.,][]{Da20b,manik25}. The solid curves show the evolutionary tracks of \citet{Hardcastle2018} for two-sided jet powers $Q_{\rm jet}=10^{35}$--$10^{39}\,{\rm W}$, calculated for an external halo with $M_{500} = 2.5\times10^{13}\,M_\odot$. The tracks illustrate the expected coupled evolution of radio power and source size for different jet powers under the adopted external pressure profile. The GRGs span a broad range of radio powers at large physical sizes, highlighting that giant dimensions are not restricted to the most radio-luminous systems.}
    \label{fig:pd_diagram}
\end{figure*}

The radio power--linear size ($P$--$D$) diagram provides a useful framework for examining the evolutionary state of the GRG population. Figure~\ref{fig:pd_diagram} shows the 2333 high-confidence GRGs together with the broader radio-AGN population from \citet{Hardcastle2019} and previously reported giant and extended radio sources \citep[e.g.,][]{Da20b,manik25}. Our GRGs occupy the large-size regime by construction, but span a broad range of radio powers, demonstrating that reaching megaparsec scales does not require a uniquely high present-day radio luminosity. This is consistent with the evolutionary picture in which a source can continue to increase in physical extent while its radio luminosity evolves as a result of adiabatic expansion, radiative losses, and changes in the energy supplied by the jets.

To provide a physical reference for the observed distribution, we overplot the evolutionary tracks of \citet{Hardcastle2018}. These tracks describe the coupled evolution of source size and radio luminosity for different jet powers while accounting for the dynamics of the lobes and the evolution of their synchrotron-emitting particles. We consider two-sided jet powers of $Q_{\rm jet}=10^{35}$--$10^{39}\,{\rm W}$ and adopt $M_{500}=2.5\times10^{13}\,M_\odot$ for the external halo. Thus, the tracks should be regarded as reference evolutionary trajectories rather than unique age estimates for individual GRGs.

An important ingredient in these models is the pressure profile of the external medium. Following the approach adopted for the evolutionary tracks, the external pressure can be described using the universal pressure-profile formalism of \citet{Arnaud10}, in which the pressure is expressed as a function of radius through $x=r/R_{500}$ and scaled primarily by the halo mass $M_{500}$:

\begin{equation}
P(r)=P_{500}
\left[\frac{M_{500}}
{3\times10^{14}h_{70}^{-1}M_\odot}\right]^{\alpha_p+\alpha'_p(x)}
p(x),
\end{equation}
where $p(x)$ is a generalized Navarro--Frenk--White profile. The usefulness of this prescription is that it provides a physically motivated pressure profile with the halo mass setting the overall normalization. The adopted $M_{500}$ therefore determines, within the model, the confining pressure encountered by the expanding radio lobes. We stress, however, that this is a reference environmental model and not a direct measurement of the pressure profiles of the individual GRG environments.

The broad distribution of our sources across the $P$--$D$ plane can then be understood as the combined result of jet power and source evolution. At fixed $Q_{\rm jet}$, continued energy injection drives the lobes to progressively larger sizes, while the radio luminosity can decline as the plasma expands and loses energy. Conversely, sources with similar sizes can have substantially different radio powers because their jet powers, particle populations, magnetic fields, and external pressure histories are not identical. The presence of relatively low-power GRGs therefore does not require exceptionally powerful jets at the present epoch; they may instead represent systems that have undergone prolonged expansion and have evolved beyond their peak radio luminosity.

The comparison with the general radio-AGN population reinforces this interpretation. GRGs extend the radio-AGN population toward much larger physical sizes rather than simply occupying its most luminous end. Their existence therefore requires not only sufficient jet power but also sustained activity and sufficiently favourable propagation conditions for the lobes to grow over hundreds of kiloparsecs to megaparsec scales. The large scatter in radio power at a given size further indicates that source extent is not a direct proxy for instantaneous jet power.

The evolutionary tracks also highlight the importance of the environment. A change in the external pressure or density profile alters the rate of lobe expansion and therefore changes the trajectory of a source through the $P$--$D$ plane. The single halo model adopted here cannot capture the full range of environments represented by our GRG sample, which includes sources associated with different halo masses and large-scale structures. Consequently, departures of individual GRGs from the reference tracks should not necessarily be interpreted as differences in jet power alone. They may also reflect differences in the external medium, source history, or radiative evolution.

Overall, the $P$--$D$ distribution supports a picture in which GRG formation is governed by the interplay between jet energetics, source evolution, and the external environment. The radio power primarily traces the current energetic state of the lobes, whereas the linear size records the integrated history of jet propagation. This distinction is important when interpreting the physical origin of giant radio structures and motivates combining the $P$--$D$ distribution with the jet-power, lobe-pressure, and environmental measurements considered below.

\subsection{Equipartition Pressure and Source Size}
\label{subsec:pressure_size}

\begin{figure*}[htpb]
\vbox{
\centerline{
\includegraphics[scale=0.6]{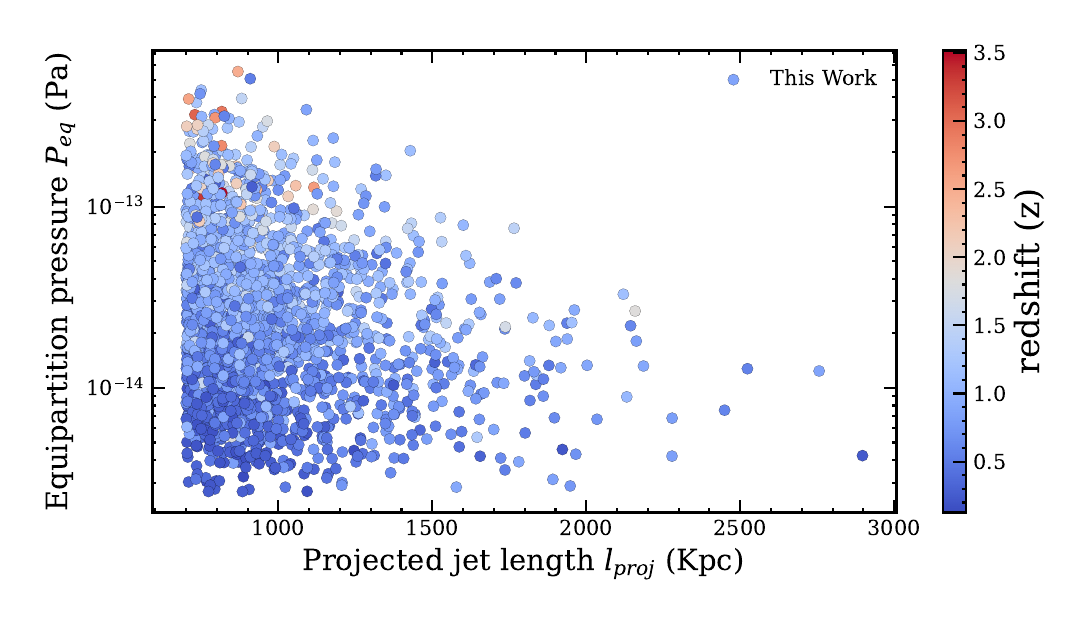}
}}
\caption{Equipartition pressure, $P_{\rm eq}$, as a function of projected linear size, $l_{\rm proj}$, for the GRGs in our sample. The sources are colour-coded by redshift.}
\label{fig:peqd}
\end{figure*}

Figure~\ref{fig:peqd} presents the equipartition pressure, $P_{\rm eq}$, as a function of the projected linear size, $l_{\rm proj}$, with the sources colour-coded by their redshift. The distribution shows a pronounced decrease in $P_{\rm eq}$ toward larger projected sizes. The most compact systems reach pressures of $\sim10^{-13} - 10^{-12}$ Pa, whereas the largest GRGs, with $l_{\rm proj}\gtrsim2$ Mpc, predominantly occupy the $\sim10^{-15} - 10^{-14}$ Pa regime. Although the scatter is substantial, this behaviour is qualitatively consistent with the expected evolution of expanding radio lobes.

In the standard dynamical picture of powerful radio galaxies, the jets deposit energy and momentum into an overpressured cocoon whose expansion is governed by the internal lobe pressure and the density and pressure of the surrounding medium \citep{KaiserAlexander1997}. As the source expands, the cocoon volume increases and the injected energy is distributed over a progressively larger volume. Adiabatic expansion, together with radiative losses and the properties of the external density profile, therefore leads to a decline in lobe pressure as the source grows. The evolutionary models of \citet{KaiserDennettThorpeAlexander1997} similarly predict coupled evolution of radio luminosity and source size as the lobes expand and their radiative and adiabatic losses evolve.

The observed decline in $P_{\rm eq}$ with $l_{\rm proj}$ is therefore consistent with the expansion and evolutionary development of radio lobes. As the lobes expand, their internal pressure can progressively approach that of the surrounding circumgalactic and intergalactic medium. However, $P_{\rm eq}$ is an inferred internal pressure and should not be interpreted as a direct measurement of the external pressure. The detailed pressure evolution is expected to depend on the jet power, external density profile, expansion history, and radiative losses. Direct equipartition estimates for nearby GRGs have also yielded lobe pressures of $\sim10^{-15}-10^{-14}$ Pa \citep[e.g.,][]{Alcyoneus}, illustrating that extended radio lobes can probe the low-pressure environments beyond their host haloes.

An additional feature of Figure~\ref{fig:peqd} is the redshift distribution. Higher-redshift objects are preferentially located toward the higher-pressure part of the $P_{\rm eq}$ distribution, whereas the largest and lowest-pressure systems are predominantly found at lower redshift. A qualitatively similar distribution has been reported for GRGs by \citet{Alcyoneus}. However, because radio power, source size, and redshift are strongly coupled in flux-limited samples, the observed $P_{\rm eq}$--$z$ relation cannot be interpreted directly as cosmic evolution. Selection effects are particularly important for diffuse radio sources because their synchrotron surface brightness decreases as the lobes expand, making low-pressure, extended structures increasingly difficult to detect at larger distances \citep{Schoenmakers2000}. Consequently, a flux- and surface-brightness-limited survey can preferentially select brighter and higher-pressure systems at higher redshift. The very low pressures measured in nearby GRGs may likewise be facilitated by their more favourable surface-brightness detectability \citep{Alcyoneus}.

We therefore tested whether the observed redshift dependence remains after accounting for radio power and source size. For the 2273 GRGs with complete measurements, $P_{\rm eq}$ shows a strong positive correlation with redshift ($\rho_{\rm S}=0.633$, $p<10^{-250}$), while the correlation with 150-MHz radio power is even stronger ($\rho_{\rm S}=0.815$, $p<10^{-250}$). The dependence on projected size is weaker but remains significant, with $\rho_{\rm S}=-0.155$ ($p=1.0\times10^{-13}$). Importantly, a partial Spearman analysis shows that the correlation between $P_{\rm eq}$ and redshift remains significant after controlling simultaneously for $P_{150}$ and $l_{\rm proj}$, yielding $\rho_{\rm S,partial}=0.273$ ($p=3.0\times10^{-40}$). Bootstrap resampling gives a 95\% confidence interval of $0.230-0.315$ for the partial correlation. Thus, the redshift dependence of the inferred equipartition pressure is not simply a consequence of the mutual correlations among radio power, source size, and redshift.

The persistence of this relation within individual radio-power intervals provides a further check. The $P_{\rm eq}-z$ correlation remains positive in all sufficiently populated $P_{150}$ bins, with $\rho_{\rm S}=0.38$, $0.47$, $0.38$, and $0.22$ for $24\leq\log P_{150}<25$, $25\leq\log P_{150}<26$, $26\leq\log P_{150}<27$, and $27\leq\log P_{150}<28$, respectively. These results indicate that the higher inferred pressures of the high-redshift population cannot be removed simply by restricting the comparison to similar radio-power ranges.

A multivariate regression provides a consistent result. Modelling $\log P_{\rm eq}$ as a function of $\log l_{\rm proj}$, $\log P_{150}$, and $\log(1+z)$ gives a positive coefficient for $\log(1+z)$ of $0.793 \pm 0.066$ ($p=6.5\times10^{-32}$), while the coefficients of $\log l_{\rm proj}$ and $\log P_{150}$ are $-1.037 \pm 0.047$ and $0.442 \pm 0.009$, respectively. This independently supports the presence of a residual redshift dependence after accounting for source size and radio power.

Nevertheless, the residual correlation should not yet be interpreted as direct evidence for intrinsic cosmological evolution of lobe pressure. Equipartition pressure is an inferred quantity that depends on the radio luminosity, assumed lobe volume, particle-energy distribution, and equipartition assumptions. In addition, the detectability of extended, low-surface-brightness emission becomes increasingly difficult with redshift, so the observed population remains subject to redshift-dependent surface-brightness and angular-size selection effects \citep{Schoenmakers2000, Alcyoneus}. Such effects may contribute to the residual relation even after controlling for the measured radio power and physical size.

The remaining redshift dependence may therefore reflect a combination of selection effects and genuine differences in the physical conditions of the observed GRG population at different cosmic epochs. Environmental variations, differences in jet power and evolutionary stage, and the properties of the detectable population may all contribute to the observed scatter and redshift dependence. Distinguishing these possibilities requires explicit modelling of the survey selection function and, ideally, comparisons of GRGs matched in radio power, physical size, and observational sensitivity. The persistence of the $P_{\rm eq}$--$z$ relation after controlling for $P_{150}$ and $l_{\rm proj}$ nevertheless provides evidence that the higher inferred pressures of high-redshift GRGs warrant further investigation.

\subsection{FR Dichotomy in the Black Hole Mass--Jet Kinetic Power Plane}
\label{subsec:fr_bh_jet}

The Fanaroff--Riley (FR) dichotomy provides an important framework for understanding how the central engine and surrounding medium jointly shape the large-scale morphology of radio galaxies. Although the original FR-I/FR-II division was defined in terms of radio luminosity, the dependence of the transition on host-galaxy optical luminosity established by the Ledlow--Owen relation \citep{LedlowOwen1996} demonstrates that radio luminosity alone cannot determine the morphological class. Large LOFAR samples have further shown that the classical dichotomy is not sharp, with substantial overlap between FR-I and FR-II populations and the presence of low-luminosity FR-IIs \citep{mingo19}. Subsequent studies have emphasized the roles of host properties and environment in determining whether jets remain collimated or become disrupted near the FR transition \citep{mingo19,Clews2025}. The identification of FR-I structures among quasars further demonstrates that optical excitation class and radio morphology are not uniquely coupled \citep{Gurkan2022}.

In this context, the $M_{\rm BH}$--$Q_{\rm jet}$ plane provides a physically motivated framework for examining the FR transition. Following \citet{WuCao2008}, we compare our estimated jet powers with the empirical FR-I/FR-II division,

\begin{equation}
\log Q_{\rm jet}({\rm W}) =
1.13\log\left(\frac{M_{\rm BH}}{M_\odot}\right)
+26.42+1.50\log f,
\label{eq:wucao}
\end{equation}
where $f$ parametrizes the uncertainty in converting radio luminosity to kinetic jet power. We adopt $f=10$ and $f=20$ to define a transition region around the nominal dividing relation.

For the 631 GRGs with available black-hole mass estimates and optical classifications, 490 sources lie below the $f=10$ boundary, 78 within the $f=10$--20 transition region, and 63 above the $f=20$ boundary. The lower region is dominated by galaxies, with 343 (70.0\%) galaxies and 147 (30.0\%) quasars. In contrast, quasars constitute 65.4\% (51/78) of the transition-region sources and 58.7\% (37/63) of those above the $f=20$ boundary. Thus, galaxies preferentially occupy the lower-power side of the empirical division, while quasars become more prominent toward and above the transition. This is broadly consistent with the traditional association of FR-I systems with radio galaxies and FR-II systems with quasars, but the substantial overlap shows that the Wu--Cao relation cannot be regarded as a direct morphological classifier.

\begin{figure*}
    \vbox{
\centerline{
    \includegraphics[scale=0.5]{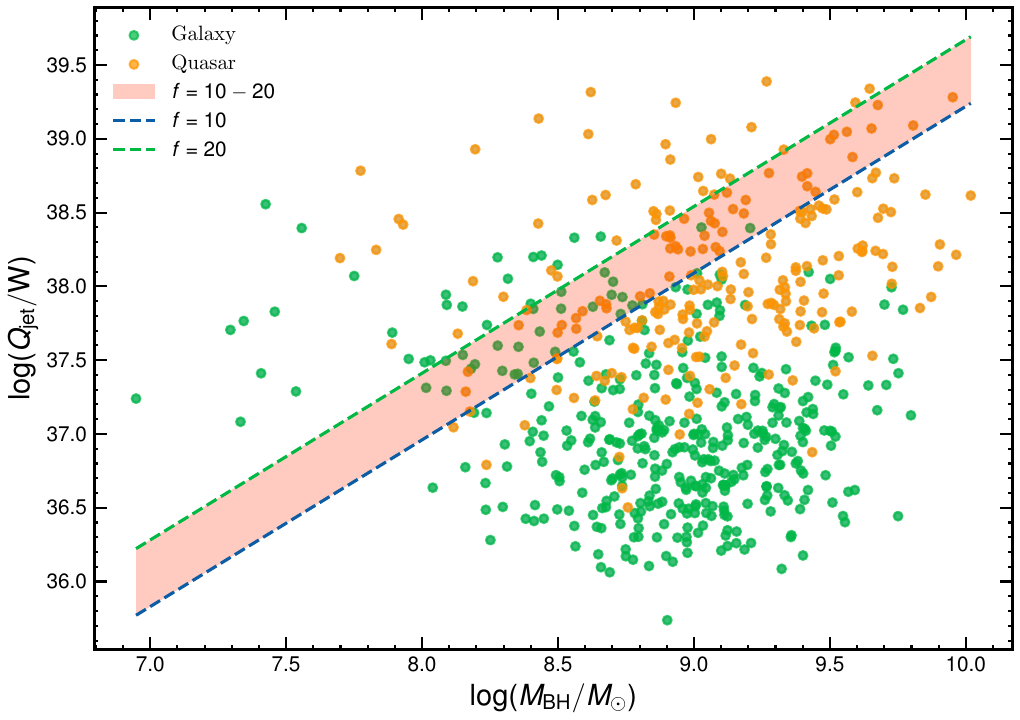}}}
     \caption{Black-hole mass--jet kinetic power ($M_{\rm BH}$--$Q_{\rm jet}$) plane for the GRGs with available black-hole mass estimates and optical classifications. The solid lines show the empirical FR-I/FR-II division of \citet{WuCao2008} for $f=10$ and $f=20$, defining a transition region between the two boundaries. Sources are distinguished according to their optical classification as galaxies and quasars. The distribution illustrates the increasing prominence of quasars toward and above the empirical transition, while substantial overlap remains between the two populations.}
    \label{fig:MBH_Qjet_WuCao_FR_division}
    \label{fig:Mbhqjet}
\end{figure*}

The physical origin of this behaviour can be considered in terms of the central engine. In the model of \citet{WuCao2008}, the maximum jet power produced by a hybrid Blandford--Znajek-type mechanism operating in an advection-dominated accretion flow (ADAF) can approximately reproduce the empirical FR division for an accretion rate $\dot{m}\sim0.01$ and a rapidly rotating black hole with $j\sim0.9$--$0.99$. The model therefore links the FR transition to black-hole mass, accretion state, and spin rather than to black-hole mass alone. In particular, the strong dependence of the jet power on black-hole spin implies that black holes of similar mass can, in principle, produce substantially different jet powers. The increasing Wu--Cao dividing power with $M_{\rm BH}$ likewise indicates that the relevant quantity is not $Q_{\rm jet}$ alone, but the jet power relative to the characteristic scale set by the central engine.

Our comparison with the Willott et al. relation illustrates the systematic uncertainty associated with the radio-to-jet-power conversion. For 2332 GRGs, we obtain a mean $\Delta\log Q=-0.259$ dex and a median of $-0.261$ dex, corresponding to a mean ratio $Q_{\rm jet}/Q_{\rm Willott}\simeq0.55$. The offset is somewhat larger for galaxies ($-0.324$ dex) than for quasars ($-0.227$ dex). Given the systematic uncertainties in radio-based jet-power estimates, these differences should not be interpreted as evidence for intrinsically different jet-power calibrations between the two optical classes. Rather, they emphasize that the exact position of an individual source relative to the Wu--Cao boundaries depends on the adopted jet-power prescription.

The remaining overlap between the expected low- and high-power regimes is naturally explained by jet--environment coupling. Even for comparable $M_{\rm BH}$ and $Q_{\rm jet}$, differences in the ambient gas density and pressure can affect entrainment, jet deceleration, reconfinement, and disruption. A jet that remains collimated can develop an FR-II-like edge-brightened structure, whereas stronger interaction with the ambient medium can produce an FR-I-like edge-darkened morphology. This environmental contribution is particularly relevant for GRGs, whose jets propagate over hundreds of kiloparsecs to megaparsec scales and therefore integrate the effects of the external medium over long timescales. The $M_{\rm BH}$--$Q_{\rm jet}$ plane consequently captures an important aspect of the central-engine physics, but does not uniquely determine the final radio morphology. Our results favour a picture in which the FR transition emerges from the interplay of black-hole mass, accretion and spin, jet power, and subsequent jet--environment interaction. This naturally connects the central-engine properties  to the environmental regulation of GRG growth discussed below.   

\subsection{GRGs at High Redshift}
\label{subsec:highz_grgs}

The identification of 726 GRGs at $z\geq1$, corresponding to $\sim31\%$ of our high-confidence sample, and the presence of sources extending to $z=3.517$, demonstrates that megaparsec-scale radio structures were already established at relatively early cosmic epochs. The existence of such systems provides an important constraint on the timescales over which radio jets can remain active and propagate to hundreds of kiloparsecs or megaparsec scales. Their presence at high redshift also suggests that sufficiently powerful and long-lived AGN activity capable of producing giant radio structures was already operating during the earlier stages of galaxy and black-hole evolution.

Detecting GRGs at high redshift is nevertheless challenging. The observed surface brightness of extended synchrotron emission is reduced by cosmological $(1+z)^4$ surface-brightness dimming, while inverse-Compton losses of relativistic electrons against the cosmic microwave background increase rapidly with redshift because its energy density scales as $U_{\rm CMB}\propto(1+z)^4$. These effects progressively suppress the observable extended radio emission from distant sources. Consequently, the high-redshift GRGs identified in LoTSS-DR3 are unlikely to represent an unbiased census of the underlying population and are expected to be preferentially weighted toward sources with sufficiently high radio surface brightness and favourable evolutionary conditions.

The distribution of our GRGs in the radio power--linear size plane provides further evidence that giant structures can be produced across a broad range of radio powers. In particular, the presence of high-redshift sources at large projected sizes demonstrates that reaching giant scales at early cosmic epochs does not require all GRGs to occupy the extreme high-power end of the radio-AGN population. However, the combined effects of flux limits, surface-brightness sensitivity, angular-size selection, and enhanced radiative losses must be considered before comparing the intrinsic size or power distributions of GRGs across redshift.

The higher equipartition pressures preferentially observed among the high-redshift GRGs, discussed in Section~\ref{subsec:pressure_size}, are consistent with these observational constraints. Although a residual $P_{\rm eq}$--$z$ relation remains after accounting for radio power and projected size, this should not be interpreted as direct evidence for cosmic evolution of lobe pressure. Instead, the high-redshift population may reflect a combination of intrinsic differences in jet power, evolutionary stage, and environment with the redshift-dependent detectability of diffuse radio emission.

The existence of GRGs at $z>1$ therefore highlights the ability of powerful AGN jets to develop large-scale structures within a relatively young Universe, while simultaneously emphasizing the importance of selection effects when interpreting their observed population. Deeper low-frequency observations with improved surface-brightness sensitivity, together with well-defined selection functions, will be essential for determining the true abundance and physical properties of GRGs at high redshift.
\subsection{Environmental Dependence of the Projected Extent of BCG-hosted GRGs}
\label{subsec:environment1}

We investigate the relationship between the projected linear sizes of BCG-hosted giant radio galaxies (GRGs) and their host-cluster properties using a sample of 382 sources with available measurements of cluster mass ($M_{500}$), characteristic radius ($r_{500}$), and richness ($\lambda_{500}$). The projected linear sizes span approximately $0.70$--$2.90$ Mpc, with 94 sources (24.6\%) having projected extents of at least 1 Mpc. To examine how the radio-source extent compares with the characteristic scale of the host cluster, we additionally consider the dimensionless quantity $D/(2r_{500})$. Here, the factor of two provides an approximate measure of the one-sided projected extent relative to $r_{500}$, under the assumption that the BCG lies near the geometrical midpoint of the radio source. This quantity should therefore be regarded as an approximate characteristic extent rather than a direct measurement of the distance between the BCG and either radio lobe. In our sample, 365 sources (95.5\%) have $D/(2r_{500})\geq0.5$, while 68 (17.8\%) have $D/(2r_{500})\geq1$, indicating that the projected radio structures of a substantial fraction of these systems extend over a considerable fraction of the characteristic cluster radius.

We first examine whether the physical projected size of BCG-hosted GRGs depends on the mass of their host clusters. The bivariate correlations show no significant association between $D$ and either $M_{500}$ or $\lambda_{500}$. We further investigate this relationship using a multivariable regression of the form
\begin{equation}
\log_{10}\left(\frac{D}{\mathrm{Mpc}}\right)
=
a+
b\log_{10}\left(\frac{M_{500}}{10^{14}\,M_\odot}\right)
+
c\log_{10}E(z),
\label{eq:bcg_grg_size_mass}
\end{equation}
where $E(z)=H(z)/H_0$ accounts for the redshift dependence of the critical density. The regression yields $b=-0.025\pm0.025$ ($p=0.314$), providing no statistically significant evidence for a dependence of physical GRG size on cluster mass after accounting for redshift. The corresponding coefficient for redshift is $c=0.299\pm0.087$ ($p=6.6\times10^{-4}$), indicating a positive conditional association between projected size and $E(z)$ within the sample. The fitted mass coefficient is also strongly inconsistent with the specific scaling $D\propto M_{500}^{1/3}$ expected if the physical radio-source extent directly followed the self-similar mass scaling of $r_{500}$ at fixed redshift. Thus, within the mass and redshift ranges probed by our sample, the physical sizes of BCG-hosted GRGs do not exhibit a measurable systematic dependence on host-cluster mass. This result does not, however, establish the absence of an underlying relationship, particularly given the effects of sample selection and the limited number of systems at the highest cluster masses.

\begin{figure*}
    \vbox{
\centerline{
    \includegraphics[scale=0.6]{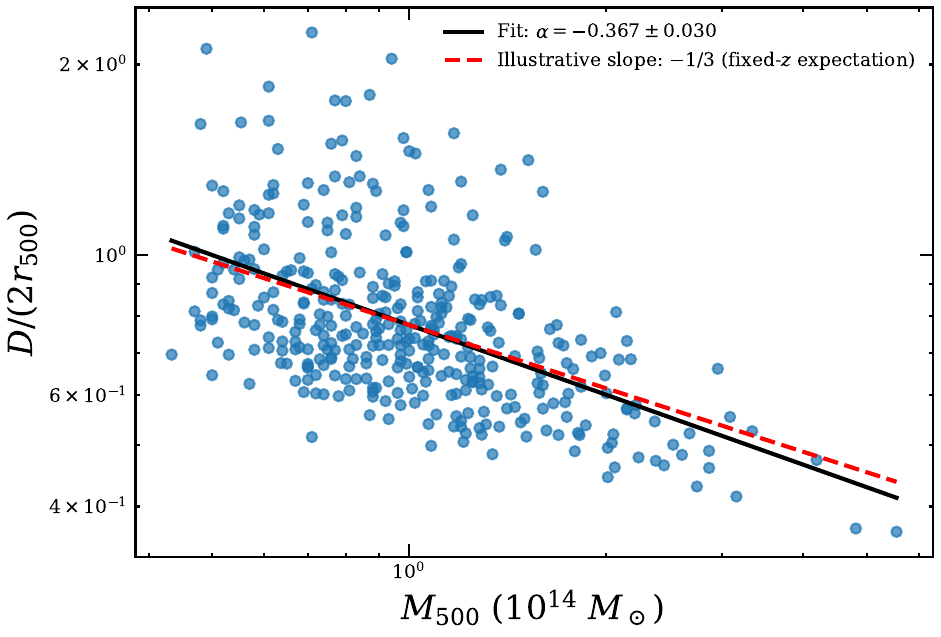}}}
    \caption{Approximate one-sided projected extent, $D/(2r_{500})$, of BCG-hosted giant radio galaxies as a function of host-cluster mass, $M_{500}$. The points represent individual GRGs, with colours indicating redshift. The solid line shows the best-fitting bivariate power-law relation, with a slope of $-0.367\pm0.030$, while the dashed line indicates the reference slope of $-1/3$ expected from the self-similar scaling of $r_{500}$ with $M_{500}$ at fixed redshift. The bivariate fit does not control for redshift; therefore, the reference slope is intended as a visual guide rather than a formal statistical comparison.}
    \label{fig:bcg_grg_normalized_extent_mass}
\end{figure*}

The behaviour of the normalized extent is different. We find a strong anticorrelation between $D/(2r_{500})$ and $M_{500}$, with Pearson and Spearman correlation coefficients of $-0.417$ and $-0.527$, respectively (both with $p<10^{-16}$). A bivariate power-law fit, without controlling for redshift, gives
\begin{equation}
\frac{D}{2r_{500}} =
0.775
\left(\frac{M_{500}}{10^{14}\,M_\odot}\right)^{-0.367\pm0.030}.
\label{eq:bcg_grg_normalized_size}
\end{equation}
The resulting distribution and the corresponding power-law fit are shown in Fig.~\ref{fig:bcg_grg_normalized_extent_mass}. This relation provides a descriptive characterization of the observed normalized-size distribution, with $R^2=0.282$. The observed slope is close to the value expected from the mass dependence of the cluster radius. In particular, for a radius defined by an overdensity of 500 relative to the critical density,
\begin{equation}
r_{500}\propto M_{500}^{1/3}E(z)^{-2/3}.
\end{equation}
Consequently, even if the physical GRG size has little or no intrinsic dependence on cluster mass, its ratio to $r_{500}$ is expected to decrease with increasing $M_{500}$ at fixed redshift. We therefore further examine the normalized extent using a redshift-controlled multivariable regression. This yields a mass coefficient of $-0.372\pm0.025$ for $\log[D/(2r_{500})]$, close to the shift of $-1/3$ expected from the mass dependence of the radius. The corresponding redshift coefficient is $0.905\pm0.090$, also close to the additional $2/3$ contribution expected from the $E(z)^{-2/3}$ dependence of $r_{500}$. The strong anticorrelation in normalized extent should therefore not be interpreted independently as evidence that GRGs are physically smaller or more strongly confined in more massive clusters. Instead, it is consistent with the expected mass and redshift scaling of the adopted cluster radius, given the comparatively weak dependence of the physical GRG size on cluster mass.

The mass-binned distributions provide a complementary illustration of this behaviour. The median physical sizes remain broadly comparable across the well-populated mass bins, whereas the median normalized extent decreases from approximately 0.82 in clusters with $M_{500}<10^{14}\,M_\odot$ to 0.69 and 0.52 in the $10^{14}$--$2\times10^{14}\,M_\odot$ and $2$--$4\times10^{14}\,M_\odot$ bins, respectively. The highest-mass bins contain relatively few sources, including only three objects in the $4$--$8\times10^{14}\,M_\odot$ interval, and therefore do not support strong conclusions about the behaviour of the population at the massive end. Nevertheless, the broadly comparable physical sizes across the better-populated bins are consistent with the regression result that cluster mass does not show a significant association with the projected physical extent of these sources.

We also find a positive correlation between projected GRG size and redshift. Although the redshift term remains statistically significant in the multivariable regression, this trend should be interpreted cautiously. The redshift distribution of a radio-selected sample can be affected by surface-brightness sensitivity, angular-resolution limitations, angular-size selection, source detectability, and the luminosity and size distributions of the underlying population. These effects may influence the observed size distribution with redshift and prevent a direct interpretation of the fitted redshift coefficient as intrinsic physical evolution. A more detailed assessment of the selection function would be required to establish whether the observed trend reflects genuine evolution in the characteristic sizes of BCG-hosted GRGs.

Overall, our analysis does not reveal a statistically significant dependence of the physical projected sizes of BCG-hosted GRGs on host-cluster mass. The strong anticorrelation observed when the sizes are normalized by $r_{500}$ is consistent with the expected mass and redshift scaling of the cluster radius, rather than providing independent evidence for a physical suppression of radio-source growth in massive clusters. These results suggest that the projected physical extent of the radio structures in our sample is not simply determined by the global mass of the host cluster. However, the present analysis does not directly establish the role of the intracluster medium in regulating jet propagation or radio-source evolution. Such an interpretation would require spatially resolved information on the gas density, pressure, temperature, and dynamical state of the host environments. In addition, the regression uncertainties reported here account for heteroscedasticity through the HC3 covariance estimator, but do not explicitly incorporate measurement uncertainties in the fitted quantities or possible correlations between multiple GRGs associated with the same physical cluster. These limitations should be considered when interpreting the statistical significance and physical implications of the observed trends.

\subsection{Comparison of BCG and Non-BCG GRGs}
\label{subsec:bcg_nonbcg_comparison}

We investigate whether giant radio galaxies (GRGs) hosted by brightest cluster galaxies (BCGs) differ systematically from those hosted by non-BCG systems in their projected linear sizes ($D$), 150-MHz radio powers ($P_{150}$), and estimated jet kinetic powers ($Q_{\rm jet}$). The comparison includes 382 BCG-hosted and 1951 non-BCG GRGs. Since the two populations have different redshift distributions, we examine their physical properties using both unmatched-sample comparisons and redshift-controlled analyses.

We first compare the unmatched populations using two-sample Kolmogorov--Smirnov (KS) and Anderson--Darling (AD) tests, applying the Benjamini--Hochberg false-discovery-rate (FDR) correction separately to the KS and AD tests across the three physical properties. Redshift is examined separately and is not included in these correction families. Effect sizes are quantified using Cliff's delta, with 95\% confidence intervals estimated through bootstrap resampling. The median redshift is 0.467 for BCG-hosted GRGs and 0.886 for non-BCG GRGs, indicating a substantial difference between their redshift distributions (KS statistic $D_{\rm KS}=0.452$, $p=1.44\times10^{-59}$). The median projected sizes are comparable (849.8 and 872.0 kpc, respectively), with no statistically significant distributional difference after FDR correction ($p_{\rm KS,FDR}=0.114$; $p_{\rm AD,FDR}=0.069$; Cliff's $\delta=-0.064$). In contrast, BCG-hosted GRGs exhibit substantially lower median radio powers ($6.40\times10^{25}$ versus $3.29\times10^{26}\ {\rm W\,Hz^{-1}}$) and estimated jet kinetic powers ($1.29\times10^{37}$ versus $6.29\times10^{37}\ {\rm W}$). The differences in both properties are highly significant. For $P_{150}$, the KS test gives $D_{\rm KS}=0.398$ ($p_{\rm FDR}=3.88\times10^{-45}$), while the AD test also indicates a strong difference ($p_{\rm FDR}=0.0015$); the corresponding Cliff's delta is $-0.476$. For $Q_{\rm jet}$, the KS test gives $D_{\rm KS}=0.415$ ($p_{\rm FDR}=9.46\times10^{-49}$), and the AD test yields $p_{\rm FDR}=0.0015$, with Cliff's delta of $-0.505$. The AD-test probabilities reach the lower reporting limit of the test. These unmatched comparisons, however, do not account for the substantially different redshift distributions of the two populations.

\begin{figure*}[t]
    \centering
    \begin{minipage}[t]{0.485\textwidth}
        \centering
        \includegraphics[width=\linewidth,height=4cm]{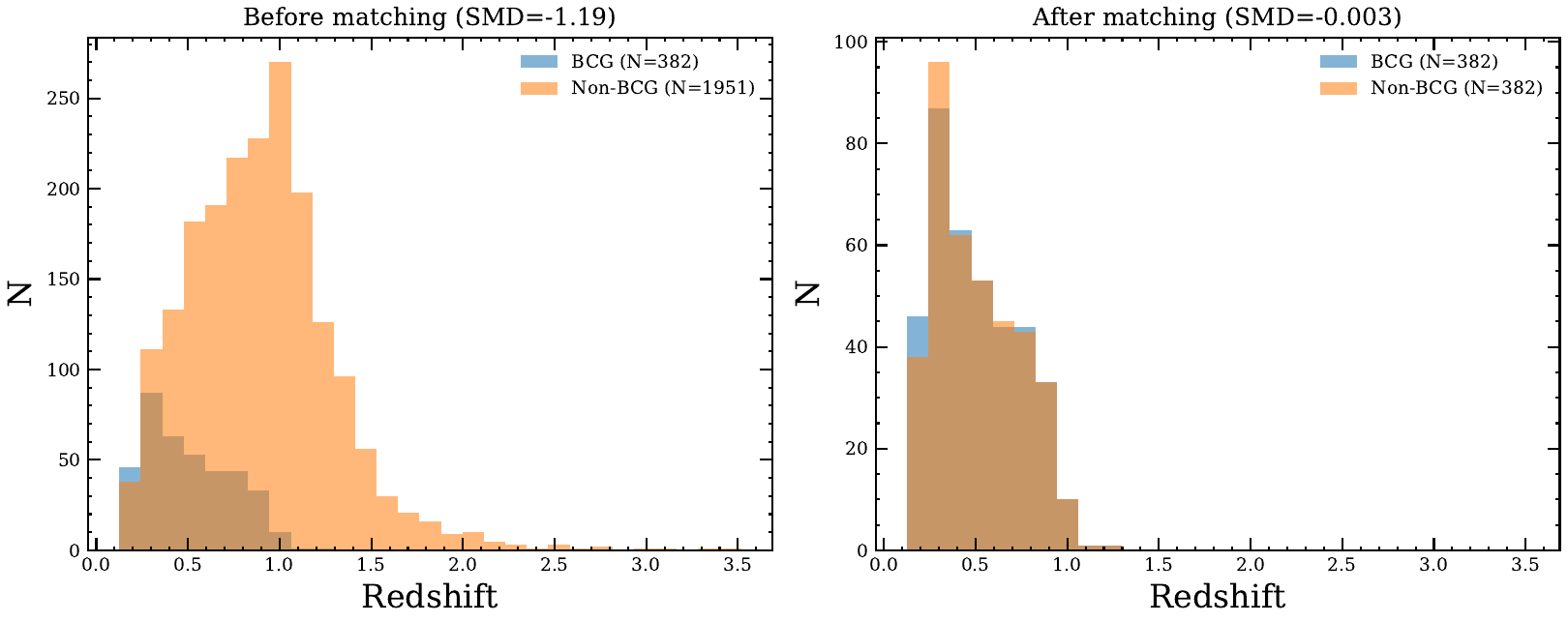}
        \par\vspace{2pt}
        \textbf{(a)}
    \end{minipage}
    \hfill
    \begin{minipage}[t]{0.485\textwidth}
        \centering
        \includegraphics[width=\linewidth,height=4cm]{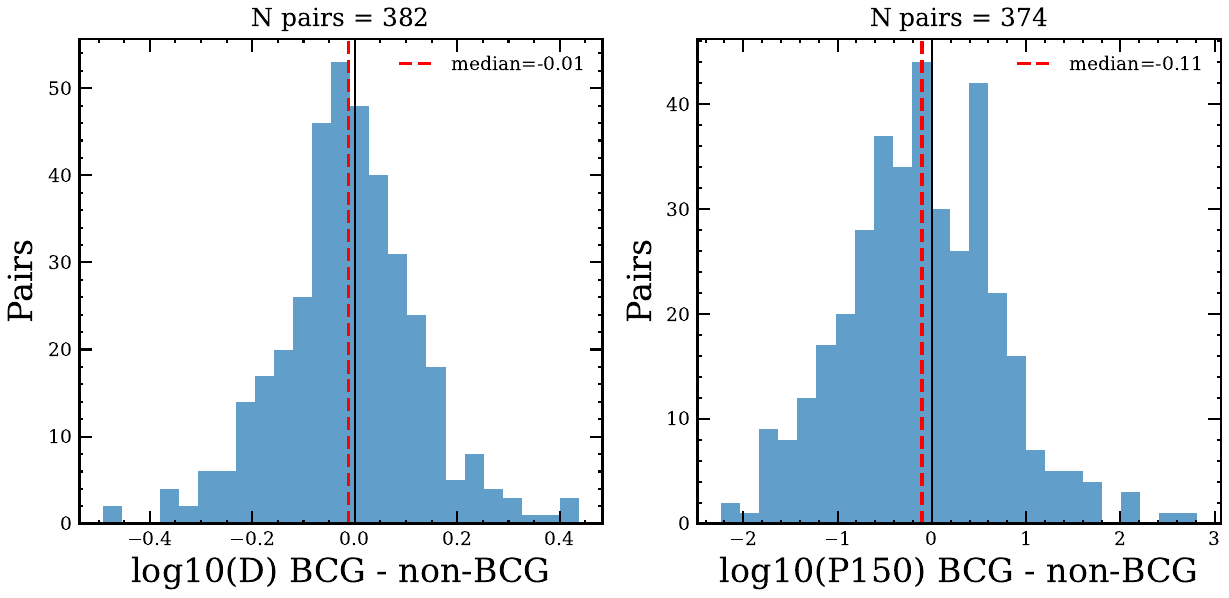}
        \par\vspace{2pt}
        \textbf{(b)}
    \end{minipage}
    \caption{(a) Redshift distributions of BCG-hosted and non-BCG GRGs before and after one-to-one matching. The left panel shows the original unmatched samples, while the right panel shows the matched samples. The matching procedure substantially reduces the redshift imbalance between the two populations. (b) Distributions of the paired logarithmic differences in projected linear size and 150-MHz radio power for the redshift-matched BCG-hosted and non-BCG GRGs. The logarithmic differences are defined as $\log_{10}(X_{\rm BCG}/X_{\rm non-BCG})$. The vertical dashed line indicates equality between the matched populations.}
    \label{fig:matching_analysis}
\end{figure*}

To control for this imbalance, we perform one-to-one optimal matching without replacement, using the absolute redshift separation as the matching cost and adopting a maximum redshift difference of $\Delta z=0.05$. All 382 BCG-hosted GRGs are matched to distinct non-BCG sources. The median absolute redshift difference between matched pairs is $3.60\times10^{-4}$, while the redshift standardized mean difference decreases from $-1.187$ before matching to $-0.0033$ after matching, indicating excellent balance in redshift. The redshift distributions before and after matching are compared in Figure~\ref{fig:matching_analysis} (a), illustrating the substantial improvement in redshift balance achieved by the matching procedure. We then compare the paired physical properties using the Wilcoxon signed-rank test, with FDR correction across the two matched properties, $D$ and $P_{150}$. The matched comparison again reveals no statistically significant difference in projected size, with a median BCG-to-non-BCG ratio of $0.972$ (95\% confidence interval: $0.946$--$1.004$; $p_{\rm FDR}=0.101$). In contrast, the median paired radio-power ratio is $0.785$ (95\% confidence interval: $0.610$--$0.925$), corresponding to significantly lower radio power among BCG-hosted GRGs ($p_{\rm FDR}=0.01245$). This result is based on 374 valid matched pairs. Repeating the matching with $\Delta z=0.02$ and $0.10$ produces similar radio-power ratios of 0.791 and 0.785, respectively, indicating that the result is not strongly dependent on the adopted matching caliper. The distributions of the paired logarithmic differences in projected size and radio power are shown in Figure~\ref{fig:matching_analysis}.

As a complementary analysis, we perform ordinary least-squares regression of $\log_{10}P_{150}$ against BCG status and redshift, including quadratic redshift terms and adopting heteroscedasticity-consistent HC3 standard errors. Here, BCG status is represented by a binary indicator that equals unity for BCG-hosted GRGs and zero for non-BCG GRGs. Our primary analysis is restricted to the common redshift interval $0.138\leq z\leq0.955$, where both populations are represented. The resulting sample comprises 1487 sources, including 364 BCG-hosted and 1123 non-BCG GRGs. The BCG coefficient is $-0.119$ dex, corresponding to a conditional geometric-mean radio-power ratio of $0.760$ (95\% confidence interval: $0.647$--$0.893$; $p_{\rm FDR}=8.49\times10^{-4}$). Thus, at a given redshift within this interval, BCG-hosted GRGs have an approximately 24\% lower conditional geometric-mean radio power than non-BCG GRGs. As a sensitivity analysis, we extend the regression to the full common redshift interval ($0.138\leq z\leq1.261$). This sample contains 2014 sources, including 374 BCG-hosted and 1640 non-BCG GRGs. Despite the similar estimated offset, the redshift distributions remain substantially imbalanced in this extended sample, with a redshift standardized mean difference of $-1.064$. In particular, only 10 BCG-hosted GRGs lie at $z>0.955$, compared with 530 non-BCG GRGs. In particular, only 10 BCG-hosted GRGs lie at $z>0.955$, compared with 530 non-BCG GRGs. The fitted BCG coefficient is $-0.102$ dex, corresponding to a conditional geometric-mean radio-power ratio of $0.791$ (95\% confidence interval: $0.675$--$0.927$; $p_{\rm FDR}=0.00385$). We therefore consider the restricted-redshift regression our primary result.

We further investigate whether the radio-power difference depends on redshift by including BCG--redshift interaction terms. Neither the linear nor quadratic interaction is statistically significant in the primary regression ($p=0.709$ and $p=0.556$, respectively). The spline-based interaction test also shows no evidence for a redshift-dependent difference in the restricted sample ($p=0.954$). Although the corresponding test in the extended sample yields a nominally significant result ($p=0.044$), this is not supported by the quadratic interaction tests and must be interpreted cautiously given the poor representation of BCG-hosted GRGs at higher redshifts. We therefore find no robust evidence that the BCG-associated radio-power difference varies systematically with redshift. Influence diagnostics indicate that the negative BCG coefficient persists after excluding influential observations, although its magnitude becomes more negative. Thus, the qualitative result is not reversed by these sources, but the estimated magnitude is sensitive to their inclusion. The principal results from the redshift-matched and common-support regression analyses are summarized in Table~\ref{tab:bcg_nonbcg_summary}.

\begin{table*}
\centering
\small
\setlength{\tabcolsep}{5pt}
\renewcommand{\arraystretch}{1.12}
\caption{Redshift-matched and redshift-adjusted comparison of BCG-hosted and non-BCG giant radio galaxies.}
\label{tab:bcg_nonbcg_summary}
\begin{tabular}{llcccc}
\toprule
Analysis & Property / sample & $N$ & BCG/non-BCG ratio & 95\% CI & $q_{\rm FDR}$ \\
\midrule
Redshift-matched & $D$ (projected size) & 382 pairs & 0.972 & 0.946--1.004 & 0.101 \\
Redshift-matched & $P_{150}$ (150-MHz power) & 374 pairs & 0.785 & 0.610--0.925 & 0.01245 \\
Restricted (primary) & $P_{150}$; $z=0.138$--$0.955$ & 364/1,123 & 0.760 & 0.647--0.893 & $8.49\times10^{-4}$ \\
Full overlap (sensitivity) & $P_{150}$; $z=0.138$--$1.261$ & 374/1,640 & 0.791 & 0.675--0.927 & 0.00385 \\
\bottomrule
\end{tabular}
\begin{minipage}{0.98\textwidth}
\vspace{3pt}
\footnotesize
For the matched analysis, $N$ denotes the number of valid one-to-one redshift-matched pairs. Matching was performed without replacement using a maximum redshift separation of $\Delta z=0.05$. The matched ratio is defined as $10^{\mathrm{median}[\log_{10}(X_{\rm BCG})-\log_{10}(X_{\rm non-BCG})]}$, and $q_{\rm FDR}$ denotes the Benjamini--Hochberg FDR-adjusted Wilcoxon signed-rank $p$-value. For the regression rows, $N$ gives the numbers of BCG-hosted and non-BCG sources, respectively. The reported ratio is $10^{\beta_{\rm BCG}}$ from the quadratic-redshift model $\log_{10}P_{150}=\alpha+\beta_{\rm BCG}I_{\rm BCG} +\beta_1(z-z_c)+\beta_2(z-z_c)^2$, where $I_{\rm BCG}$ is a binary indicator equal to unity for BCG-hosted GRGs and zero for non-BCG GRGs, and $z_c$ is the redshift value used to centre the redshift terms. The model is fitted using ordinary least squares with HC3 heteroscedasticity-consistent standard errors. Here, $q_{\rm FDR}$ is the FDR-adjusted $p$-value for the BCG coefficient. The restricted common-support sample is the primary regression analysis, while the full-overlap sample is included as a sensitivity analysis. Ratios below unity indicate lower values for BCG-hosted GRGs. Regression ratios are conditional model estimates, not ratios of sample medians. The HC3 standard errors account for heteroscedasticity but do not account for possible within-cluster dependence. The matched comparison includes only $D$ and $P_{150}$; $Q_{\rm jet}$ is not included because it is derived from radio properties and therefore does not provide an independent confirmation of the radio-power result.
\end{minipage}
\end{table*}

Overall, the unmatched, redshift-matched, and regression analyses consistently indicate that BCG-hosted GRGs have lower radio powers than redshift-comparable non-BCG GRGs, whereas their projected linear sizes are statistically consistent. The persistence of this difference across the adopted matching and regression approaches suggests that it cannot be explained solely by the different redshift distributions of the two populations. However, our analysis primarily controls for redshift and does not account for possible differences in host-galaxy properties, black-hole masses, accretion states, or detailed environments. Furthermore, $Q_{\rm jet}$ is estimated from radio properties and therefore does not provide an independent confirmation of the radio-power difference. These results should consequently be interpreted as population-level associations rather than evidence that BCG association itself causes a reduction in radio or jet power.

\section{Conclusion}
\label{sec:conclusion}

We have presented a sample of 2333 newly identified giant radio galaxies (GRGs) from the third data release of the LOFAR Two-metre Sky Survey (LoTSS-DR3). By combining low-frequency radio observations with optical, infrared, and spectroscopic information, we investigated their radio properties, lobe energetics, equipartition pressures, black hole masses, host-galaxy characteristics, and cluster environments. We also examined the differences between BCG-hosted and non-BCG GRGs using redshift-controlled statistical analyses. Our principal findings are summarised below.

\begin{enumerate}

\item The final sample spans a redshift range of $0.126\leq z\leq3.517$, with 726 sources ($\sim31\%$) at $z\geq1$. Their 150-MHz radio luminosities range from $1.72\times10^{24}$ to $3.04\times10^{28}$ W\,Hz$^{-1}$, demonstrating the wide range of radio powers associated with megaparsec-scale radio structures. The substantial number of high-redshift GRGs provides an opportunity to investigate the population of giant radio sources across a broad range of cosmic epochs, although cosmological surface-brightness dimming and enhanced inverse-Compton losses against the cosmic microwave background may limit the detection of extended emission at high redshift.

\item The GRGs occupy a broad region of the radio power--linear size plane, indicating that large physical extent is not determined by radio luminosity alone. Their equipartition pressures exhibit a general decrease with increasing projected linear size, consistent with the expansion and associated reduction in energy density of the radio lobes. We also find a positive correlation between equipartition pressure and redshift that persists after controlling for radio power and projected size. However, this relation should be interpreted cautiously because of possible redshift-dependent selection effects and systematic uncertainties associated with the equipartition assumptions.

\item The distribution of 631 GRGs in the $M_{\rm BH}$--$Q_{\rm jet}$ plane reveals a systematic shift in the relative proportions of galaxies and quasars across the empirical FR transition proposed by \citet{WuCao2008}. Galaxies dominate below the $f=10$ boundary (70.0\%), whereas quasars dominate in the intermediate ($10<f<20$; 65.4\%) and higher-$f$ ($f>20$; 58.7\%) regions. However, substantial overlap persists between the two classes, indicating that black-hole mass and jet power alone do not uniquely determine radio morphology.

\item The WISE mid-infrared colour analysis reveals diversity in the host-galaxy properties of GRGs, with sources predominantly occupying the empirical AGN/HERG colour region. Approximately 74.6\% of the 862 sources with available WISE photometry occupy the empirical AGN/HERG colour region, compared with 52.6\% in the reference GRG population of \citet{mos24}. The remaining sources populate the LERG-like elliptical, star-forming galaxy, and ULIRG regions. These colour-based classifications provide useful insights into the mid-infrared properties of GRG host galaxies; however, they should not be interpreted as definitive spectroscopic classifications of their AGN.

\item For the 382 BCG-hosted GRGs with available cluster properties measurements, the physical projected sizes show no statistically significant dependence on cluster mass. However, their normalized projected half-sizes, $D/(2r_{500})$, exhibit a strong anticorrelation with $M_{500}$. A redshift-controlled multivariable regression gives a mass coefficient of $-0.372\pm0.025$, consistent with the $-1/3$ dependence expected from the self-similar scaling of the cluster radius. The corresponding redshift coefficient is also consistent with the expected contribution from the redshift dependence of $r_{500}$. Thus, the observed anticorrelation in normalized extent does not independently establish that GRGs are physically smaller or more strongly confined in more massive clusters. Instead, the broadly comparable physical sizes across the better-populated cluster-mass bins suggest that global cluster mass alone is not a sufficient predictor of GRG extent.

\item The comparison between BCG-hosted and non-BCG GRGs reveals an important difference in their radio properties. Although the unmatched samples show substantially different redshift distributions, one-to-one redshift matching and regression analyses consistently indicate that their projected linear sizes are statistically comparable. In contrast, the primary redshift-restricted regression shows that BCG-hosted GRGs have approximately 24\% lower conditional geometric-mean 150-MHz radio power than non-BCG GRGs, corresponding to a BCG-to-non-BCG ratio of $0.760$ (95\% confidence interval: $0.647$--$0.893$). The matched analysis independently supports this difference, yielding a median paired radio-power ratio of $0.785$. We find no robust evidence that the radio-power difference varies systematically with redshift. Since the present analysis does not control for all differences in host-galaxy properties, black hole masses, accretion states, or detailed environments, this result should be interpreted as a population-level association rather than a causal effect of BCG status.

\end{enumerate}

Overall, our results demonstrate that the observed properties of GRGs reflect a complex interplay between radio-source energetics, evolutionary history, central-engine properties, and the surrounding environment. In particular, the comparable physical sizes of BCG-hosted and non-BCG GRGs, despite their different radio-power distributions, suggest that the mechanisms governing the large-scale expansion of radio sources cannot be understood through radio luminosity or global cluster mass alone. The large and broad-redshift LoTSS-DR3 sample provides an important foundation for further investigations of GRG evolution. Future studies incorporating spatially resolved measurements of the thermal gas, radio spectral ageing, and more detailed characterisation of host galaxies and large-scale environments will be essential for understanding the physical processes that enable radio galaxies to develop megaparsec-scale structures.

\section*{Acknowledgements}
SM \& SP acknowledge the use of the Pegasus high-performance computing facility at IUCAA for this work. The LOFAR is the Low-Frequency Array designed and constructed by ASTRON.
It has observing, data processing, and data storage facilities in several countries, which are owned by various parties (each with their own funding sources), and which are collectively operated by the ILT Foundation under a joint scientific policy.
The ILT resources have benefited from the following recent major funding sources: CNRS--INSU, Observatoire de Paris and Université d'Orléans, France; BMBF, MIWF--NRW, MPG, Germany; Science Foundation Ireland (SFI), Department of Business, Enterprise and Innovation (DBEI), Ireland; NWO, The Netherlands; the Science and Technology Facilities Council, UK; Ministry of Science and Higher Education, Poland; the Istituto Nazionale di Astrofisica (INAF), Italy.\\
This research made use of the Dutch national e-infrastructure with support of the SURF Cooperative (e-infra 180169) and the LOFAR e-infra group.
The J\"ulich LOFAR Long Term Archive and the German LOFAR network are both coordinated and operated by the J\"ulich Supercomputing Centre (JSC), and computing resources on the supercomputer JUWELS at JSC were provided by the Gauss Centre for Supercomputing e.V. (grant CHTB00) through the John von Neumann Institute for Computing (NIC).\\
This research made use of the University of Hertfordshire high-performance computing facility and the LOFAR-UK computing facility located at the University of Hertfordshire and supported by STFC [ST/P000096/1], and of the Italian LOFAR IT computing infrastructure supported and operated by INAF, and by the Physics Department of the University of Turin (under an agreement with Consorzio Interuniversitario per la Fisica Spaziale) at the C3S Supercomputing Centre, Italy.\\
%


\bibliographystyle{aasjournal}
\bibliography{references}

@ARTICLE{Blandford2019,
       author = {{Blandford}, Roger and {Meier}, David and {Readhead}, Anthony},
        title = "{Relativistic Jets from Active Galactic Nuclei}",
      journal = {\araa},
         year = 2019,
        month = aug,
       volume = {57},
        pages = {467-509},
          doi = {10.1146/annurev-astro-081817-051948},
archivePrefix = {arXiv},
       eprint = {1812.06025},
 primaryClass = {astro-ph.HE},
       adsurl = {https://ui.adsabs.harvard.edu/abs/2019ARA&A..57..467B}
}

@ARTICLE{Magliocchetti2022,
       author = {{Magliocchetti}, Manuela},
        title = "{Hosts and environments: a (large-scale) radio history of AGN and star-forming galaxies}",
      journal = {\aapr},
         year = 2022,
        month = dec,
       volume = {30},
       number = {1},
          eid = {6},
        pages = {6},
          doi = {10.1007/s00159-022-00142-1},
archivePrefix = {arXiv},
       eprint = {2206.15286},
 primaryClass = {astro-ph.CO},
       adsurl = {https://ui.adsabs.harvard.edu/abs/2022A&ARv..30....6M}
}

@ARTICLE{FR74,
       author = {{Fanaroff}, B.~L. and {Riley}, J.~M.},
        title = "{The morphology of extragalactic radio sources of high and low luminosity}",
      journal = {\mnras},
         year = 1974,
        month = may,
       volume = {167},
        pages = {31P-36P},
          doi = {10.1093/mnras/167.1.31P},
       adsurl = {https://ui.adsabs.harvard.edu/abs/1974MNRAS.167P..31F}
}

@ARTICLE{Willis1974,
       author = {{Willis}, A.~G. and {Strom}, R.~G. and {Wilson}, A.~S.},
        title = "{3C236, DA240; the largest radio sources known}",
      journal = {\nat},
         year = 1974,
        month = aug,
       volume = {250},
       number = {5468},
        pages = {625-630},
          doi = {10.1038/250625a0},
       adsurl = {https://ui.adsabs.harvard.edu/abs/1974Natur.250..625W}
}

@ARTICLE{Kuzmicz2018,
       author = {{Ku{\'z}micz}, Agnieszka and {Jamrozy}, Marek and {Bronarska}, Katarzyna and {Janda-Boczar}, Katarzyna and {Saikia}, D.~J.},
        title = "{An Updated Catalog of Giant Radio Sources}",
      journal = {\apjs},
         year = 2018,
        month = sep,
       volume = {238},
       number = {1},
          eid = {9},
        pages = {9},
          doi = {10.3847/1538-4365/aad9ff},
archivePrefix = {arXiv},
       eprint = {1809.09008},
 primaryClass = {astro-ph.GA},
       adsurl = {https://ui.adsabs.harvard.edu/abs/2018ApJS..238....9K}
}

@ARTICLE{FIRST1995,
       author = {{Becker}, Robert H. and {White}, Richard L. and {Helfand}, David J.},
        title = "{The FIRST Survey: Faint Images of the Radio Sky at Twenty Centimeters}",
      journal = {\apj},
         year = 1995,
        month = sep,
       volume = {450},
        pages = {559},
          doi = {10.1086/176166},
       adsurl = {https://ui.adsabs.harvard.edu/abs/1995ApJ...450..559B}
}

@ARTICLE{WENSS1997,
       author = {{Rengelink}, R.~B. and {Tang}, Y. and {de Bruyn}, A.~G. and {Miley}, G.~K. and {Bremer}, M.~N. and {Roettgering}, H.~J.~A. and {Bremer}, M.~A.~R.},
        title = "{The Westerbork Northern Sky Survey (WENSS), I. A 570 square degree Mini-Survey around the North Ecliptic Pole}",
      journal = {\aaps},
         year = 1997,
        month = aug,
       volume = {124},
        pages = {259-280},
          doi = {10.1051/aas:1997358},
       adsurl = {https://ui.adsabs.harvard.edu/abs/1997A&AS..124..259R}
}

@ARTICLE{NVSS1998,
       author = {{Condon}, J.~J. and {Cotton}, W.~D. and {Greisen}, E.~W. and {Yin}, Q.~F. and {Perley}, R.~A. and {Taylor}, G.~B. and {Broderick}, J.~J.},
        title = "{The NRAO VLA Sky Survey}",
      journal = {\aj},
         year = 1998,
        month = may,
       volume = {115},
       number = {5},
        pages = {1693-1716},
          doi = {10.1086/300337},
       adsurl = {https://ui.adsabs.harvard.edu/abs/1998AJ....115.1693C}
}

@ARTICLE{SUMSS2003,
       author = {{Mauch}, T. and {Murphy}, T. and {Buttery}, H.~J. and {Curran}, J. and {Hunstead}, R.~W. and {Piestrzynski}, B. and {Robertson}, J.~G. and {Sadler}, E.~M.},
        title = "{SUMSS: a wide-field radio imaging survey of the southern sky - II. The source catalogue}",
      journal = {\mnras},
         year = 2003,
        month = jul,
       volume = {342},
       number = {4},
        pages = {1117-1130},
          doi = {10.1046/j.1365-8711.2003.06605.x},
archivePrefix = {arXiv},
       eprint = {astro-ph/0303188},
 primaryClass = {astro-ph},
       adsurl = {https://ui.adsabs.harvard.edu/abs/2003MNRAS.342.1117M}
}

@ARTICLE{intematgss,
       author = {{Intema}, H.~T. and {Jagannathan}, P. and {Mooley}, K.~P. and {Frail}, D.~A.},
        title = "{The GMRT 150 MHz all-sky radio survey. First alternative data release TGSS ADR1}",
      journal = {\aap},
         year = 2017,
        month = feb,
       volume = {598},
          eid = {A78},
        pages = {A78},
          doi = {10.1051/0004-6361/201628536},
archivePrefix = {arXiv},
       eprint = {1603.04368},
 primaryClass = {astro-ph.CO},
       adsurl = {https://ui.adsabs.harvard.edu/abs/2017A&A...598A..78I}
}

@ARTICLE{RACS2020,
       author = {{McConnell}, D. and {Hale}, C.~L. and {Lenc}, E. and {Banfield}, J.~K. and {Heald}, George and {Hotan}, A.~W. and {Leung}, James K. and {Moss}, Vanessa A. and {Murphy}, Tara and {O'Brien}, Andrew and {Pritchard}, Joshua and {Raja}, Wasim and {Sadler}, Elaine M. and {Stewart}, Adam and {Thomson}, Alec J.~M. and {Whiting}, M. and {Allison}, James R. and {Amy}, S.~W. and {Anderson}, C. and {Ball}, Lewis and {Bannister}, Keith W. and {Bell}, Martin and {Bock}, Douglas C.-J. and {Bolton}, Russ and {Bunton}, J.~D. and {Chippendale}, A.~P. and {Collier}, J.~D. and {Cooray}, F.~R. and {Cornwell}, T.~J. and {Diamond}, P.~J. and {Edwards}, P.~G. and {Gupta}, N. and {Hayman}, Douglas B. and {Heywood}, Ian and {Jackson}, C.~A. and {Koribalski}, B{\"a}rbel S. and {Lee-Waddell}, Karen and {McClure-Griffiths}, N.~M. and {Ng}, Alan and {Norris}, Ray P. and {Phillips}, Chris and {Reynolds}, John E. and {Roxby}, Daniel N. and {Schinckel}, Antony E.~T. and {Shields}, Matt and {Tremblay}, Chenoa and {Tzioumis}, A. and {Voronkov}, M.~A. and {Westmeier}, Tobias},
        title = "{The Rapid ASKAP Continuum Survey I: Design and first results}",
      journal = {\pasa},
         year = 2020,
        month = nov,
       volume = {37},
          eid = {e048},
        pages = {e048},
          doi = {10.1017/pasa.2020.41},
archivePrefix = {arXiv},
       eprint = {2012.00747},
 primaryClass = {astro-ph.IM},
       adsurl = {https://ui.adsabs.harvard.edu/abs/2020PASA...37...48M}
}

@ARTICLE{Is99,
       author = {{Ishwara-Chandra}, C.~H. and {Saikia}, D.~J.},
        title = "{Giant radio sources}",
      journal = {\mnras},
         year = 1999,
        month = oct,
       volume = {309},
       number = {1},
        pages = {100-112},
          doi = {10.1046/j.1365-8711.1999.02835.x},
archivePrefix = {arXiv},
       eprint = {astro-ph/9902252},
 primaryClass = {astro-ph},
       adsurl = {https://ui.adsabs.harvard.edu/abs/1999MNRAS.309..100I}
}

@ARTICLE{Ku12,
       author = {{Ku{\'z}micz}, A. and {Jamrozy}, M.},
        title = "{Optical and radio properties of giant radio quasars: central black hole characteristics}",
      journal = {\mnras},
         year = 2012,
        month = oct,
       volume = {426},
       number = {2},
        pages = {851-867},
          doi = {10.1111/j.1365-2966.2012.21576.x},
archivePrefix = {arXiv},
       eprint = {1210.1706},
 primaryClass = {astro-ph.CO},
       adsurl = {https://ui.adsabs.harvard.edu/abs/2012MNRAS.426..851K}
}

@ARTICLE{reviewgiant,
       author = {{Dabhade}, Pratik and {Saikia}, D.~J. and {Mahato}, Mousumi},
        title = "{Decoding the giant extragalactic radio sources}",
      journal = {Journal of Astrophysics and Astronomy},
         year = 2023,
        month = jun,
       volume = {44},
       number = {1},
          eid = {13},
        pages = {13},
          doi = {10.1007/s12036-022-09898-5},
archivePrefix = {arXiv},
       eprint = {2208.02130},
 primaryClass = {astro-ph.GA},
       adsurl = {https://ui.adsabs.harvard.edu/abs/2023JApA...44...13D}
}

@ARTICLE{Dab17,
       author = {{Dabhade}, Pratik and {Gaikwad}, Madhuri and {Bagchi}, Joydeep and {Pandey-Pommier}, M. and {Sankhyayan}, Shishir and {Raychaudhury}, Somak},
        title = "{Discovery of giant radio galaxies from NVSS: radio and infrared properties}",
      journal = {\mnras},
         year = 2017,
        month = aug,
       volume = {469},
       number = {3},
        pages = {2886-2906},
          doi = {10.1093/mnras/stx860},
archivePrefix = {arXiv},
       eprint = {1704.00516},
 primaryClass = {astro-ph.GA},
       adsurl = {https://ui.adsabs.harvard.edu/abs/2017MNRAS.469.2886D}
}

@ARTICLE{Da20a,
       author = {{Dabhade}, P. and {R{\"o}ttgering}, H.~J.~A. and {Bagchi}, J. and {Shimwell}, T.~W. and {Hardcastle}, M.~J. and {Sankhyayan}, S. and {Morganti}, R. and {Jamrozy}, M. and {Shulevski}, A. and {Duncan}, K.~J.},
        title = "{Giant radio galaxies in the LOFAR Two-metre Sky Survey. I. Radio and environmental properties}",
      journal = {\aap},
         year = 2020,
        month = mar,
       volume = {635},
          eid = {A5},
        pages = {A5},
          doi = {10.1051/0004-6361/201935589},
archivePrefix = {arXiv},
       eprint = {1904.00409},
 primaryClass = {astro-ph.GA},
       adsurl = {https://ui.adsabs.harvard.edu/abs/2020A&A...635A...5D}
}

@ARTICLE{Da20b,
       author = {{Dabhade}, P. and {Mahato}, M. and {Bagchi}, J. and {Saikia}, D.~J. and {Combes}, F. and {Sankhyayan}, S. and {R{\"o}ttgering}, H.~J.~A. and {Ho}, L.~C. and {Gaikwad}, M. and {Raychaudhury}, S. and {Vaidya}, B. and {Guiderdoni}, B.},
        title = "{Search and analysis of giant radio galaxies with associated nuclei (SAGAN). I. New sample and multi-wavelength studies}",
      journal = {\aap},
         year = 2020,
        month = oct,
       volume = {642},
          eid = {A153},
        pages = {A153},
          doi = {10.1051/0004-6361/202038344},
archivePrefix = {arXiv},
       eprint = {2005.03708},
 primaryClass = {astro-ph.GA},
       adsurl = {https://ui.adsabs.harvard.edu/abs/2020A&A...642A.153D}
}

@ARTICLE{Ku21,
       author = {{Ku{\'z}micz}, Agnieszka and {Jamrozy}, Marek},
        title = "{Giant Radio Quasars: Sample and Basic Properties}",
      journal = {\apjs},
         year = 2021,
        month = mar,
       volume = {253},
       number = {1},
          eid = {25},
        pages = {25},
          doi = {10.3847/1538-4365/abd483},
archivePrefix = {arXiv},
       eprint = {2012.08857},
 primaryClass = {astro-ph.GA},
       adsurl = {https://ui.adsabs.harvard.edu/abs/2021ApJS..253...25K}
}

@ARTICLE{bhukta24,
       author = {{Bhukta}, Netai and {Manik}, Souvik and {Pal}, Sabyasachi and {Mondal}, Sushanta K.},
        title = "{Discovery of Giant Radio Sources from TGSS Alternative Data Release 1: Radio, Optical, and Infrared Properties}",
      journal = {\apjs},
         year = 2024,
        month = aug,
       volume = {273},
       number = {2},
          eid = {30},
        pages = {30},
          doi = {10.3847/1538-4365/ad5184},
archivePrefix = {arXiv},
       eprint = {2201.12353},
 primaryClass = {astro-ph.GA},
       adsurl = {https://ui.adsabs.harvard.edu/abs/2024ApJS..273...30B}
}

@ARTICLE{manik25,
       author = {{Manik}, Souvik and {Bhukta}, Netai and {Pal}, Sabyasachi and {Mondal}, Sushanta K.},
        title = "{Unveiling New Giant Radio Quasars from the TGSS Sky and Their Large-scale Environment}",
      journal = {\apjs},
         year = 2025,
        month = dec,
       volume = {281},
       number = {2},
          eid = {34},
        pages = {34},
          doi = {10.3847/1538-4365/ae0b52},
       adsurl = {https://ui.adsabs.harvard.edu/abs/2025ApJS..281...34M}
}

@ARTICLE{An22,
       author = {{Andernach}, Heinz and {Jim{\'e}nez-Andrade}, Eric F. and {Willis}, Anthony G.},
        title = "{Discovery of 178 Giant Radio Galaxies in 1059 deg$^{2}$ of the Rapid ASKAP Continuum Survey at 888 MHz}",
      journal = {Galaxies},
         year = 2021,
        month = nov,
       volume = {9},
       number = {4},
          eid = {99},
        pages = {99},
          doi = {10.3390/galaxies9040099},
archivePrefix = {arXiv},
       eprint = {2111.08807},
 primaryClass = {astro-ph.GA},
       adsurl = {https://ui.adsabs.harvard.edu/abs/2021Galax...9...99A}
}

@ARTICLE{Si22,
       author = {{Simonte}, M. and {Andernach}, H. and {Br{\"u}ggen}, M. and {Schwarz}, D.~J. and {Prandoni}, I. and {Willis}, A.~G.},
        title = "{Giant radio galaxies in the LOw-Frequency ARray Two-metre Sky Survey Bo{\"o}tes deep field}",
      journal = {\mnras},
         year = 2022,
        month = sep,
       volume = {515},
       number = {2},
        pages = {2032-2052},
          doi = {10.1093/mnras/stac1911},
archivePrefix = {arXiv},
       eprint = {2207.01648},
 primaryClass = {astro-ph.GA},
       adsurl = {https://ui.adsabs.harvard.edu/abs/2022MNRAS.515.2032S}
}

@ARTICLE{Andernach25,
       author = {{Andernach}, H. and {Br{\"u}ggen}, M.},
        title = "{Properties of giant radio galaxies larger than 3 Mpc}",
      journal = {\aap},
         year = 2025,
        month = jul,
       volume = {699},
          eid = {A257},
        pages = {A257},
          doi = {10.1051/0004-6361/202452961},
archivePrefix = {arXiv},
       eprint = {2505.09181},
 primaryClass = {astro-ph.GA},
       adsurl = {https://ui.adsabs.harvard.edu/abs/2025A&A...699A.257A}
}

@ARTICLE{LOFAR2013,
       author = {{van Haarlem}, M.~P. and {Wise}, M.~W. and {Gunst}, A.~W. and {Heald}, G. and {McKean}, J.~P. and {Hessels}, J.~W.~T. and {de Bruyn}, A.~G. and {Nijboer}, R. and {Swinbank}, J. and {Fallows}, R. and {Brentjens}, M. and {Nelles}, A. and {Beck}, R. and {Falcke}, H. and {Fender}, R. and {H{\"o}randel}, J. and {Koopmans}, L.~V.~E. and {Mann}, G. and {Miley}, G. and {R{\"o}ttgering}, H. and {Stappers}, B.~W. and {Wijers}, R.~A.~M.~J. and {Zaroubi}, S. and {van den Akker}, M. and {Alexov}, A. and {Anderson}, J. and {Anderson}, K. and {van Ardenne}, A. and {Arts}, M. and {Asgekar}, A. and {Avruch}, I.~M. and {Batejat}, F. and {B{\"a}hren}, L. and {Bell}, M.~E. and {Bell}, M.~R. and {van Bemmel}, I. and {Bennema}, P. and {Bentum}, M.~J. and {Bernardi}, G. and {Best}, P. and {B{\^\i}rzan}, L. and {Bonafede}, A. and {Boonstra}, A.-J. and {Braun}, R. and {Bregman}, J. and {Breitling}, F. and {van de Brink}, R.~H. and {Broderick}, J. and {Broekema}, P.~C. and {Brouw}, W.~N. and {Br{\"u}ggen}, M. and {Butcher}, H.~R. and {van Cappellen}, W. and {Ciardi}, B. and {Coenen}, T. and {Conway}, J. and {Coolen}, A. and {Corstanje}, A. and {Damstra}, S. and {Davies}, O. and {Deller}, A.~T. and {Dettmar}, R.-J. and {van Diepen}, G. and {Dijkstra}, K. and {Donker}, P. and {Doorduin}, A. and {Dromer}, J. and {Drost}, M. and {van Duin}, A. and {Eisl{\"o}ffel}, J. and {van Enst}, J. and {Ferrari}, C. and {Frieswijk}, W. and {Gankema}, H. and {Garrett}, M.~A. and {de Gasperin}, F. and {Gerbers}, M. and {de Geus}, E. and {Grie{\ss}meier}, J.-M. and {Grit}, T. and {Gruppen}, P. and {Hamaker}, J.~P. and {Hassall}, T. and {Hoeft}, M. and {Holties}, H.~A. and {Horneffer}, A. and {van der Horst}, A. and {van Houwelingen}, A. and {Huijgen}, A. and {Iacobelli}, M. and {Intema}, H. and {Jackson}, N. and {Jelic}, V. and {de Jong}, A. and {Juette}, E. and {Kant}, D. and {Karastergiou}, A. and {Koers}, A. and {Kollen}, H. and {Kondratiev}, V.~I. and {Kooistra}, E. and {Koopman}, Y. and {Koster}, A. and {Kuniyoshi}, M. and {Kramer}, M. and {Kuper}, G. and {Lambropoulos}, P. and {Law}, C. and {van Leeuwen}, J. and {Lemaitre}, J. and {Loose}, M. and {Maat}, P. and {Macario}, G. and {Markoff}, S. and {Masters}, J. and {McFadden}, R.~A. and {McKay-Bukowski}, D. and {Meijering}, H. and {Meulman}, H. and {Mevius}, M. and {Middelberg}, E. and {Millenaar}, R. and {Miller-Jones}, J.~C.~A. and {Mohan}, R.~N. and {Mol}, J.~D. and {Morawietz}, J. and {Morganti}, R. and {Mulcahy}, D.~D. and {Mulder}, E. and {Munk}, H. and {Nieuwenhuis}, L. and {van Nieuwpoort}, R. and {Noordam}, J.~E. and {Norden}, M. and {Noutsos}, A. and {Offringa}, A.~R. and {Olofsson}, H. and {Omar}, A. and {Orr{\'u}}, E. and {Overeem}, R. and {Paas}, H. and {Pandey-Pommier}, M. and {Pandey}, V.~N. and {Pizzo}, R. and {Polatidis}, A. and {Rafferty}, D. and {Rawlings}, S. and {Reich}, W. and {de Reijer}, J.-P. and {Reitsma}, J. and {Renting}, G.~A. and {Riemers}, P. and {Rol}, E. and {Romein}, J.~W. and {Roosjen}, J. and {Ruiter}, M. and {Scaife}, A. and {van der Schaaf}, K. and {Scheers}, B. and {Schellart}, P. and {Schoenmakers}, A. and {Schoonderbeek}, G. and {Serylak}, M. and {Shulevski}, A. and {Sluman}, J. and {Smirnov}, O. and {Sobey}, C. and {Spreeuw}, H. and {Steinmetz}, M. and {Sterks}, C.~G.~M. and {Stiepel}, H.-J. and {Stuurwold}, K. and {Tagger}, M. and {Tang}, Y. and {Tasse}, C. and {Thomas}, I. and {Thoudam}, S. and {Toribio}, M.~C. and {van der Tol}, B. and {Usov}, O. and {van Veelen}, M. and {van der Veen}, A.-J. and {ter Veen}, S. and {Verbiest}, J.~P.~W. and {Vermeulen}, R. and {Vermaas}, N. and {Vocks}, C. and {Vogt}, C. and {de Vos}, M. and {van der Wal}, E. and {van Weeren}, R. and {Weggemans}, H. and {Weltevrede}, P. and {White}, S. and {Wijnholds}, S.~J. and {Wilhelmsson}, T. and {Wucknitz}, O. and {Yatawatta}, S. and {Zarka}, P. and {Zensus}, A.},
        title = "{LOFAR: The LOw-Frequency ARray}",
      journal = {\aap},
         year = 2013,
        month = aug,
       volume = {556},
          eid = {A2},
        pages = {A2},
          doi = {10.1051/0004-6361/201220873},
archivePrefix = {arXiv},
       eprint = {1305.3550},
 primaryClass = {astro-ph.IM},
       adsurl = {https://ui.adsabs.harvard.edu/abs/2013A&A...556A...2V}
}

@ARTICLE{Shimwell2019,
       author = {{Shimwell}, T.~W. and {Tasse}, C. and {Hardcastle}, M.~J. and {Mechev}, A.~P. and {Williams}, W.~L. and {Best}, P.~N. and {R{\"o}ttgering}, H.~J.~A. and {Callingham}, J.~R. and {Dijkema}, T.~J. and {de Gasperin}, F. and {Hoang}, D.~N. and {Hugo}, B. and {Mirmont}, M. and {Oonk}, J.~B.~R. and {Prandoni}, I. and {Rafferty}, D. and {Sabater}, J. and {Smirnov}, O. and {van Weeren}, R.~J. and {White}, G.~J. and {Atemkeng}, M. and {Bester}, L. and {Bonnassieux}, E. and {Br{\"u}ggen}, M. and {Brunetti}, G. and {Chy{\.z}y}, K.~T. and {Cochrane}, R. and {Conway}, J.~E. and {Croston}, J.~H. and {Danezi}, A. and {Duncan}, K. and {Haverkorn}, M. and {Heald}, G.~H. and {Iacobelli}, M. and {Intema}, H.~T. and {Jackson}, N. and {Jamrozy}, M. and {Jarvis}, M.~J. and {Lakhoo}, R. and {Mevius}, M. and {Miley}, G.~K. and {Morabito}, L. and {Morganti}, R. and {Nisbet}, D. and {Orr{\'u}}, E. and {Perkins}, S. and {Pizzo}, R.~F. and {Schrijvers}, C. and {Smith}, D.~J.~B. and {Vermeulen}, R. and {Wise}, M.~W. and {Alegre}, L. and {Bacon}, D.~J. and {van Bemmel}, I.~M. and {Beswick}, R.~J. and {Bonafede}, A. and {Botteon}, A. and {Bourke}, S. and {Brienza}, M. and {Calistro Rivera}, G. and {Cassano}, R. and {Clarke}, A.~O. and {Conselice}, C.~J. and {Dettmar}, R.~J. and {Drabent}, A. and {Dumba}, C. and {Emig}, K.~L. and {En{\ss}lin}, T.~A. and {Ferrari}, C. and {Garrett}, M.~A. and {G{\'e}nova-Santos}, R.~T. and {Goyal}, A. and {G{\"u}rkan}, G. and {Hale}, C. and {Harwood}, J.~J. and {Heesen}, V. and {Hoeft}, M. and {Horellou}, C. and {Jackson}, C. and {Kokotanekov}, G. and {Kondapally}, R. and {Kunert-Bajraszewska}, M. and {Mahatma}, V. and {Mahony}, E.~K. and {Mandal}, S. and {McKean}, J.~P. and {Merloni}, A. and {Mingo}, B. and {Miskolczi}, A. and {Mooney}, S. and {Nikiel-Wroczy{\'n}ski}, B. and {O'Sullivan}, S.~P. and {Quinn}, J. and {Reich}, W. and {Roskowi{\'n}ski}, C. and {Rowlinson}, A. and {Savini}, F. and {Saxena}, A. and {Schwarz}, D.~J. and {Shulevski}, A. and {Sridhar}, S.~S. and {Stacey}, H.~R. and {Urquhart}, S. and {van der Wiel}, M.~H.~D. and {Varenius}, E. and {Webster}, B. and {Wilber}, A.},
        title = "{The LOFAR Two-metre Sky Survey. II. First data release}",
      journal = {\aap},
         year = 2019,
        month = feb,
       volume = {622},
          eid = {A1},
        pages = {A1},
          doi = {10.1051/0004-6361/201833559},
archivePrefix = {arXiv},
       eprint = {1811.07926},
 primaryClass = {astro-ph.GA},
       adsurl = {https://ui.adsabs.harvard.edu/abs/2019A&A...622A...1S}
}

@ARTICLE{Shimwell2022,
       author = {{Shimwell}, T.~W. and {Hardcastle}, M.~J. and {Tasse}, C. and {Best}, P.~N. and {R{\"o}ttgering}, H.~J.~A. and {Williams}, W.~L. and {Botteon}, A. and {Drabent}, A. and {Mechev}, A. and {Shulevski}, A. and {van Weeren}, R.~J. and {Bester}, L. and {Br{\"u}ggen}, M. and {Brunetti}, G. and {Callingham}, J.~R. and {Chy{\.z}y}, K.~T. and {Conway}, J.~E. and {Dijkema}, T.~J. and {Duncan}, K. and {de Gasperin}, F. and {Hale}, C.~L. and {Haverkorn}, M. and {Hugo}, B. and {Jackson}, N. and {Mevius}, M. and {Miley}, G.~K. and {Morabito}, L.~K. and {Morganti}, R. and {Offringa}, A. and {Oonk}, J.~B.~R. and {Rafferty}, D. and {Sabater}, J. and {Smith}, D.~J.~B. and {Schwarz}, D.~J. and {Smirnov}, O. and {O'Sullivan}, S.~P. and {Vedantham}, H. and {White}, G.~J. and {Albert}, J.~G. and {Alegre}, L. and {Asabere}, B. and {Bacon}, D.~J. and {Bonafede}, A. and {Bonnassieux}, E. and {Brienza}, M. and {Bilicki}, M. and {Bonato}, M. and {Calistro Rivera}, G. and {Cassano}, R. and {Cochrane}, R. and {Croston}, J.~H. and {Cuciti}, V. and {Dallacasa}, D. and {Danezi}, A. and {Dettmar}, R.~J. and {Di Gennaro}, G. and {Edler}, H.~W. and {En{\ss}lin}, T.~A. and {Emig}, K.~L. and {Franzen}, T.~M.~O. and {Garc{\'\i}a-Vergara}, C. and {Grange}, Y.~G. and {G{\"u}rkan}, G. and {Hajduk}, M. and {Heald}, G. and {Heesen}, V. and {Hoang}, D.~N. and {Hoeft}, M. and {Horellou}, C. and {Iacobelli}, M. and {Jamrozy}, M. and {Jeli{\'c}}, V. and {Kondapally}, R. and {Kukreti}, P. and {Kunert-Bajraszewska}, M. and {Magliocchetti}, M. and {Mahatma}, V. and {Ma{\l}ek}, K. and {Mandal}, S. and {Massaro}, F. and {Meyer-Zhao}, Z. and {Mingo}, B. and {Mostert}, R.~I.~J. and {Nair}, D.~G. and {Nakoneczny}, S.~J. and {Nikiel-Wroczy{\'n}ski}, B. and {Orr{\'u}}, E. and {Pajdosz-{\'S}mierciak}, U. and {Pasini}, T. and {Prandoni}, I. and {van Piggelen}, H.~E. and {Rajpurohit}, K. and {Retana-Montenegro}, E. and {Riseley}, C.~J. and {Rowlinson}, A. and {Saxena}, A. and {Schrijvers}, C. and {Sweijen}, F. and {Siewert}, T.~M. and {Timmerman}, R. and {Vaccari}, M. and {Vink}, J. and {West}, J.~L. and {Wo{\l}owska}, A. and {Zhang}, X. and {Zheng}, J.},
        title = "{The LOFAR Two-metre Sky Survey. V. Second data release}",
      journal = {\aap},
         year = 2022,
        month = mar,
       volume = {659},
          eid = {A1},
        pages = {A1},
          doi = {10.1051/0004-6361/202142484},
archivePrefix = {arXiv},
       eprint = {2202.11733},
 primaryClass = {astro-ph.GA},
       adsurl = {https://ui.adsabs.harvard.edu/abs/2022A&A...659A...1S}
}

@ARTICLE{Shimwell2026,
       author = {{Shimwell}, T.~W. and {Hardcastle}, M.~J. and {Tasse}, C. and {Drabent}, A. and {Botteon}, A. and {Williams}, W.~L. and {Best}, P.~N. and {R{\"o}ttgering}, H.~J.~A. and {Br{\"u}ggen}, M. and {Brunetti}, G. and {Callingham}, J.~R. and {Chy{\.z}y}, K.~T. and {Conway}, J.~E. and {De Gasperin}, F. and {Haverkorn}, M. and {Horellou}, C. and {Jackson}, N. and {Miley}, G.~K. and {Morabito}, L.~K. and {Morganti}, R. and {O'Sullivan}, S.~P. and {Schwarz}, D.~J. and {Smith}, D.~J.~B. and {van Weeren}, R.~J. and {Vedantham}, H.~K. and {White}, G.~J. and {Ahmadi}, A. and {Alegre}, L. and {Arias}, M. and {Asabere}, B. and {Bahr-Kalus}, B. and {Barkus}, B. and {Bilicki}, M. and {B{\"o}hme}, L. and {Brentjens}, M. and {Brienza}, M. and {Bomans}, D.~J. and {Bonafede}, A. and {Bonato}, M. and {Bonnassieux}, E. and {Boxelaar}, J.~M. and {Camera}, S. and {Cassano}, R. and {Chilufya}, J. and {Cianfaglione}, M. and {Croston}, J.~H. and {Cuciti}, V. and {Dabhade}, P. and {De Rubeis}, E. and {de Jong}, J.~M.~G.~H.~J. and {Dallacasa}, D. and {Dettmar}, R.~J. and {Duncan}, K.~J. and {Di Gennaro}, G. and {Edler}, H.~W. and {Groeneveld}, C. and {G{\"u}rkan}, G. and {Hajduk}, M. and {Hale}, C.~L. and {Heesen}, V. and {Hoang}, D.~N. and {Hoeft}, M. and {Holties}, H. and {Horton}, M.~A. and {Iacobelli}, M. and {Jamrozy}, M. and {Jarvis}, M.~J. and {Jelic}, V. and {Kadler}, M. and {Kondapally}, R. and {Kunert-Bajraszewska}, M. and {Loose}, M. and {Magliocchetti}, M. and {Ma{\l}ek}, K. and {Manzano}, C. and {McKean}, J.~P. and {Mevius}, M. and {Mingo}, B. and {Miskolczi}, A. and {Misra}, A. and {Mold{\'o}n}, J. and {Nair}, D.~G. and {Nakoneczny}, S.~J. and {Orru}, E. and {Pashapour-Ahmadabadi}, M. and {Pasini}, T. and {Petley}, J. and {Pierce}, J.~C.~S. and {Prandoni}, I. and {Rafferty}, D. and {Rajpurohit}, K. and {Riseley}, C.~J. and {Roberts}, I.~D. and {Sethi}, S. and {Shulevski}, A. and {Stein}, M. and {Stuardi}, C. and {Sweijen}, F. and {ter Veen}, S. and {Timmerman}, R. and {Vaccari}, M. and {Wijnholds}, S.},
        title = "{The LOFAR Two-metre Sky Survey: VII. Third Data Release}",
      journal = {\aap},
         year = 2026,
        month = mar,
       volume = {707},
          eid = {A198},
        pages = {A198},
          doi = {10.1051/0004-6361/202557749},
archivePrefix = {arXiv},
       eprint = {2602.15949},
 primaryClass = {astro-ph.GA},
       adsurl = {https://ui.adsabs.harvard.edu/abs/2026A&A...707A.198S}
}

@ARTICLE{Oei2023a,
       author = {{Oei}, Martijn S.~S.~L. and {van Weeren}, Reinout J. and {Gast}, Aivin R.~D.~J.~G.~I.~B. and {Botteon}, Andrea and {Hardcastle}, Martin J. and {Dabhade}, Pratik and {Shimwell}, Tim W. and {R{\"o}ttgering}, Huub J.~A. and {Drabent}, Alexander},
        title = "{Measuring the giant radio galaxy length distribution with the LoTSS}",
      journal = {\aap},
         year = 2023,
        month = apr,
       volume = {672},
          eid = {A163},
        pages = {A163},
          doi = {10.1051/0004-6361/202243572},
archivePrefix = {arXiv},
       eprint = {2210.10234},
 primaryClass = {astro-ph.GA},
       adsurl = {https://ui.adsabs.harvard.edu/abs/2023A&A...672A.163O}
}

@ARTICLE{mos24,
       author = {{Mostert}, R.~I.~J. and {Oei}, M.~S.~S.~L. and {Barkus}, B. and {Alegre}, L. and {Hardcastle}, M.~J. and {Duncan}, K.~J. and {R{\"o}ttgering}, H.~J.~A. and {van Weeren}, R.~J. and {Horton}, M.},
        title = "{Constraining the giant radio galaxy population with machine learning and Bayesian inference}",
      journal = {\aap},
         year = 2024,
        month = nov,
       volume = {691},
          eid = {A185},
        pages = {A185},
          doi = {10.1051/0004-6361/202348897},
archivePrefix = {arXiv},
       eprint = {2405.00232},
 primaryClass = {astro-ph.GA},
       adsurl = {https://ui.adsabs.harvard.edu/abs/2024A&A...691A.185M}
}

@ARTICLE{Porphyrion,
       author = {{Oei}, Martijn S.~S.~L. and {Hardcastle}, Martin J. and {Timmerman}, Roland and {Gast}, Aivin R.~D.~J.~G.~I.~B. and {Botteon}, Andrea and {Rodriguez}, Antonio C. and {Stern}, Daniel and {Calistro Rivera}, Gabriela and {van Weeren}, Reinout J. and {R{\"o}ttgering}, Huub J.~A. and {Intema}, Huib T. and {de Gasperin}, Francesco and {Djorgovski}, S.~G.},
        title = "{Black hole jets on the scale of the cosmic web}",
      journal = {\nat},
         year = 2024,
        month = sep,
       volume = {633},
       number = {8030},
        pages = {537-541},
          doi = {10.1038/s41586-024-07879-y},
archivePrefix = {arXiv},
       eprint = {2411.08630},
 primaryClass = {astro-ph.HE},
       adsurl = {https://ui.adsabs.harvard.edu/abs/2024Natur.633..537O}
}

@ARTICLE{Alcyoneus,
       author = {{Oei}, Martijn S.~S.~L. and {van Weeren}, Reinout J. and {Hardcastle}, Martin J. and {Botteon}, Andrea and {Shimwell}, Tim W. and {Dabhade}, Pratik and {Gast}, Aivin R.~D.~J.~G.~I.~B. and {R{\"o}ttgering}, Huub J.~A. and {Br{\"u}ggen}, Marcus and {Tasse}, Cyril and {Williams}, Wendy L. and {Shulevski}, Aleksandar},
        title = "{The discovery of a radio galaxy of at least 5 Mpc}",
      journal = {\aap},
         year = 2022,
        month = apr,
       volume = {660},
          eid = {A2},
        pages = {A2},
          doi = {10.1051/0004-6361/202142778},
archivePrefix = {arXiv},
       eprint = {2202.05427},
 primaryClass = {astro-ph.GA},
       adsurl = {https://ui.adsabs.harvard.edu/abs/2022A&A...660A...2O}
}

@ARTICLE{Dabhade2025ddrg,
       author = {{Dabhade}, P. and {Chavan}, K. and {Saikia}, D.~J. and {Oei}, M.~S.~S.~L. and {R{\"o}ttgering}, H.~J.~A.},
        title = "{Search and analysis of giant radio galaxies with associated nuclei (SAGAN): V. Study of giant double-double radio galaxies from LoTSS DR2}",
      journal = {\aap},
         year = 2025,
        month = apr,
       volume = {696},
          eid = {A97},
        pages = {A97},
          doi = {10.1051/0004-6361/202451918},
archivePrefix = {arXiv},
       eprint = {2408.13607},
 primaryClass = {astro-ph.GA},
       adsurl = {https://ui.adsabs.harvard.edu/abs/2025A&A...696A..97D}
}

@ARTICLE{Turner2015,
       author = {{Turner}, Ross J. and {Shabala}, Stanislav S.},
        title = "{Energetics and Lifetimes of Local Radio Active Galactic Nuclei}",
      journal = {\apj},
         year = 2015,
        month = jun,
       volume = {806},
       number = {1},
          eid = {59},
        pages = {59},
          doi = {10.1088/0004-637X/806/1/59},
archivePrefix = {arXiv},
       eprint = {1504.05204},
 primaryClass = {astro-ph.GA},
       adsurl = {https://ui.adsabs.harvard.edu/abs/2015ApJ...806...59T}
}

@ARTICLE{Hardcastle2018,
       author = {{Hardcastle}, M.~J.},
        title = "{A simulation-based analytic model of radio galaxies}",
      journal = {\mnras},
         year = 2018,
        month = apr,
       volume = {475},
       number = {2},
        pages = {2768-2786},
          doi = {10.1093/mnras/stx3358},
archivePrefix = {arXiv},
       eprint = {1801.00667},
 primaryClass = {astro-ph.HE},
       adsurl = {https://ui.adsabs.harvard.edu/abs/2018MNRAS.475.2768H}
}

@ARTICLE{Turner2023,
       author = {{Turner}, Ross J. and {Yates-Jones}, Patrick M. and {Shabala}, Stanislav S. and {Quici}, Benjamin and {Stewart}, Georgia S.~C.},
        title = "{RAiSE: simulation-based analytical model of AGN jets and lobes}",
      journal = {\mnras},
         year = 2023,
        month = jan,
       volume = {518},
       number = {1},
        pages = {945-964},
          doi = {10.1093/mnras/stac2998},
archivePrefix = {arXiv},
       eprint = {2206.09573},
 primaryClass = {astro-ph.HE},
       adsurl = {https://ui.adsabs.harvard.edu/abs/2023MNRAS.518..945T}
}

@ARTICLE{Ma08,
       author = {{Machalski}, J. and {Kozie{\l}-Wierzbowska}, D. and {Jamrozy}, M. and {Saikia}, D.~J.},
        title = "{J1420-0545: The Radio Galaxy Larger than 3C 236}",
      journal = {\apj},
         year = 2008,
        month = may,
       volume = {679},
       number = {1},
        pages = {149-155},
          doi = {10.1086/586703},
archivePrefix = {arXiv},
       eprint = {0808.2742},
 primaryClass = {astro-ph},
       adsurl = {https://ui.adsabs.harvard.edu/abs/2008ApJ...679..149M}
}

@ARTICLE{Giri2025a,
       author = {{Giri}, Gourab and {Bagchi}, Joydeep and {Thorat}, Kshitij and {Deane}, Roger P. and {Delhaize}, Jacinta and {Saikia}, Dhruba J.},
        title = "{Probing the formation of megaparsec-scale giant radio galaxies: I. Dynamical insights from magnetohydrodynamic simulations}",
      journal = {\aap},
         year = 2025,
        month = jan,
       volume = {693},
          eid = {A77},
        pages = {A77},
          doi = {10.1051/0004-6361/202451812},
archivePrefix = {arXiv},
       eprint = {2411.10864},
 primaryClass = {astro-ph.GA},
       adsurl = {https://ui.adsabs.harvard.edu/abs/2025A&A...693A..77G}
}

@ARTICLE{Giri2025b,
       author = {{Giri}, Gourab and {Fendt}, Christian and {Bagchi}, Joydeep and {Thorat}, Kshitij and {Saikia}, Dhruba J. and {Deane}, Roger P. and {Delhaize}, Jacinta},
        title = "{Probing the formation of megaparsec-scale giant radio galaxies: II. Continuum and polarization behavior from magneto-hydrodynamic simulations}",
      journal = {\aap},
         year = 2025,
        month = nov,
       volume = {703},
          eid = {A214},
        pages = {A214},
          doi = {10.1051/0004-6361/202554872},
archivePrefix = {arXiv},
       eprint = {2510.03037},
 primaryClass = {astro-ph.GA},
       adsurl = {https://ui.adsabs.harvard.edu/abs/2025A&A...703A.214G}
}

@INCOLLECTION{Wiita1989,
       author = {{Wiita}, Paul J. and {Rosen}, Alexander and {Gopal-Krishna} and {Saripalli}, L.},
        title = "{Giant Radio Galaxies via Inverse Compton Weakened Jets}",
    booktitle = {Hot Spots in Extragalactic Radio Sources},
         year = 1989,
       editor = {{Meisenheimer}, Klaus and {Roeser}, Hermann-Josef},
       volume = {327},
        pages = {173},
          doi = {10.1007/BFb0036027},
       adsurl = {https://ui.adsabs.harvard.edu/abs/1989LNP...327..173W}
}

@ARTICLE{Lan2021,
       author = {{Lan}, Ting-Wen and {Prochaska}, J. Xavier},
        title = "{On the environments of giant radio galaxies}",
      journal = {\mnras},
         year = 2021,
        month = apr,
       volume = {502},
       number = {4},
        pages = {5104-5114},
          doi = {10.1093/mnras/stab297},
archivePrefix = {arXiv},
       eprint = {2009.04482},
 primaryClass = {astro-ph.GA},
       adsurl = {https://ui.adsabs.harvard.edu/abs/2021MNRAS.502.5104L}
}

@ARTICLE{oei24,
       author = {{Oei}, Martijn S.~S.~L. and {van Weeren}, Reinout J. and {Hardcastle}, Martin J. and {Gast}, Aivin R.~D.~J.~G.~I.~B. and {Leclercq}, Florent and {R{\"o}ttgering}, Huub J.~A. and {Dabhade}, Pratik and {Shimwell}, Tim W. and {Botteon}, Andrea},
        title = "{Luminous giants populate the dense Cosmic Web. The radio luminosity-environmental density relation for radio galaxies in action}",
      journal = {\aap},
         year = 2024,
        month = jun,
       volume = {686},
          eid = {A137},
        pages = {A137},
          doi = {10.1051/0004-6361/202347115},
archivePrefix = {arXiv},
       eprint = {2404.17776},
 primaryClass = {astro-ph.CO},
       adsurl = {https://ui.adsabs.harvard.edu/abs/2024A&A...686A.137O}
}

@ARTICLE{Machalski2011,
       author = {{Machalski}, J.},
        title = "{A multifrequency study of giant radio sources - IV. Length/luminosity asymmetries and dynamical evolution of the lobes of selected sources}",
      journal = {\mnras},
         year = 2011,
        month = jun,
       volume = {413},
       number = {4},
        pages = {2429-2438},
          doi = {10.1111/j.1365-2966.2011.18314.x},
       adsurl = {https://ui.adsabs.harvard.edu/abs/2011MNRAS.413.2429M}
}

@ARTICLE{Malarecki2013,
       author = {{Malarecki}, J.~M. and {Staveley-Smith}, L. and {Saripalli}, L. and {Subrahmanyan}, R. and {Jones}, D.~H. and {Duffy}, A.~R. and {Rioja}, M.},
        title = "{Giant radio galaxies - I. Intergalactic barometers}",
      journal = {\mnras},
         year = 2013,
        month = jun,
       volume = {432},
       number = {1},
        pages = {200-224},
          doi = {10.1093/mnras/stt471},
archivePrefix = {arXiv},
       eprint = {1303.4825},
 primaryClass = {astro-ph.CO},
       adsurl = {https://ui.adsabs.harvard.edu/abs/2013MNRAS.432..200M}
}

@ARTICLE{Malarecki2015,
       author = {{Malarecki}, J.~M. and {Jones}, D.~H. and {Saripalli}, L. and {Staveley-Smith}, L. and {Subrahmanyan}, R.},
        title = "{Giant radio galaxies - II. Tracers of large-scale structure}",
      journal = {\mnras},
         year = 2015,
        month = may,
       volume = {449},
       number = {1},
        pages = {955-986},
          doi = {10.1093/mnras/stv273},
archivePrefix = {arXiv},
       eprint = {1502.03954},
 primaryClass = {astro-ph.GA},
       adsurl = {https://ui.adsabs.harvard.edu/abs/2015MNRAS.449..955M}
}

@ARTICLE{Mahato2026,
       author = {{Mahato}, Mousumi and {Tempel}, Elmo and {Sankhyayan}, Shishir and {Dabhade}, Pratik and {Chavan}, Kshitij},
        title = "{Search and analysis of giant radio galaxies with associated nuclei (SAGAN): VI: When jets meet filaments ─ Environmental imprints on the growth of giant radio galaxies}",
      journal = {\aap},
         year = 2026,
        month = feb,
       volume = {706},
          eid = {A310},
        pages = {A310},
          doi = {10.1051/0004-6361/202557646},
archivePrefix = {arXiv},
       eprint = {2512.07985},
 primaryClass = {astro-ph.GA},
       adsurl = {https://ui.adsabs.harvard.edu/abs/2026A&A...706A.310M}
}

@ARTICLE{Shimwell2017,
       author = {{Shimwell}, T.~W. and {R{\"o}ttgering}, H.~J.~A. and {Best}, P.~N. and {Williams}, W.~L. and {Dijkema}, T.~J. and {de Gasperin}, F. and {Hardcastle}, M.~J. and {Heald}, G.~H. and {Hoang}, D.~N. and {Horneffer}, A. and {Intema}, H. and {Mahony}, E.~K. and {Mandal}, S. and {Mechev}, A.~P. and {Morabito}, L. and {Oonk}, J.~B.~R. and {Rafferty}, D. and {Retana-Montenegro}, E. and {Sabater}, J. and {Tasse}, C. and {van Weeren}, R.~J. and {Br{\"u}ggen}, M. and {Brunetti}, G. and {Chy{\.z}y}, K.~T. and {Conway}, J.~E. and {Haverkorn}, M. and {Jackson}, N. and {Jarvis}, M.~J. and {McKean}, J.~P. and {Miley}, G.~K. and {Morganti}, R. and {White}, G.~J. and {Wise}, M.~W. and {van Bemmel}, I.~M. and {Beck}, R. and {Brienza}, M. and {Bonafede}, A. and {Calistro Rivera}, G. and {Cassano}, R. and {Clarke}, A.~O. and {Cseh}, D. and {Deller}, A. and {Drabent}, A. and {van Driel}, W. and {Engels}, D. and {Falcke}, H. and {Ferrari}, C. and {Fr{\"o}hlich}, S. and {Garrett}, M.~A. and {Harwood}, J.~J. and {Heesen}, V. and {Hoeft}, M. and {Horellou}, C. and {Israel}, F.~P. and {Kapi{\'n}ska}, A.~D. and {Kunert-Bajraszewska}, M. and {McKay}, D.~J. and {Mohan}, N.~R. and {Orr{\'u}}, E. and {Pizzo}, R.~F. and {Prandoni}, I. and {Schwarz}, D.~J. and {Shulevski}, A. and {Sipior}, M. and {Smith}, D.~J.~B. and {Sridhar}, S.~S. and {Steinmetz}, M. and {Stroe}, A. and {Varenius}, E. and {van der Werf}, P.~P. and {Zensus}, J.~A. and {Zwart}, J.~T.~L.},
        title = "{The LOFAR Two-metre Sky Survey. I. Survey description and preliminary data release}",
      journal = {\aap},
         year = 2017,
        month = feb,
       volume = {598},
          eid = {A104},
        pages = {A104},
          doi = {10.1051/0004-6361/201629313},
archivePrefix = {arXiv},
       eprint = {1611.02700},
 primaryClass = {astro-ph.IM},
       adsurl = {https://ui.adsabs.harvard.edu/abs/2017A&A...598A.104S}
}

@article{giri2026,
       author = {{Giri}, Gourab and {Borgogno}, Dario and {Tavani}, Marco and {Mignone}, Andrea and {Sharma}, Prateek and {Gheller}, Claudio and {Vittorini}, Valerio and {Wiita}, Paul J. and {Fanaroff}, Bernie and {Suriano}, Alessio and {Saikia}, D.~J. and {Brunetti}, Gianfranco},
        title = "{Simulating megaparsec-scale jets of radio galaxies: magnetohydrodynamics of jets reaching 5 Mpc}",
      journal = {\mnras},
         year = 2026,
        month = aug,
       volume = {550},
       number = {3},
          eid = {stag1243},
        pages = {stag1243},
          doi = {10.1093/mnras/stag1243},
archivePrefix = {arXiv},
       eprint = {2606.28019},
 primaryClass = {astro-ph.GA},
       adsurl = {https://ui.adsabs.harvard.edu/abs/2026MNRAS.550g1243G}
}

@article{ShabalaGodfrey2013,
       author = {{Shabala}, S.~S. and {Godfrey}, L.~E.~H.},
        title = "{Size Dependence of the Radio-luminosity-Mechanical-power Correlation in Radio Galaxies}",
      journal = {\apj},
         year = 2013,
        month = jun,
       volume = {769},
       number = {2},
          eid = {129},
        pages = {129},
          doi = {10.1088/0004-637X/769/2/129},
archivePrefix = {arXiv},
       eprint = {1304.6089},
 primaryClass = {astro-ph.CO},
       adsurl = {https://ui.adsabs.harvard.edu/abs/2013ApJ...769..129S}
}

@ARTICLE{sdss16,
       author = {{Ahumada}, Romina and {Allende Prieto}, Carlos and {Almeida}, Andr{\'e}s and {Anders}, Friedrich and {Anderson}, Scott F. and {Andrews}, Brett H. and {Anguiano}, Borja and {Arcodia}, Riccardo and {Armengaud}, Eric and {Aubert}, Marie and {Avila}, Santiago and {Avila-Reese}, Vladimir and {Badenes}, Carles and {Balland}, Christophe and {Barger}, Kat and {Barrera-Ballesteros}, Jorge K. and {Basu}, Sarbani and {Bautista}, Julian and {Beaton}, Rachael L. and {Beers}, Timothy C. and {Benavides}, B. Izamar T. and {Bender}, Chad F. and {Bernardi}, Mariangela and {Bershady}, Matthew and {Beutler}, Florian and {Bidin}, Christian Moni and {Bird}, Jonathan and {Bizyaev}, Dmitry and {Blanc}, Guillermo A. and {Blanton}, Michael R. and {Boquien}, M{\'e}d{\'e}ric and {Borissova}, Jura and {Bovy}, Jo and {Brandt}, W.~N. and {Brinkmann}, Jonathan and {Brownstein}, Joel R. and {Bundy}, Kevin and {Bureau}, Martin and {Burgasser}, Adam and {Burtin}, Etienne and {Cano-D{\'\i}az}, Mariana and {Capasso}, Raffaella and {Cappellari}, Michele and {Carrera}, Ricardo and {Chabanier}, Sol{\`e}ne and {Chaplin}, William and {Chapman}, Michael and {Cherinka}, Brian and {Chiappini}, Cristina and {Doohyun Choi}, Peter and {Chojnowski}, S. Drew and {Chung}, Haeun and {Clerc}, Nicolas and {Coffey}, Damien and {Comerford}, Julia M. and {Comparat}, Johan and {da Costa}, Luiz and {Cousinou}, Marie-Claude and {Covey}, Kevin and {Crane}, Jeffrey D. and {Cunha}, Katia and {Ilha}, Gabriele da Silva and {Dai}, Yu Sophia and {Damsted}, Sanna B. and {Darling}, Jeremy and {Davidson}, Jr., James W. and {Davies}, Roger and {Dawson}, Kyle and {De}, Nikhil and {de la Macorra}, Axel and {De Lee}, Nathan and {Queiroz}, Anna B{\'a}rbara de Andrade and {Deconto Machado}, Alice and {de la Torre}, Sylvain and {Dell'Agli}, Flavia and {du Mas des Bourboux}, H{\'e}lion and {Diamond-Stanic}, Aleksandar M. and {Dillon}, Sean and {Donor}, John and {Drory}, Niv and {Duckworth}, Chris and {Dwelly}, Tom and {Ebelke}, Garrett and {Eftekharzadeh}, Sarah and {Davis Eigenbrot}, Arthur and {Elsworth}, Yvonne P. and {Eracleous}, Mike and {Erfanianfar}, Ghazaleh and {Escoffier}, Stephanie and {Fan}, Xiaohui and {Farr}, Emily and {Fern{\'a}ndez-Trincado}, Jos{\'e} G. and {Feuillet}, Diane and {Finoguenov}, Alexis and {Fofie}, Patricia and {Fraser-McKelvie}, Amelia and {Frinchaboy}, Peter M. and {Fromenteau}, Sebastien and {Fu}, Hai and {Galbany}, Llu{\'\i}s and {Garcia}, Rafael A. and {Garc{\'\i}a-Hern{\'a}ndez}, D.~A. and {Garma Oehmichen}, Luis Alberto and {Ge}, Junqiang and {Geimba Maia}, Marcio Antonio and {Geisler}, Doug and {Gelfand}, Joseph and {Goddy}, Julian and {Gonzalez-Perez}, Violeta and {Grabowski}, Kathleen and {Green}, Paul and {Grier}, Catherine J. and {Guo}, Hong and {Guy}, Julien and {Harding}, Paul and {Hasselquist}, Sten and {Hawken}, Adam James and {Hayes}, Christian R. and {Hearty}, Fred and {Hekker}, S. and {Hogg}, David W. and {Holtzman}, Jon A. and {Horta}, Danny and {Hou}, Jiamin and {Hsieh}, Bau-Ching and {Huber}, Daniel and {Hunt}, Jason A.~S. and {Ider Chitham}, J. and {Imig}, Julie and {Jaber}, Mariana and {Jimenez Angel}, Camilo Eduardo and {Johnson}, Jennifer A. and {Jones}, Amy M. and {J{\"o}nsson}, Henrik and {Jullo}, Eric and {Kim}, Yerim and {Kinemuchi}, Karen and {Kirkpatrick}, IV, Charles C. and {Kite}, George W. and {Klaene}, Mark and {Kneib}, Jean-Paul and {Kollmeier}, Juna A. and {Kong}, Hui and {Kounkel}, Marina and {Krishnarao}, Dhanesh and {Lacerna}, Ivan and {Lan}, Ting-Wen and {Lane}, Richard R. and {Law}, David R. and {Le Goff}, Jean-Marc and {Leung}, Henry W. and {Lewis}, Hannah and {Li}, Cheng and {Lian}, Jianhui and {Lin}, Lihwai and {Long}, Dan and {Longa-Pe{\~n}a}, Pen{\'e}lope and {Lundgren}, Britt and {Lyke}, Brad W. and {Mackereth}, J. Ted and {MacLeod}, Chelsea L. and {Majewski}, Steven R. and {Manchado}, Arturo and {Maraston}, Claudia and {Martini}, Paul and {Masseron}, Thomas and {Masters}, Karen L. and {Mathur}, Savita and {McDermid}, Richard M. and {Merloni}, Andrea and {Merrifield}, Michael and {M{\'e}sz{\'a}ros}, Szabolcs and {Miglio}, Andrea and {Minniti}, Dante and {Minsley}, Rebecca and {Miyaji}, Takamitsu and {Mohammad}, Faizan Gohar and {Mosser}, Benoit and {Mueller}, Eva-Maria and {Muna}, Demitri and {Mu{\~n}oz-Guti{\'e}rrez}, Andrea and {Myers}, Adam D. and {Nadathur}, Seshadri and {Nair}, Preethi and {Nandra}, Kirpal and {Correa do Nascimento}, Janaina and {Nevin}, Rebecca Jean and {Newman}, Jeffrey A. and {Nidever}, David L. and {Nitschelm}, Christian and {Noterdaeme}, Pasquier and {O'Connell}, Julia E. and {Olmstead}, Matthew D. and {Oravetz}, Daniel and {Oravetz}, Audrey and {Osorio}, Yeisson and {Pace}, Zachary J. and {Padilla}, Nelson and {Palanque-Delabrouille}, Nathalie and {Palicio}, Pedro A.},
        title = "{The 16th Data Release of the Sloan Digital Sky Surveys: First Release from the APOGEE-2 Southern Survey and Full Release of eBOSS Spectra}",
      journal = {\apjs},
         year = 2020,
        month = jul,
       volume = {249},
       number = {1},
          eid = {3},
        pages = {3},
          doi = {10.3847/1538-4365/ab929e},
archivePrefix = {arXiv},
       eprint = {1912.02905},
 primaryClass = {astro-ph.GA},
       adsurl = {https://ui.adsabs.harvard.edu/abs/2020ApJS..249....3A}
}

@article{duncan22,
       author = {{Duncan}, Kenneth J.},
        title = "{All-purpose, all-sky photometric redshifts for the Legacy Imaging Surveys Data Release 8}",
      journal = {\mnras},
         year = 2022,
        month = may,
       volume = {512},
       number = {3},
        pages = {3662-3683},
          doi = {10.1093/mnras/stac608},
archivePrefix = {arXiv},
       eprint = {2203.01949},
 primaryClass = {astro-ph.GA},
       adsurl = {https://ui.adsabs.harvard.edu/abs/2022MNRAS.512.3662D}
}

@ARTICLE{Hardcastle1998,
       author = {{Hardcastle}, M.~J. and {Worrall}, D.~M. and {Birkinshaw}, M.},
        title = "{Dynamics of the radio galaxy 3C449}",
      journal = {\mnras},
         year = 1998,
        month = jun,
       volume = {296},
       number = {4},
        pages = {1098-1104},
          doi = {10.1046/j.1365-8711.1998.01535.x},
archivePrefix = {arXiv},
       eprint = {astro-ph/9802175},
 primaryClass = {astro-ph},
       adsurl = {https://ui.adsabs.harvard.edu/abs/1998MNRAS.296.1098H}
}

@ARTICLE{BeckKrause2005,
       author = {{Beck}, R. and {Krause}, M.},
        title = "{Revised equipartition and minimum energy formula for magnetic field strength estimates from radio synchrotron observations}",
      journal = {Astronomische Nachrichten},
         year = 2005,
        month = jul,
       volume = {326},
       number = {6},
        pages = {414-427},
          doi = {10.1002/asna.200510366},
archivePrefix = {arXiv},
       eprint = {astro-ph/0507367},
 primaryClass = {astro-ph},
       adsurl = {https://ui.adsabs.harvard.edu/abs/2005AN....326..414B}
}

@ARTICLE{Zheng17skymodel,
       author = {{Zheng}, H. and {Tegmark}, M. and {Dillon}, J.~S. and {Kim}, D.~A. and {Liu}, A. and {Neben}, A.~R. and {Jonas}, J. and {Reich}, P. and {Reich}, W.},
        title = "{An improved model of diffuse galactic radio emission from 10 MHz to 5 THz}",
      journal = {\mnras},
         year = 2017,
        month = jan,
       volume = {464},
       number = {3},
        pages = {3486-3497},
          doi = {10.1093/mnras/stw2525},
archivePrefix = {arXiv},
       eprint = {1605.04920},
 primaryClass = {astro-ph.CO},
       adsurl = {https://ui.adsabs.harvard.edu/abs/2017MNRAS.464.3486Z}
}

@ARTICLE{KH13,
       author = {{Kormendy}, John and {Ho}, Luis C.},
        title = "{Coevolution (Or Not) of Supermassive Black Holes and Host Galaxies}",
      journal = {\araa},
         year = 2013,
        month = aug,
       volume = {51},
       number = {1},
        pages = {511-653},
          doi = {10.1146/annurev-astro-082708-101811},
archivePrefix = {arXiv},
       eprint = {1304.7762},
 primaryClass = {astro-ph.CO},
       adsurl = {https://ui.adsabs.harvard.edu/abs/2013ARA&A..51..511K}
}

@ARTICLE{si24,
       author = {{Simonte}, M. and {Andernach}, H. and {Br{\"u}ggen}, M. and {Miley}, G.~K. and {Barthel}, P.},
        title = "{Giant radio galaxies in the LOFAR deep fields}",
      journal = {\aap},
         year = 2024,
        month = jun,
       volume = {686},
          eid = {A21},
        pages = {A21},
          doi = {10.1051/0004-6361/202348904},
archivePrefix = {arXiv},
       eprint = {2403.08037},
 primaryClass = {astro-ph.GA},
       adsurl = {https://ui.adsabs.harvard.edu/abs/2024A&A...686A..21S}
}

@ARTICLE{mingo19,
       author = {{Mingo}, B. and {Croston}, J.~H. and {Hardcastle}, M.~J. and {Best}, P.~N. and {Duncan}, K.~J. and {Morganti}, R. and {Rottgering}, H.~J.~A. and {Sabater}, J. and {Shimwell}, T.~W. and {Williams}, W.~L. and {Brienza}, M. and {Gurkan}, G. and {Mahatma}, V.~H. and {Morabito}, L.~K. and {Prandoni}, I. and {Bondi}, M. and {Ineson}, J. and {Mooney}, S.},
        title = "{Revisiting the Fanaroff-Riley dichotomy and radio-galaxy morphology with the LOFAR Two-Metre Sky Survey (LoTSS)}",
      journal = {\mnras},
         year = 2019,
        month = sep,
       volume = {488},
       number = {2},
        pages = {2701-2721},
          doi = {10.1093/mnras/stz1901},
archivePrefix = {arXiv},
       eprint = {1907.03726},
 primaryClass = {astro-ph.GA},
       adsurl = {https://ui.adsabs.harvard.edu/abs/2019MNRAS.488.2701M}
}

@ARTICLE{Ursini2018,
       author = {{Ursini}, F. and {Bassani}, L. and {Panessa}, F. and {Bird}, A.~J. and {Bruni}, G. and {Fiocchi}, M. and {Malizia}, A. and {Saripalli}, L. and {Ubertini}, P.},
        title = "{Hard X-ray-selected giant radio galaxies - I. The X-ray properties and radio connection}",
      journal = {\mnras},
         year = 2018,
        month = dec,
       volume = {481},
       number = {3},
        pages = {4250-4260},
          doi = {10.1093/mnras/sty2547},
archivePrefix = {arXiv},
       eprint = {1809.05297},
 primaryClass = {astro-ph.HE},
       adsurl = {https://ui.adsabs.harvard.edu/abs/2018MNRAS.481.4250U}
}

@ARTICLE{oei23,
       author = {{Oei}, Martijn S.~S.~L. and {van Weeren}, Reinout J. and {Hardcastle}, Martin J. and {Vazza}, Franco and {Shimwell}, Tim W. and {Leclercq}, Florent and {Br{\"u}ggen}, Marcus and {R{\"o}ttgering}, Huub J.~A.},
        title = "{An intergalactic medium temperature from a giant radio galaxy}",
      journal = {\mnras},
         year = 2023,
        month = jan,
       volume = {518},
       number = {1},
        pages = {240-256},
          doi = {10.1093/mnras/stac2948},
archivePrefix = {arXiv},
       eprint = {2210.10156},
 primaryClass = {astro-ph.GA},
       adsurl = {https://ui.adsabs.harvard.edu/abs/2023MNRAS.518..240O}
}

@ARTICLE{Konar2004,
       author = {{Konar}, C. and {Saikia}, D.~J. and {Ishwara-Chandra}, C.~H. and {Kulkarni}, V.~K.},
        title = "{Radio observations of a few giant sources}",
      journal = {\mnras},
         year = 2004,
        month = dec,
       volume = {355},
       number = {3},
        pages = {845-854},
          doi = {10.1111/j.1365-2966.2004.08362.x},
archivePrefix = {arXiv},
       eprint = {astro-ph/0408534},
 primaryClass = {astro-ph},
       adsurl = {https://ui.adsabs.harvard.edu/abs/2004MNRAS.355..845K}
}

@ARTICLE{wu22,
       author = {{Wu}, Qiaoya and {Shen}, Yue},
        title = "{A Catalog of Quasar Properties from Sloan Digital Sky Survey Data Release 16}",
      journal = {\apjs},
         year = 2022,
        month = dec,
       volume = {263},
       number = {2},
          eid = {42},
        pages = {42},
          doi = {10.3847/1538-4365/ac9ead},
archivePrefix = {arXiv},
       eprint = {2209.03987},
 primaryClass = {astro-ph.GA},
       adsurl = {https://ui.adsabs.harvard.edu/abs/2022ApJS..263...42W}
}

@ARTICLE{gurkan14,
       author = {{G{\"u}rkan}, G. and {Hardcastle}, M.~J. and {Jarvis}, M.~J.},
        title = "{The Wide-field Infrared Survey Explorer properties of complete samples of radio-loud active galactic nucleus}",
      journal = {\mnras},
         year = 2014,
        month = feb,
       volume = {438},
       number = {2},
        pages = {1149-1161},
          doi = {10.1093/mnras/stt2264},
archivePrefix = {arXiv},
       eprint = {1308.4843},
 primaryClass = {astro-ph.CO},
       adsurl = {https://ui.adsabs.harvard.edu/abs/2014MNRAS.438.1149G}
}

@ARTICLE{Manik26,
       author = {{Manik}, Souvik and {Kumari}, Shobha and {Bhukta}, Netai and {Pal}, Sabyasachi and {Mondal}, Sushanta K.},
        title = "{Hybrid Morphology Radio Sources from the MeerKAT Absorption Line Survey (MALS): Radio, Mid-infrared, and Environmental Characteristics}",
      journal = {\apj},
         year = 2026,
        month = feb,
       volume = {997},
       number = {2},
          eid = {157},
        pages = {157},
          doi = {10.3847/1538-4357/ae1f85},
archivePrefix = {arXiv},
       eprint = {2506.20211},
 primaryClass = {astro-ph.GA},
       adsurl = {https://ui.adsabs.harvard.edu/abs/2026ApJ...997..157M}
}

@ARTICLE{Hardcastle2019,
       author = {{Hardcastle}, M.~J. and {Williams}, W.~L. and {Best}, P.~N. and {Croston}, J.~H. and {Duncan}, K.~J. and {R{\"o}ttgering}, H.~J.~A. and {Sabater}, J. and {Shimwell}, T.~W. and {Tasse}, C. and {Callingham}, J.~R. and {Cochrane}, R.~K. and {de Gasperin}, F. and {G{\"u}rkan}, G. and {Jarvis}, M.~J. and {Mahatma}, V. and {Miley}, G.~K. and {Mingo}, B. and {Mooney}, S. and {Morabito}, L.~K. and {O'Sullivan}, S.~P. and {Prandoni}, I. and {Shulevski}, A. and {Smith}, D.~J.~B.},
        title = "{Radio-loud AGN in the first LoTSS data release. The lifetimes and environmental impact of jet-driven sources}",
      journal = {\aap},
         year = 2019,
        month = feb,
       volume = {622},
          eid = {A12},
        pages = {A12},
          doi = {10.1051/0004-6361/201833893},
archivePrefix = {arXiv},
       eprint = {1811.07943},
 primaryClass = {astro-ph.GA},
       adsurl = {https://ui.adsabs.harvard.edu/abs/2019A&A...622A..12H}
}

@ARTICLE{Arnaud10,
       author = {{Arnaud}, M. and {Pratt}, G.~W. and {Piffaretti}, R. and {B{\"o}hringer}, H. and {Croston}, J.~H. and {Pointecouteau}, E.},
        title = "{The universal galaxy cluster pressure profile from a representative sample of nearby systems (REXCESS) and the Y$_{SZ}$ - M$_{500}$ relation}",
      journal = {\aap},
         year = 2010,
        month = jul,
       volume = {517},
          eid = {A92},
        pages = {A92},
          doi = {10.1051/0004-6361/200913416},
archivePrefix = {arXiv},
       eprint = {0910.1234},
 primaryClass = {astro-ph.CO},
       adsurl = {https://ui.adsabs.harvard.edu/abs/2010A&A...517A..92A}
}

@ARTICLE{KaiserDennettThorpeAlexander1997,
       author = {{Kaiser}, Christian R. and {Dennett-Thorpe}, Jane and {Alexander}, Paul},
        title = "{Evolutionary tracks of FRII sources through the P-D diagram}",
      journal = {\mnras},
         year = 1997,
        month = dec,
       volume = {292},
       number = {3},
        pages = {723-732},
          doi = {10.1093/mnras/292.3.723},
archivePrefix = {arXiv},
       eprint = {astro-ph/9710104},
 primaryClass = {astro-ph},
       adsurl = {https://ui.adsabs.harvard.edu/abs/1997MNRAS.292..723K}
}

@ARTICLE{KaiserAlexander1997,
       author = {{Kaiser}, Christian R. and {Alexander}, Paul},
        title = "{A self-similar model for extragalactic radio sources}",
      journal = {\mnras},
         year = 1997,
        month = mar,
       volume = {286},
       number = {1},
        pages = {215-222},
          doi = {10.1093/mnras/286.1.215},
       adsurl = {https://ui.adsabs.harvard.edu/abs/1997MNRAS.286..215K}
}

@ARTICLE{Schoenmakers2000,
       author = {{Schoenmakers}, Arno P. and {de Bruyn}, A.~G. and {R{\"o}ttgering}, H.~J.~A. and {van der Laan}, H. and {Kaiser}, C.~R.},
        title = "{Radio galaxies with a `double-double morphology' - I. Analysis of the radio properties and evidence for interrupted activity in active galactic nuclei}",
      journal = {\mnras},
         year = 2000,
        month = jun,
       volume = {315},
       number = {2},
        pages = {371-380},
          doi = {10.1046/j.1365-8711.2000.03430.x},
archivePrefix = {arXiv},
       eprint = {astro-ph/9912141},
 primaryClass = {astro-ph},
       adsurl = {https://ui.adsabs.harvard.edu/abs/2000MNRAS.315..371S}
}

@ARTICLE{LedlowOwen1996,
       author = {{Ledlow}, Michael J. and {Owen}, Frazer N.},
        title = "{20 CM VLA Survey of Abell Clusters of Galaxies. VI. Radio/Optical Luminosity Functions}",
      journal = {\aj},
         year = 1996,
        month = jul,
       volume = {112},
        pages = {9},
          doi = {10.1086/117985},
archivePrefix = {arXiv},
       eprint = {astro-ph/9607014},
 primaryClass = {astro-ph},
       adsurl = {https://ui.adsabs.harvard.edu/abs/1996AJ....112....9L}
}

@ARTICLE{Clews2025,
       author = {{Clews}, L. and {Croston}, J.~H. and {Dickinson}, H. and {Mingo}, B. and {Hardcastle}, M.~J. and {Barkus}, B. and {de Jong}, J.~M.~G.~H.~J. and {R{\"o}ttgering}, H.~J.~A.},
        title = "{Radio-loud AGN morphology and host-galaxy properties in the LOFAR Two-metre Sky Survey Data Release 2}",
      journal = {\mnras},
         year = 2025,
        month = aug,
       volume = {541},
       number = {4},
        pages = {3452-3467},
          doi = {10.1093/mnras/staf966},
archivePrefix = {arXiv},
       eprint = {2506.08878},
 primaryClass = {astro-ph.GA},
       adsurl = {https://ui.adsabs.harvard.edu/abs/2025MNRAS.541.3452C}
}

@ARTICLE{Gurkan2022,
       author = {{G{\"u}rkan}, G{\"u}lay and {Prandoni}, I. and {O'Brien}, A. and {Raja}, W. and {Marchetti}, L. and {Vaccari}, M. and {Driver}, S. and {Taylor}, E. and {Franzen}, T. and {Brown}, M.~J.~I. and {Shabala}, S. and {Andernach}, H. and {Hopkins}, A.~M. and {Norris}, R.~P. and {Leahy}, D. and {Bilicki}, M. and {Farajollahi}, H. and {Galvin}, T. and {Heald}, G. and {Koribalski}, B.~S. and {An}, T. and {Warhurst}, K.},
        title = "{Deep ASKAP EMU Survey of the GAMA23 field: properties of radio sources}",
      journal = {\mnras},
         year = 2022,
        month = jun,
       volume = {512},
       number = {4},
        pages = {6104-6121},
          doi = {10.1093/mnras/stac880},
archivePrefix = {arXiv},
       eprint = {2203.14727},
 primaryClass = {astro-ph.GA},
       adsurl = {https://ui.adsabs.harvard.edu/abs/2022MNRAS.512.6104G}
}

@ARTICLE{WuCao2008,
       author = {{Wu}, Qingwen and {Cao}, Xinwu},
        title = "{Jet Power Extracted from ADAF and the Applications to X-Ray Binaries and the Radio Galaxy FR Dichotomy}",
      journal = {\apj},
         year = 2008,
        month = nov,
       volume = {687},
       number = {1},
        pages = {156-161},
          doi = {10.1086/591933},
archivePrefix = {arXiv},
       eprint = {0807.2288},
 primaryClass = {astro-ph},
       adsurl = {https://ui.adsabs.harvard.edu/abs/2008ApJ...687..156W}
}

@ARTICLE{WH22,
       author = {{Wen}, Z.~L. and {Han}, J.~L.},
        title = "{Clusters of galaxies up to z = 1.5 identified from photometric data of the Dark Energy Survey and unWISE}",
      journal = {\mnras},
         year = 2022,
        month = jul,
       volume = {513},
       number = {3},
        pages = {3946-3959},
          doi = {10.1093/mnras/stac1149},
archivePrefix = {arXiv},
       eprint = {2204.11215},
 primaryClass = {astro-ph.CO},
       adsurl = {https://ui.adsabs.harvard.edu/abs/2022MNRAS.513.3946W}
}

@ARTICLE{WH15,
       author = {{Wen}, Z.~L. and {Han}, J.~L.},
        title = "{Calibration of the Optical Mass Proxy for Clusters of Galaxies and an Update of the WHL12 Cluster Catalog}",
      journal = {\apj},
         year = 2015,
        month = jul,
       volume = {807},
       number = {2},
          eid = {178},
        pages = {178},
          doi = {10.1088/0004-637X/807/2/178},
archivePrefix = {arXiv},
       eprint = {1506.04503},
 primaryClass = {astro-ph.GA},
       adsurl = {https://ui.adsabs.harvard.edu/abs/2015ApJ...807..178W}
}

@ARTICLE{WH24,
       author = {{Wen}, Z.~L. and {Han}, J.~L.},
        title = "{A Catalog of 1.58 Million Clusters of Galaxies Identified from the DESI Legacy Imaging Surveys}",
      journal = {\apjs},
         year = 2024,
        month = jun,
       volume = {272},
       number = {2},
          eid = {39},
        pages = {39},
          doi = {10.3847/1538-4365/ad409d},
archivePrefix = {arXiv},
       eprint = {2404.02002},
 primaryClass = {astro-ph.CO},
       adsurl = {https://ui.adsabs.harvard.edu/abs/2024ApJS..272...39W}
}

@ARTICLE{Ag18,
       author = {{Planck Collaboration} and {Aghanim}, N. and {Akrami}, Y. and {Ashdown}, M. and {Aumont}, J. and {Baccigalupi}, C. and {Ballardini}, M. and {Banday}, A.~J. and {Barreiro}, R.~B. and {Bartolo}, N. and {Basak}, S. and {Battye}, R. and {Benabed}, K. and {Bernard}, J.-P. and {Bersanelli}, M. and {Bielewicz}, P. and {Bock}, J.~J. and {Bond}, J.~R. and {Borrill}, J. and {Bouchet}, F.~R. and {Boulanger}, F. and {Bucher}, M. and {Burigana}, C. and {Butler}, R.~C. and {Calabrese}, E. and {Cardoso}, J.-F. and {Carron}, J. and {Challinor}, A. and {Chiang}, H.~C. and {Chluba}, J. and {Colombo}, L.~P.~L. and {Combet}, C. and {Contreras}, D. and {Crill}, B.~P. and {Cuttaia}, F. and {de Bernardis}, P. and {de Zotti}, G. and {Delabrouille}, J. and {Delouis}, J.-M. and {Di Valentino}, E. and {Diego}, J.~M. and {Dor{\'e}}, O. and {Douspis}, M. and {Ducout}, A. and {Dupac}, X. and {Dusini}, S. and {Efstathiou}, G. and {Elsner}, F. and {En{\ss}lin}, T.~A. and {Eriksen}, H.~K. and {Fantaye}, Y. and {Farhang}, M. and {Fergusson}, J. and {Fernandez-Cobos}, R. and {Finelli}, F. and {Forastieri}, F. and {Frailis}, M. and {Fraisse}, A.~A. and {Franceschi}, E. and {Frolov}, A. and {Galeotta}, S. and {Galli}, S. and {Ganga}, K. and {G{\'e}nova-Santos}, R.~T. and {Gerbino}, M. and {Ghosh}, T. and {Gonz{\'a}lez-Nuevo}, J. and {G{\'o}rski}, K.~M. and {Gratton}, S. and {Gruppuso}, A. and {Gudmundsson}, J.~E. and {Hamann}, J. and {Handley}, W. and {Hansen}, F.~K. and {Herranz}, D. and {Hildebrandt}, S.~R. and {Hivon}, E. and {Huang}, Z. and {Jaffe}, A.~H. and {Jones}, W.~C. and {Karakci}, A. and {Keih{\"a}nen}, E. and {Keskitalo}, R. and {Kiiveri}, K. and {Kim}, J. and {Kisner}, T.~S. and {Knox}, L. and {Krachmalnicoff}, N. and {Kunz}, M. and {Kurki-Suonio}, H. and {Lagache}, G. and {Lamarre}, J.-M. and {Lasenby}, A. and {Lattanzi}, M. and {Lawrence}, C.~R. and {Le Jeune}, M. and {Lemos}, P. and {Lesgourgues}, J. and {Levrier}, F. and {Lewis}, A. and {Liguori}, M. and {Lilje}, P.~B. and {Lilley}, M. and {Lindholm}, V. and {L{\'o}pez-Caniego}, M. and {Lubin}, P.~M. and {Ma}, Y.-Z. and {Mac{\'\i}as-P{\'e}rez}, J.~F. and {Maggio}, G. and {Maino}, D. and {Mandolesi}, N. and {Mangilli}, A. and {Marcos-Caballero}, A. and {Maris}, M. and {Martin}, P.~G. and {Martinelli}, M. and {Mart{\'\i}nez-Gonz{\'a}lez}, E. and {Matarrese}, S. and {Mauri}, N. and {McEwen}, J.~D. and {Meinhold}, P.~R. and {Melchiorri}, A. and {Mennella}, A. and {Migliaccio}, M. and {Millea}, M. and {Mitra}, S. and {Miville-Desch{\^e}nes}, M.-A. and {Molinari}, D. and {Montier}, L. and {Morgante}, G. and {Moss}, A. and {Natoli}, P. and {N{\o}rgaard-Nielsen}, H.~U. and {Pagano}, L. and {Paoletti}, D. and {Partridge}, B. and {Patanchon}, G. and {Peiris}, H.~V. and {Perrotta}, F. and {Pettorino}, V. and {Piacentini}, F. and {Polastri}, L. and {Polenta}, G. and {Puget}, J.-L. and {Rachen}, J.~P. and {Reinecke}, M. and {Remazeilles}, M. and {Renzi}, A. and {Rocha}, G. and {Rosset}, C. and {Roudier}, G. and {Rubi{\~n}o-Mart{\'\i}n}, J.~A. and {Ruiz-Granados}, B. and {Salvati}, L. and {Sandri}, M. and {Savelainen}, M. and {Scott}, D. and {Shellard}, E.~P.~S. and {Sirignano}, C. and {Sirri}, G. and {Spencer}, L.~D. and {Sunyaev}, R. and {Suur-Uski}, A.-S. and {Tauber}, J.~A. and {Tavagnacco}, D. and {Tenti}, M. and {Toffolatti}, L. and {Tomasi}, M. and {Trombetti}, T. and {Valenziano}, L. and {Valiviita}, J. and {Van Tent}, B. and {Vibert}, L. and {Vielva}, P. and {Villa}, F. and {Vittorio}, N. and {Wandelt}, B.~D. and {Wehus}, I.~K. and {White}, M. and {White}, S.~D.~M. and {Zacchei}, A. and {Zonca}, A.},
        title = "{Planck 2018 results. VI. Cosmological parameters}",
      journal = {\aap},
         year = 2020,
        month = sep,
       volume = {641},
          eid = {A6},
        pages = {A6},
          doi = {10.1051/0004-6361/201833910},
archivePrefix = {arXiv},
       eprint = {1807.06209},
 primaryClass = {astro-ph.CO},
       adsurl = {https://ui.adsabs.harvard.edu/abs/2020A&A...641A...6P}
}
\end{multicols*}
\begingroup
\scriptsize
\setlength{\tabcolsep}{1pt}
\renewcommand{\arraystretch}{1.05}
\setlength{\LTcapwidth}{\textwidth}
\begin{longtable}{@{}cccccccccccccccccc@{}}
\caption{Radio properties of the 2333 giant radio galaxies (GRGs) identified from LoTSS-DR3. Columns list the sequential catalogue number; source name; right ascension (RA) and declination (Dec); best redshift ($z_{\rm best}$) and its uncertainty ($\sigma_z$); redshift reference (Ref.); largest angular size (LAS); projected linear size ($D$); integrated 150-MHz flux density and its uncertainty ($S_{150}\pm\sigma_{S_{150}}$); 150-MHz radio power ($P_{150}$); equipartition magnetic field ($B_{\rm eq}$), total energy ($E_{\rm eq}$), energy density ($U_{\rm eq}$), and pressure ($P_{\rm eq}$); jet power ($Q_{\rm jet}$); logarithmic black-hole mass [$\log(M_{\rm BH}/M_\odot)$]; and its uncertainty ($\sigma_{\log M_{\rm BH}}$). Units are indicated in the respective column headings. Radio power and equipartition quantities are expressed in scaled SI units. The uncertainty in logarithmic black-hole mass is given in dex. Missing measurements are indicated by --.}
\label{tab:grg_properties}\\
\toprule
No. & Source & RA & Dec & $z_{\rm best}$ & $\sigma_z$ & Ref. & LAS & $D$ & $S_{150}\pm\sigma_{S_{150}}$ & $P_{150}$ & $B_{\rm eq}$ & $E_{\rm eq}$ & $U_{\rm eq}$ & $P_{\rm eq}$ & $Q_{\rm jet}$ & $\log(M_{\rm BH}/M_\odot)$ & $\sigma_{\log M_{\rm BH}}$ \\
 &  & (deg) & (deg) &  &  &  & (arcmin) & (kpc) & (mJy) & ($10^{26}$ W Hz$^{-1}$) & ($10^{-10}$ T) & ($10^{52}$ J) & ($10^{-14}$ J m$^{-3}$) & ($10^{-14}$ Pa) & ($10^{37}$ W) &  & (dex) \\
\midrule
\endfirsthead
\multicolumn{18}{c}{\tablename\ \thetable{} -- continued from previous page}\\
\toprule
No. & Source & RA & Dec & $z_{\rm best}$ & $\sigma_z$ & Ref. & LAS & $D$ & $S_{150}\pm\sigma_{S_{150}}$ & $P_{150}$ & $B_{\rm eq}$ & $E_{\rm eq}$ & $U_{\rm eq}$ & $P_{\rm eq}$ & $Q_{\rm jet}$ & $\log(M_{\rm BH}/M_\odot)$ & $\sigma_{\log M_{\rm BH}}$ \\
 &  & (deg) & (deg) &  &  &  & (arcmin) & (kpc) & (mJy) & ($10^{26}$ W Hz$^{-1}$) & ($10^{-10}$ T) & ($10^{52}$ J) & ($10^{-14}$ J m$^{-3}$) & ($10^{-14}$ Pa) & ($10^{37}$ W) &  & (dex) \\
(1) &(2) &(3) &(4) &(5) &(6) &(7) &(8) &(9) &(10) &(11) &(12) &(13) &(14) &(15) &(16) &(17) &(18) \\
\midrule
\endhead
\midrule
\multicolumn{18}{r}{Continued on next page}\\
\endfoot
\bottomrule
\endlastfoot
1 & J000005+243658 & 0.02036 & 24.61622 & 0.866 & 0.321 & DESI & 1.82 & 865.6 & $222.11\pm1.25$ & 7.36 & 4.84 & 2.90 & 18.64 & 6.21 & 11.07 & $-$ & $-$ \\
2 & J000024+350027 & 0.09809 & 35.00740 & 0.586 & 0.000 & SDSS & 2.37 & 966.7 & $43.14\pm1.43$ & 0.57 & 1.81 & 1.33 & 2.61 & 0.87 & 1.30 & 9.19 & 0.24 \\
3 & J000111+042916 & 0.29593 & 4.48778 & 0.334 & 0.034 & DESI & 3.85 & 1142.6 & $143.62\pm3.68$ & 0.53 & 1.43 & 1.86 & 1.63 & 0.54 & 1.13 & $-$ & $-$ \\
4 & J000115+082029 & 0.31242 & 8.34142 & 0.544 & 0.000 & SDSS & 2.21 & 872.0 & $2727.35\pm4.20$ & 30.54 & 7.59 & 5.78 & 45.81 & 15.27 & 28.72 & 8.96 & 0.07 \\
5 & J000124-021247 & 0.34912 & -2.21316 & 0.887 & 0.300 & DESI & 1.99 & 952.8 & $106.50\pm2.39$ & 3.73 & 5.27 & 1.27 & 22.12 & 7.37 & 6.88 & $-$ & $-$ \\
6 & J000152+010222 & 0.46466 & 1.03933 & 1.315 & 0.001 & SDSS & 2.39 & 1233.6 & $29.22\pm1.47$ & 2.57 & 4.24 & 1.28 & 14.31 & 4.77 & 7.27 & 8.87 & 0.13 \\
7 & J000141+183508 & 0.42191 & 18.58548 & 0.902 & 0.064 & DESI & 1.89 & 909.6 & $18.59\pm0.73$ & 0.68 & 2.54 & 0.85 & 5.14 & 1.71 & 1.72 & $-$ & $-$ \\
8 & J000329+011607 & 0.86936 & 1.26857 & 1.259 & 0.092 & DESI & 2.47 & 1268.7 & $28.19\pm0.92$ & 2.24 & 6.86 & 0.48 & 37.45 & 12.48 & 6.46 & $-$ & $-$ \\
\end{longtable}
\endgroup
\begingroup
\scriptsize
\setlength{\tabcolsep}{3pt}
\renewcommand{\arraystretch}{1.05}
\setlength{\LTcapwidth}{\textwidth}
\begin{longtable}{@{}ccccccc@{}}
\caption{Stellar and cluster-scale properties of the brightest cluster galaxies (BCGs) associated with the giant radio galaxy (GRG) sample. Columns list the sequential catalogue number; name of the brightest cluster galaxy (BCG Name); logarithmic stellar mass of the BCG [$\log(M_\star/M_\odot)$]; radius of the cluster within which the mean enclosed density is 500 times the critical density of the Universe ($r_{500}$); corresponding cluster mass ($M_{500}$); cluster richness within $r_{500}$ ($\lambda_{500}$); and the corresponding cluster reference (Ref.). Units are indicated in the respective column headings. Cluster mass is expressed in scaled solar-mass units. Missing measurements are indicated by --.}
\label{tab:cluster_properties}\\
\toprule
No. & BCG Name & $\log(M_\star/M_\odot)$ & $r_{500}$ & $M_{500}$ & $\lambda_{500}$ & Ref. \\
 &  & (dex) & (Mpc) & ($10^{14}~M_{\odot}$) &  &  \\
\midrule
\endfirsthead
\multicolumn{7}{c}{\tablename\ \thetable{} -- continued from previous page}\\
\toprule
No. & BCG Name & $\log(M_\star/M_\odot)$ & $r_{500}$ & $M_{500}$ & $\lambda_{500}$ & Ref. \\
 &  & (dex) & (Mpc) & ($10^{14}~M_{\odot}$) &  &  \\
\midrule
\endhead
\midrule
\multicolumn{7}{r}{Continued on next page}\\
\endfoot
\bottomrule
\endlastfoot
1 & J000024+350027 & $-$ & 0.75 & 1.99 & 38.09 & WHL15 \\
2 & J012833+150716 & $-$ & 0.69 & 1.25 & 23.22 & WHL15 \\
3 & J091423+002857 & $-$ & 0.58 & 0.55 & 9.75 & WHL15 \\
4 & J123752+445248 & $-$ & 0.88 & 2.87 & 56.25 & WHL15 \\
5 & J141556+104805 & $-$ & 0.60 & 1.02 & 18.74 & WHL15 \\
6 & J150034+422018 & $-$ & 0.83 & 2.37 & 45.90 & WHL15 \\
7 & J153645+070528 & $-$ & 0.52 & 0.59 & 10.39 & WHL15 \\
8 & J163434+244206 & $-$ & 0.66 & 1.32 & 24.57 & WHL15 \\
9 & J205911+104200 & $-$ & 0.60 & 0.64 & 11.33 & WHL15 \\
10 & J215655+185609 & $-$ & 0.70 & 1.37 & 25.65 & WHL15 \\
11 & J220052+105638 & $-$ & 0.55 & 0.43 & 7.52 & WHL15 \\
12 & J014650+003741 & $-$ & 0.78 & 2.68 & 61.51 & WH22 \\
13 & J233724-005811 & $-$ & 0.55 & 1.20 & 26.58 & WH22 \\
14 & J000141+183508 & 11.89 & 0.63 & 1.69 & 37.71 & WH24 \\
15 & J000754+025517 & 11.58 & 0.46 & 0.69 & 14.84 & WH24 \\
16 & J001144+015758 & 11.74 & 0.73 & 2.15 & 48.15 & WH24 \\
17 & J001200+111630 & 11.83 & 0.55 & 1.33 & 29.40 & WH24 \\
18 & J001854+031700 & 11.40 & 0.46 & 0.83 & 17.99 & WH24 \\
19 & J002243+103540 & 11.96 & 0.65 & 1.45 & 32.18 & WH24 \\
20 & J002612+124943 & 11.73 & 0.58 & 0.73 & 15.88 & WH24 \\
21 & J002620+150551 & 11.32 & 0.54 & 0.50 & 10.78 & WH24 \\
22 & J002839+003540 & 11.68 & 0.63 & 0.88 & 19.16 & WH24 \\
23 & J003250+011153 & 11.53 & 0.53 & 0.63 & 13.62 & WH24 \\
24 & J003248-001937 & 11.12 & 0.43 & 0.53 & 11.40 & WH24 \\
25 & J003511+013345 & 11.41 & 0.51 & 0.84 & 18.28 & WH24 \\
26 & J003448+090419 & 12.25 & 0.76 & 2.45 & 55.18 & WH24 \\
27 & J003623+105111 & 11.37 & 0.49 & 0.62 & 13.42 & WH24 \\
28 & J003550+120508 & 11.60 & 0.49 & 0.77 & 16.67 & WH24 \\
29 & J003711+014817 & 11.38 & 0.46 & 0.87 & 18.91 & WH24 \\
30 & J003913+002320 & 11.24 & 0.38 & 0.55 & 11.82 & WH24 \\
31 & J003904+083509 & 11.33 & 0.45 & 0.60 & 12.96 & WH24 \\
32 & J003936+121652 & 11.68 & 0.53 & 0.78 & 16.92 & WH24 \\
33 & J003941-000526 & 11.82 & 0.62 & 1.08 & 23.69 & WH24 \\
34 & J004545+001646 & 11.36 & 0.53 & 0.68 & 14.68 & WH24 \\
35 & J004558+100346 & 11.68 & 0.55 & 0.76 & 16.47 & WH24 \\
36 & J004746+140850 & 11.65 & 0.63 & 1.25 & 27.57 & WH24 \\
37 & J005016+012723 & 11.26 & 0.52 & 0.81 & 17.53 & WH24 \\
38 & J005237+223548 & 11.73 & 0.68 & 1.24 & 27.37 & WH24 \\
39 & J005450+060644 & 11.40 & 0.43 & 0.50 & 10.73 & WH24 \\
40 & J010409+075422 & 11.76 & 0.55 & 1.20 & 26.47 & WH24 \\
41 & J010358+095211 & 11.96 & 0.68 & 1.48 & 32.81 & WH24 \\
42 & J010451+101828 & 11.59 & 0.54 & 0.83 & 18.11 & WH24 \\
43 & J010432-004021 & 11.68 & 0.66 & 1.05 & 23.10 & WH24 \\
44 & J010825-015449 & 11.65 & 0.47 & 0.75 & 16.36 & WH24 \\
45 & J010908+090206 & 11.62 & 0.50 & 0.80 & 17.34 & WH24 \\
46 & J011015+094116 & 11.47 & 0.54 & 0.98 & 21.54 & WH24 \\
47 & J011318+111423 & 12.08 & 0.79 & 1.58 & 35.06 & WH24 \\
48 & J011318+144450 & 11.42 & 0.56 & 0.70 & 15.11 & WH24 \\
49 & J011811+131507 & 11.94 & 0.80 & 1.78 & 39.67 & WH24 \\
50 & J012013-003838 & 11.45 & 0.80 & 1.69 & 37.58 & WH24 \\
51 & J012342-012023 & 11.71 & 0.46 & 1.02 & 22.44 & WH24 \\
52 & J012749-014105 & 11.45 & 0.60 & 0.74 & 16.14 & WH24 \\
53 & J012906+065616 & 11.86 & 0.58 & 1.20 & 26.38 & WH24 \\
54 & J012955+071243 & 11.18 & 0.47 & 0.50 & 10.82 & WH24 \\
55 & J013400+220709 & 11.54 & 0.50 & 0.59 & 12.80 & WH24 \\
56 & J013640+241240 & 11.56 & 0.47 & 0.72 & 15.53 & WH24 \\
57 & J014015+034836 & 11.47 & 0.45 & 0.74 & 16.08 & WH24 \\
58 & J014112+113854 & 11.68 & 0.54 & 0.69 & 14.89 & WH24 \\
59 & J014153+122108 & 11.32 & 0.56 & 0.70 & 15.23 & WH24 \\
60 & J014210+262132 & 11.47 & 0.45 & 0.66 & 14.32 & WH24 \\
61 & J014418+145455 & 11.83 & 0.60 & 0.91 & 19.81 & WH24 \\
62 & J014520+155232 & 11.63 & 0.63 & 1.12 & 24.58 & WH24 \\
63 & J014549+100913 & 11.66 & 0.55 & 0.74 & 16.13 & WH24 \\
64 & J014536+142846 & 12.12 & 0.90 & 2.75 & 62.20 & WH24 \\
65 & J014741+014253 & 11.60 & 0.62 & 0.77 & 16.81 & WH24 \\
66 & J014755+110649 & 11.52 & 0.45 & 0.73 & 15.83 & WH24 \\
67 & J014818+135652 & 11.68 & 0.61 & 0.96 & 21.09 & WH24 \\
68 & J014941-015215 & 11.83 & 0.55 & 1.00 & 21.85 & WH24 \\
69 & J015313+061213 & 11.78 & 0.63 & 1.02 & 22.25 & WH24 \\
70 & J015433+173729 & 11.45 & 0.52 & 0.55 & 11.84 & WH24 \\
71 & J015623+002437 & 11.74 & 0.71 & 1.21 & 26.65 & WH24 \\
72 & J015714+101307 & 11.75 & 0.53 & 0.95 & 20.68 & WH24 \\
73 & J015656+132530 & 11.42 & 0.51 & 0.91 & 19.97 & WH24 \\
74 & J015716+143926 & 11.64 & 0.80 & 2.53 & 56.88 & WH24 \\
75 & J020603+084650 & 11.41 & 0.68 & 0.49 & 10.48 & WH24 \\
76 & J020958-010057 & 11.37 & 0.53 & 0.84 & 18.21 & WH24 \\
77 & J021128+012555 & 11.74 & 0.61 & 1.28 & 28.31 & WH24 \\
78 & J021636+123101 & 11.70 & 0.59 & 0.89 & 19.42 & WH24 \\
79 & J022034+295859 & 11.54 & 0.50 & 0.58 & 12.49 & WH24 \\
80 & J022159+082705 & 11.80 & 0.73 & 1.28 & 28.16 & WH24 \\
81 & J022131+115127 & 11.61 & 0.66 & 1.52 & 33.60 & WH24 \\
82 & J022344+225030 & 11.67 & 0.52 & 1.17 & 25.83 & WH24 \\
83 & J022649+230905 & 11.77 & 0.56 & 1.37 & 30.35 & WH24 \\
84 & J023016+205939 & 11.60 & 0.51 & 0.70 & 15.14 & WH24 \\
85 & J023035+125450 & 11.67 & 0.55 & 1.09 & 23.94 & WH24 \\
86 & J023810+093529 & 11.73 & 0.54 & 0.83 & 18.11 & WH24 \\
87 & J024000+101825 & 11.82 & 0.84 & 2.15 & 48.14 & WH24 \\
88 & J024243+070214 & 11.55 & 0.58 & 0.80 & 17.33 & WH24 \\
89 & J030218+061700 & 11.25 & 0.54 & 0.97 & 21.29 & WH24 \\
90 & J032637+023252 & 11.74 & 0.61 & 0.81 & 17.69 & WH24 \\
91 & J043007-015732 & 11.81 & 0.69 & 1.23 & 26.99 & WH24 \\
92 & J061958+443732 & 11.81 & 0.58 & 1.07 & 23.51 & WH24 \\
93 & J062918+560423 & 11.29 & 0.61 & 0.81 & 17.60 & WH24 \\
94 & J063755+533305 & 11.76 & 0.74 & 1.57 & 34.74 & WH24 \\
95 & J064240+373604 & 11.28 & 0.67 & 1.22 & 26.96 & WH24 \\
96 & J064332+634032 & 11.56 & 0.93 & 2.96 & 66.92 & WH24 \\
97 & J064547+775612 & 11.45 & 0.51 & 0.88 & 19.24 & WH24 \\
98 & J065229+375708 & 11.83 & 0.83 & 2.38 & 53.56 & WH24 \\
99 & J065536+324848 & 11.33 & 0.52 & 0.50 & 10.70 & WH24 \\
100 & J070058+383948 & 11.74 & 0.67 & 1.74 & 38.75 & WH24 \\
101 & J070103+642435 & 12.00 & 0.65 & 1.83 & 40.74 & WH24 \\
102 & J070831+623526 & 11.55 & 0.58 & 0.75 & 16.38 & WH24 \\
103 & J071004+703748 & 11.50 & 0.46 & 0.79 & 17.15 & WH24 \\
104 & J071051+292435 & 11.88 & 0.71 & 1.22 & 26.97 & WH24 \\
105 & J071800+504426 & 11.68 & 0.54 & 0.95 & 20.80 & WH24 \\
106 & J072021+533729 & 11.41 & 0.62 & 1.58 & 35.15 & WH24 \\
107 & J072631+213923 & 11.78 & 0.69 & 1.13 & 24.78 & WH24 \\
108 & J073052+580217 & 11.55 & 0.53 & 0.64 & 13.71 & WH24 \\
109 & J073226+224058 & 11.21 & 0.56 & 0.57 & 12.30 & WH24 \\
110 & J073313+421157 & 11.74 & 0.61 & 1.10 & 24.12 & WH24 \\
111 & J073309+421624 & 11.73 & 0.64 & 1.10 & 24.07 & WH24 \\
112 & J073324+555953 & 11.85 & 0.61 & 1.15 & 25.35 & WH24 \\
113 & J073544+495718 & 11.46 & 0.59 & 1.06 & 23.18 & WH24 \\
114 & J073728+233113 & 11.53 & 0.50 & 0.58 & 12.48 & WH24 \\
115 & J073850+493751 & 11.96 & 0.75 & 1.96 & 43.85 & WH24 \\
116 & J073947+205421 & 11.68 & 0.62 & 1.41 & 31.14 & WH24 \\
117 & J074101+540736 & 11.66 & 0.52 & 0.99 & 21.66 & WH24 \\
118 & J074142+525458 & 11.65 & 0.57 & 1.25 & 27.51 & WH24 \\
119 & J074758+621954 & 11.83 & 0.56 & 1.13 & 24.72 & WH24 \\
120 & J074742+680617 & 11.54 & 0.49 & 1.09 & 23.86 & WH24 \\
121 & J074937+512823 & 11.69 & 0.65 & 0.88 & 19.13 & WH24 \\
122 & J074952+682708 & 11.40 & 0.51 & 1.16 & 25.44 & WH24 \\
123 & J075039+182629 & 11.57 & 0.50 & 0.83 & 18.07 & WH24 \\
124 & J075254+105607 & 11.65 & 0.57 & 0.97 & 21.23 & WH24 \\
125 & J075623+795617 & 11.47 & 0.42 & 0.52 & 11.26 & WH24 \\
126 & J075704+485106 & 11.59 & 0.43 & 0.76 & 16.42 & WH24 \\
127 & J075753+225922 & 11.84 & 0.62 & 1.36 & 29.97 & WH24 \\
128 & J075830+052642 & 11.85 & 0.84 & 2.18 & 48.83 & WH24 \\
129 & J080216+734256 & 11.53 & 0.50 & 0.61 & 13.05 & WH24 \\
130 & J080242+131715 & 11.66 & 0.46 & 0.94 & 20.53 & WH24 \\
131 & J080245+223118 & 11.62 & 0.57 & 1.24 & 27.41 & WH24 \\
132 & J080448+081254 & 11.73 & 0.57 & 0.83 & 18.04 & WH24 \\
133 & J080705+202841 & 11.99 & 0.63 & 2.00 & 44.62 & WH24 \\
134 & J080735+200753 & 11.73 & 0.52 & 1.05 & 23.07 & WH24 \\
135 & J080946+725251 & 11.50 & 0.56 & 0.92 & 20.09 & WH24 \\
136 & J081123+105430 & 11.83 & 0.76 & 1.41 & 31.09 & WH24 \\
137 & J081440+214606 & 11.70 & 0.54 & 0.88 & 19.19 & WH24 \\
138 & J081710+251328 & 11.94 & 0.68 & 2.09 & 46.72 & WH24 \\
139 & J081832+160636 & 11.74 & 0.56 & 1.08 & 23.72 & WH24 \\
140 & J082129+084728 & 11.63 & 0.62 & 0.69 & 14.91 & WH24 \\
141 & J082529+011435 & 11.32 & 0.46 & 0.47 & 10.09 & WH24 \\
142 & J082525+202456 & 11.89 & 0.56 & 1.25 & 27.46 & WH24 \\
143 & J082540+205655 & 11.52 & 0.64 & 1.64 & 36.34 & WH24 \\
144 & J082926+210444 & 11.59 & 0.61 & 1.59 & 35.23 & WH24 \\
145 & J082834+383822 & 11.38 & 0.47 & 0.68 & 14.76 & WH24 \\
146 & J083113+145903 & 11.95 & 0.81 & 1.86 & 41.52 & WH24 \\
147 & J083449+173653 & 11.50 & 0.50 & 0.70 & 15.13 & WH24 \\
148 & J083723+073832 & 11.68 & 0.56 & 1.15 & 25.22 & WH24 \\
149 & J083818+044707 & 11.73 & 0.50 & 0.79 & 17.26 & WH24 \\
150 & J084301+051326 & 11.55 & 0.48 & 0.55 & 11.83 & WH24 \\
151 & J084246+244357 & 12.01 & 1.02 & 4.81 & 110.50 & WH24 \\
152 & J084340+155653 & 11.62 & 0.66 & 1.20 & 26.38 & WH24 \\
153 & J084726+273833 & 11.65 & 0.57 & 1.05 & 23.00 & WH24 \\
154 & J085117+370429 & 11.84 & 0.69 & 1.18 & 25.89 & WH24 \\
155 & J085138+144414 & 11.53 & 0.50 & 0.88 & 19.22 & WH24 \\
156 & J085213+744813 & 11.67 & 0.64 & 1.36 & 30.10 & WH24 \\
157 & J085242+055449 & 11.56 & 0.63 & 0.97 & 21.30 & WH24 \\
158 & J085232+242345 & 11.42 & 0.43 & 0.52 & 11.06 & WH24 \\
159 & J085438+270720 & 11.67 & 0.58 & 0.86 & 18.75 & WH24 \\
160 & J085554+121325 & 11.58 & 0.56 & 0.93 & 20.33 & WH24 \\
161 & J085828+030428 & 11.81 & 0.64 & 1.32 & 29.03 & WH24 \\
162 & J085927+185705 & 11.73 & 0.68 & 1.04 & 22.81 & WH24 \\
163 & J090220+095034 & 11.93 & 0.62 & 1.66 & 36.96 & WH24 \\
164 & J090206+124425 & 11.55 & 0.57 & 0.76 & 16.38 & WH24 \\
165 & J091221+223420 & 11.67 & 0.57 & 1.08 & 23.77 & WH24 \\
166 & J091238+371848 & 11.76 & 0.52 & 1.14 & 25.00 & WH24 \\
167 & J091456+242245 & 11.60 & 0.57 & 0.67 & 14.47 & WH24 \\
168 & J091943+091502 & 11.91 & 0.74 & 1.72 & 38.19 & WH24 \\
169 & J092209+323634 & 11.77 & 0.62 & 1.12 & 24.66 & WH24 \\
170 & J094006+735911 & 11.56 & 0.80 & 1.51 & 33.44 & WH24 \\
171 & J094104+203400 & 11.74 & 0.58 & 0.99 & 21.66 & WH24 \\
172 & J094230+144521 & 11.29 & 0.46 & 0.75 & 16.21 & WH24 \\
173 & J094317+062711 & 11.83 & 0.58 & 1.26 & 27.70 & WH24 \\
174 & J094254+074446 & 11.11 & 0.48 & 0.56 & 12.10 & WH24 \\
175 & J094311+125702 & 11.62 & 0.56 & 1.03 & 22.65 & WH24 \\
176 & J094252+251355 & 11.67 & 0.48 & 0.97 & 21.14 & WH24 \\
177 & J094443+024754 & 11.95 & 1.22 & 5.55 & 128.10 & WH24 \\
178 & J094638+121932 & 11.73 & 0.67 & 2.15 & 48.17 & WH24 \\
179 & J094717+224239 & 11.71 & 0.50 & 1.18 & 25.99 & WH24 \\
180 & J094722+370723 & 11.66 & 0.60 & 1.11 & 24.37 & WH24 \\
181 & J094737+090741 & 11.65 & 0.71 & 1.08 & 23.59 & WH24 \\
182 & J095002+124900 & 11.64 & 0.55 & 0.65 & 14.10 & WH24 \\
183 & J095348+220334 & 11.53 & 0.42 & 0.71 & 15.44 & WH24 \\
184 & J095444+021437 & 11.15 & 0.51 & 0.80 & 17.41 & WH24 \\
185 & J095528+193727 & 11.97 & 0.79 & 2.21 & 49.64 & WH24 \\
186 & J095656+123039 & 11.68 & 0.65 & 1.16 & 25.43 & WH24 \\
187 & J095834+060015 & 11.63 & 0.44 & 0.58 & 12.54 & WH24 \\
188 & J095928+300547 & 11.51 & 0.59 & 0.81 & 17.57 & WH24 \\
189 & J100709+284919 & 11.25 & 0.49 & 0.51 & 10.83 & WH24 \\
190 & J100757+280148 & 11.74 & 0.67 & 0.87 & 18.91 & WH24 \\
191 & J101605+005522 & 11.84 & 0.65 & 0.97 & 21.32 & WH24 \\
192 & J101541+122030 & 11.75 & 0.69 & 1.81 & 40.37 & WH24 \\
193 & J101719+124300 & 11.68 & 0.55 & 0.77 & 16.69 & WH24 \\
194 & J101659+192221 & 11.92 & 0.53 & 1.07 & 23.52 & WH24 \\
195 & J101807+091252 & 11.60 & 0.54 & 0.74 & 15.98 & WH24 \\
196 & J102050+002728 & 11.50 & 0.73 & 2.05 & 45.90 & WH24 \\
197 & J102500+104433 & 11.79 & 0.66 & 1.23 & 27.12 & WH24 \\
198 & J102455+564724 & 11.82 & 0.57 & 0.92 & 20.02 & WH24 \\
199 & J102957+020218 & 11.61 & 0.66 & 0.93 & 20.32 & WH24 \\
200 & J102940+154102 & 11.69 & 0.51 & 1.00 & 21.94 & WH24 \\
201 & J102939+451600 & 11.60 & 0.66 & 1.57 & 34.79 & WH24 \\
202 & J102946+814834 & 11.23 & 0.46 & 0.62 & 13.44 & WH24 \\
203 & J103108+335312 & 11.41 & 0.54 & 0.70 & 15.09 & WH24 \\
204 & J103250+023829 & 11.53 & 0.69 & 1.07 & 23.39 & WH24 \\
205 & J103524+065138 & 11.42 & 0.47 & 0.75 & 16.33 & WH24 \\
206 & J103516+285531 & 11.69 & 0.54 & 1.12 & 24.50 & WH24 \\
207 & J103629+160222 & 11.58 & 0.60 & 0.81 & 17.55 & WH24 \\
208 & J103837+142916 & 11.26 & 0.45 & 0.60 & 12.92 & WH24 \\
209 & J103946+280057 & 11.51 & 0.54 & 0.79 & 17.12 & WH24 \\
210 & J103942+784603 & 11.60 & 0.54 & 0.97 & 21.11 & WH24 \\
211 & J104216+181714 & 11.46 & 0.66 & 1.56 & 34.54 & WH24 \\
212 & J104841+042736 & 11.18 & 0.41 & 0.61 & 13.09 & WH24 \\
213 & J105242+111434 & 11.45 & 0.62 & 0.72 & 15.50 & WH24 \\
214 & J105239+715944 & 11.82 & 0.66 & 1.18 & 25.92 & WH24 \\
215 & J105421+022755 & 11.60 & 0.49 & 0.70 & 15.18 & WH24 \\
216 & J105544+303607 & 11.53 & 0.45 & 0.61 & 13.24 & WH24 \\
217 & J105645+161825 & 11.69 & 0.62 & 0.92 & 20.00 & WH24 \\
218 & J110041+145838 & 11.70 & 0.50 & 0.89 & 19.36 & WH24 \\
219 & J110305+254731 & 11.54 & 0.49 & 0.64 & 13.72 & WH24 \\
220 & J110526+094350 & 11.87 & 0.70 & 1.68 & 37.38 & WH24 \\
221 & J110500+151400 & 11.61 & 0.68 & 1.28 & 28.19 & WH24 \\
222 & J111157+715521 & 11.44 & 0.44 & 0.73 & 15.79 & WH24 \\
223 & J111441-001211 & 11.41 & 0.50 & 0.55 & 11.73 & WH24 \\
224 & J111614+024609 & 11.94 & 0.67 & 1.90 & 42.51 & WH24 \\
225 & J111757+645711 & 11.67 & 0.79 & 2.06 & 46.11 & WH24 \\
226 & J112107+162722 & 11.95 & 0.57 & 1.38 & 30.52 & WH24 \\
227 & J112444+191243 & 11.55 & 0.60 & 1.33 & 29.31 & WH24 \\
228 & J112701+204219 & 11.80 & 0.68 & 1.31 & 29.02 & WH24 \\
229 & J113507+134650 & 11.64 & 0.71 & 1.82 & 40.54 & WH24 \\
230 & J113642+223244 & 11.75 & 0.87 & 2.24 & 50.19 & WH24 \\
231 & J113813+323718 & 12.13 & 0.89 & 3.34 & 75.99 & WH24 \\
232 & J113829+610016 & 11.37 & 0.49 & 0.50 & 10.78 & WH24 \\
233 & J114009+143347 & 11.80 & 0.53 & 1.20 & 26.38 & WH24 \\
234 & J114005+145042 & 11.70 & 0.51 & 0.85 & 18.62 & WH24 \\
235 & J114152+171249 & 11.46 & 0.50 & 0.65 & 14.03 & WH24 \\
236 & J114958+160732 & 11.55 & 0.64 & 1.18 & 25.90 & WH24 \\
237 & J115051+202207 & 12.11 & 0.86 & 2.87 & 65.01 & WH24 \\
238 & J115253+232124 & 11.51 & 0.49 & 0.51 & 10.84 & WH24 \\
239 & J115510+100332 & 11.73 & 0.62 & 1.03 & 22.58 & WH24 \\
240 & J115553+762201 & 11.98 & 0.70 & 1.28 & 28.19 & WH24 \\
241 & J115900+282613 & 11.53 & 0.46 & 0.52 & 11.26 & WH24 \\
242 & J120024+024630 & 11.47 & 0.56 & 1.20 & 26.44 & WH24 \\
243 & J115954-012108 & 11.57 & 0.48 & 0.62 & 13.30 & WH24 \\
244 & J121118+013727 & 11.63 & 0.51 & 0.97 & 21.17 & WH24 \\
245 & J121344+404004 & 11.89 & 0.68 & 1.29 & 28.36 & WH24 \\
246 & J121434+004728 & 11.90 & 0.83 & 1.45 & 32.03 & WH24 \\
247 & J121613+274142 & 11.76 & 0.55 & 0.89 & 19.48 & WH24 \\
248 & J121814+203943 & 11.92 & 0.64 & 1.50 & 33.26 & WH24 \\
249 & J122708+245156 & 11.78 & 0.52 & 1.16 & 25.51 & WH24 \\
250 & J123614+501939 & 11.65 & 0.57 & 0.67 & 14.51 & WH24 \\
251 & J123717+425604 & 11.55 & 0.46 & 0.53 & 11.28 & WH24 \\
252 & J124810+510804 & 11.90 & 0.66 & 1.40 & 30.95 & WH24 \\
253 & J125655+021905 & 11.35 & 0.40 & 0.47 & 10.15 & WH24 \\
254 & J125638+351638 & 11.75 & 0.58 & 1.09 & 23.99 & WH24 \\
255 & J125814+552756 & 11.39 & 0.56 & 0.58 & 12.56 & WH24 \\
256 & J125920+801424 & 11.54 & 0.50 & 0.58 & 12.40 & WH24 \\
257 & J130404+205236 & 11.98 & 0.81 & 2.04 & 45.61 & WH24 \\
258 & J130344+213519 & 11.76 & 0.60 & 1.07 & 23.44 & WH24 \\
259 & J130332+522002 & 11.69 & 0.54 & 0.70 & 15.23 & WH24 \\
260 & J130632+113633 & 11.47 & 0.49 & 0.80 & 17.45 & WH24 \\
261 & J130826+323014 & 11.61 & 0.58 & 0.96 & 21.07 & WH24 \\
262 & J131658+171630 & 11.40 & 0.43 & 0.53 & 11.47 & WH24 \\
263 & J132244+192748 & 11.87 & 0.60 & 0.98 & 21.46 & WH24 \\
264 & J132510+200106 & 11.81 & 0.67 & 1.11 & 24.32 & WH24 \\
265 & J132749+042156 & 11.32 & 0.51 & 0.67 & 14.57 & WH24 \\
266 & J133106+573019 & 11.45 & 0.67 & 1.20 & 26.43 & WH24 \\
267 & J133203+024757 & 11.44 & 0.49 & 0.67 & 14.40 & WH24 \\
268 & J133254+203054 & 11.44 & 0.61 & 1.01 & 22.13 & WH24 \\
269 & J134711+262600 & 11.42 & 0.42 & 0.52 & 11.06 & WH24 \\
270 & J135503+395102 & 11.15 & 0.52 & 0.76 & 16.54 & WH24 \\
271 & J140527+682355 & 11.58 & 0.59 & 1.12 & 24.49 & WH24 \\
272 & J140535+253006 & 11.43 & 0.47 & 0.54 & 11.55 & WH24 \\
273 & J140750+174600 & 11.49 & 0.54 & 0.70 & 15.20 & WH24 \\
274 & J140906+112740 & 11.62 & 0.55 & 0.99 & 21.67 & WH24 \\
275 & J142551-005443 & 11.48 & 0.53 & 0.79 & 17.21 & WH24 \\
276 & J142634-010942 & 11.49 & 0.51 & 0.61 & 13.12 & WH24 \\
277 & J142959+063034 & 11.73 & 0.50 & 0.90 & 19.60 & WH24 \\
278 & J143022+190821 & 12.14 & 0.97 & 3.16 & 71.72 & WH24 \\
279 & J143051+114409 & 11.78 & 0.62 & 0.96 & 21.04 & WH24 \\
280 & J144241+282703 & 11.76 & 0.56 & 1.03 & 22.54 & WH24 \\
281 & J144616+061018 & 11.81 & 0.69 & 1.29 & 28.50 & WH24 \\
282 & J144556+523010 & 11.47 & 0.55 & 1.03 & 22.59 & WH24 \\
283 & J145330+142112 & 11.61 & 0.72 & 1.18 & 26.07 & WH24 \\
284 & J145348+572645 & 11.80 & 0.67 & 1.47 & 32.66 & WH24 \\
285 & J145522+031355 & 11.53 & 0.71 & 2.07 & 46.33 & WH24 \\
286 & J145440+262129 & 11.77 & 0.84 & 2.01 & 44.99 & WH24 \\
287 & J145534+053516 & 11.44 & 0.47 & 0.79 & 17.08 & WH24 \\
288 & J145919+755854 & 11.76 & 0.68 & 1.83 & 40.90 & WH24 \\
289 & J150602+064449 & 11.98 & 0.63 & 1.19 & 26.29 & WH24 \\
290 & J150538+080320 & 11.80 & 0.68 & 1.47 & 32.67 & WH24 \\
291 & J150707+152020 & 11.52 & 0.57 & 0.93 & 20.32 & WH24 \\
292 & J150805+133540 & 11.52 & 0.60 & 1.29 & 28.42 & WH24 \\
293 & J150735+543055 & 11.49 & 0.63 & 0.82 & 17.82 & WH24 \\
294 & J150925+691744 & 11.95 & 0.63 & 1.28 & 28.32 & WH24 \\
295 & J151008+253308 & 11.53 & 0.70 & 0.71 & 15.36 & WH24 \\
296 & J151357+123908 & 11.71 & 0.74 & 1.73 & 38.45 & WH24 \\
297 & J151434+042515 & 11.57 & 0.50 & 0.64 & 13.73 & WH24 \\
298 & J151628+180637 & 11.97 & 0.74 & 1.79 & 39.81 & WH24 \\
299 & J152014+231056 & 11.48 & 0.56 & 1.15 & 25.20 & WH24 \\
300 & J152318+172149 & 11.55 & 0.51 & 0.68 & 14.60 & WH24 \\
301 & J152428+255756 & 11.66 & 0.55 & 0.76 & 16.50 & WH24 \\
302 & J152457+190940 & 11.71 & 0.59 & 0.88 & 19.10 & WH24 \\
303 & J152501+215707 & 11.75 & 0.63 & 0.88 & 19.08 & WH24 \\
304 & J153001+143509 & 11.76 & 0.56 & 1.38 & 30.45 & WH24 \\
305 & J153204+141333 & 11.65 & 0.52 & 0.74 & 16.11 & WH24 \\
306 & J153617+155838 & 11.69 & 0.60 & 1.08 & 23.76 & WH24 \\
307 & J154534+293344 & 11.45 & 0.58 & 0.68 & 14.76 & WH24 \\
308 & J154906+264331 & 11.50 & 0.56 & 1.28 & 28.21 & WH24 \\
309 & J155054+473450 & 11.56 & 0.49 & 0.63 & 13.68 & WH24 \\
310 & J155048+510045 & 11.60 & 0.56 & 0.77 & 16.71 & WH24 \\
311 & J155352+194756 & 11.49 & 0.51 & 0.88 & 19.18 & WH24 \\
312 & J155455+770726 & 11.67 & 0.78 & 1.74 & 38.78 & WH24 \\
313 & J160708+695016 & 11.30 & 0.54 & 0.99 & 21.75 & WH24 \\
314 & J161139+153255 & 11.71 & 0.72 & 1.34 & 29.55 & WH24 \\
315 & J161206+280303 & 11.71 & 0.60 & 0.72 & 15.67 & WH24 \\
316 & J161940+194454 & 11.29 & 0.40 & 0.52 & 11.25 & WH24 \\
317 & J161952+671224 & 11.43 & 0.55 & 0.90 & 19.60 & WH24 \\
318 & J162230+254431 & 11.73 & 0.60 & 1.07 & 23.38 & WH24 \\
319 & J163017+143511 & 11.94 & 0.69 & 1.30 & 28.70 & WH24 \\
320 & J163254+205655 & 11.70 & 0.59 & 1.03 & 22.57 & WH24 \\
321 & J163906+811511 & 11.38 & 0.50 & 0.61 & 13.12 & WH24 \\
322 & J164339+195534 & 11.69 & 0.75 & 1.34 & 29.52 & WH24 \\
323 & J165223+174034 & 11.64 & 0.59 & 0.91 & 19.91 & WH24 \\
324 & J165300+254722 & 11.45 & 0.51 & 0.64 & 13.89 & WH24 \\
325 & J165230+740828 & 11.95 & 0.90 & 2.61 & 58.79 & WH24 \\
326 & J165458+233333 & 11.57 & 0.53 & 0.57 & 12.26 & WH24 \\
327 & J170130+292044 & 11.88 & 0.82 & 2.01 & 44.91 & WH24 \\
328 & J170732+104951 & 11.52 & 0.76 & 3.09 & 70.03 & WH24 \\
329 & J171516+113318 & 11.47 & 0.45 & 0.69 & 14.93 & WH24 \\
330 & J172508+103312 & 11.92 & 0.76 & 1.60 & 35.51 & WH24 \\
331 & J172627+743643 & 11.64 & 0.55 & 0.53 & 11.48 & WH24 \\
332 & J173614+340254 & 11.61 & 0.57 & 0.74 & 15.94 & WH24 \\
333 & J173829+512243 & 12.06 & 0.66 & 1.60 & 35.64 & WH24 \\
334 & J174150+625912 & 11.64 & 0.51 & 0.92 & 20.13 & WH24 \\
335 & J175055+134857 & 11.70 & 0.70 & 1.22 & 26.77 & WH24 \\
336 & J175140+255401 & 11.71 & 0.56 & 1.24 & 27.25 & WH24 \\
337 & J175149+310316 & 11.62 & 0.58 & 0.72 & 15.58 & WH24 \\
338 & J175451+544408 & 11.63 & 0.53 & 0.69 & 14.91 & WH24 \\
339 & J175807+210242 & 11.86 & 0.59 & 1.06 & 23.17 & WH24 \\
340 & J175925+261116 & 11.29 & 0.37 & 0.48 & 10.25 & WH24 \\
341 & J175948+425400 & 11.39 & 0.40 & 0.71 & 15.44 & WH24 \\
342 & J180242+703040 & 11.69 & 0.60 & 0.93 & 20.21 & WH24 \\
343 & J181408+502609 & 11.46 & 0.50 & 0.56 & 11.98 & WH24 \\
344 & J181922+543447 & 11.52 & 0.63 & 0.97 & 21.18 & WH24 \\
345 & J182944+502407 & 11.95 & 0.62 & 1.43 & 31.74 & WH24 \\
346 & J183154+395529 & 11.81 & 0.61 & 0.90 & 19.72 & WH24 \\
347 & J183415+533632 & 11.66 & 0.64 & 1.23 & 27.18 & WH24 \\
348 & J184904+692927 & 11.48 & 0.69 & 0.50 & 10.78 & WH24 \\
349 & J190906+540428 & 11.66 & 0.57 & 0.98 & 21.45 & WH24 \\
350 & J191148+713152 & 11.53 & 0.71 & 0.74 & 16.05 & WH24 \\
351 & J192438+531325 & 11.68 & 0.74 & 1.62 & 35.89 & WH24 \\
352 & J193146+575728 & 11.81 & 0.64 & 0.86 & 18.84 & WH24 \\
353 & J204357-010640 & 11.47 & 0.48 & 0.96 & 21.02 & WH24 \\
354 & J213906+015842 & 11.41 & 0.42 & 0.77 & 16.82 & WH24 \\
355 & J214836+133826 & 11.44 & 0.37 & 0.61 & 13.16 & WH24 \\
356 & J214925+190749 & 11.77 & 0.58 & 0.86 & 18.72 & WH24 \\
357 & J215740+034510 & 11.46 & 0.45 & 0.80 & 17.32 & WH24 \\
358 & J220023-021800 & 11.53 & 0.50 & 0.89 & 19.35 & WH24 \\
359 & J220853+110601 & 11.61 & 0.50 & 0.76 & 16.50 & WH24 \\
360 & J221521+030339 & 11.49 & 0.55 & 0.54 & 11.56 & WH24 \\
361 & J221601+012500 & 11.65 & 0.78 & 2.22 & 49.76 & WH24 \\
362 & J222056+001820 & 11.41 & 0.53 & 0.48 & 10.25 & WH24 \\
363 & J222210-004421 & 11.92 & 0.82 & 4.19 & 95.90 & WH24 \\
364 & J222819+180021 & 11.51 & 0.58 & 0.83 & 18.10 & WH24 \\
365 & J222746-013639 & 11.72 & 0.45 & 0.93 & 20.26 & WH24 \\
366 & J223550+030615 & 11.70 & 0.70 & 1.17 & 25.69 & WH24 \\
367 & J223713+321544 & 11.26 & 0.54 & 1.07 & 23.38 & WH24 \\
368 & J224339+111341 & 11.60 & 0.45 & 0.77 & 16.67 & WH24 \\
369 & J224833+214006 & 11.77 & 0.60 & 0.78 & 17.05 & WH24 \\
370 & J231627+061746 & 12.04 & 0.72 & 1.60 & 35.58 & WH24 \\
371 & J231824+014923 & 11.78 & 0.59 & 0.99 & 21.67 & WH24 \\
372 & J232151+114306 & 11.50 & 0.44 & 0.58 & 12.48 & WH24 \\
373 & J232352+063738 & 11.38 & 0.50 & 0.48 & 10.18 & WH24 \\
374 & J232706+344222 & 11.20 & 0.43 & 0.76 & 16.52 & WH24 \\
375 & J233243+055907 & 11.86 & 0.52 & 1.17 & 25.82 & WH24 \\
376 & J233720+154930 & 12.04 & 0.75 & 1.75 & 38.98 & WH24 \\
377 & J234239+155413 & 11.73 & 0.57 & 1.02 & 22.30 & WH24 \\
378 & J234610+015904 & 11.86 & 0.54 & 1.00 & 21.97 & WH24 \\
379 & J234713-001846 & 11.83 & 0.53 & 1.08 & 23.77 & WH24 \\
380 & J234741-004100 & 11.32 & 0.41 & 0.62 & 13.40 & WH24 \\
381 & J235008+142851 & 11.62 & 0.60 & 0.89 & 19.49 & WH24 \\
382 & J235138+250711 & 11.75 & 0.69 & 1.00 & 21.99 & WH24 \\
\end{longtable}
\endgroup
\end{doublespacing}
\end{document}